\documentclass[10pt,preprint]{aastex}

\usepackage{mathtools}
\usepackage{amsmath}
\usepackage{relsize}
\usepackage{booktabs}
\usepackage{comment}
\usepackage{natbib}
\usepackage[caption=false]{subfig}
\usepackage{wrapfig}
\usepackage{lscape}
\usepackage{rotating}
\usepackage{subfig}
\usepackage{alphalph}

\usepackage[usenames,dvipsnames]{xcolor}
\usepackage{color}

\shorttitle{Modeling radio foregrounds for EoR}
\shortauthors{Sathyanaraynara Rao et al.}

\begin{document}

\title{Modeling the Radio Foreground for detection of CMB spectral distortions from Cosmic Dawn and Epoch of Reionization}

\author{Mayuri Sathyanarayana Rao$^{1,2}$, Ravi Subrahmanyan$^{1}$, N Udaya Shankar$^{1}$, Jens Chluba$^{3}$}
\affil{{\small $^{1}$Raman Research Institute, C V Raman Avenue, Sadashivanagar, Bangalore 560080, India}}
\affil{{\small $^{2}$Australian National University, Research School for Astronomy \& Astrophysics, Mount Stromlo Observatory, Cotter Road, Weston, ACT 2611, Australia}}
\affil{{\small $^{3}$Jodrell Bank Centre for Astrophysics, University of Manchester, Oxford Road, M13 9PL, U.K.}}
\email{Email of corresponding author: mayuris@rri.res.in}

\begin{abstract}
Cosmic baryon evolution during the Cosmic Dawn and Reionization results in redshifted 21-cm spectral distortions in the cosmic microwave background (CMB). These encode information about the nature and timing of first sources over redshifts 30--6 and appear at meter wavelengths as a tiny CMB distortion along with the Galactic and extragalactic radio sky, which is orders of magnitude brighter. Therefore, detection requires precise methods to model foregrounds. We present a method of foreground fitting using maximally smooth (MS) functions. We demonstrate the usefulness of MS functions over traditionally used polynomials to separate foregrounds from the Epoch of Reionization (EoR) signal. We also examine the level of spectral complexity in plausible foregrounds using GMOSS, a physically motivated model of the radio sky, and find that they are indeed smooth and can be modeled by MS functions to levels sufficient to discern the vanilla model of the EoR signal. We show that MS functions are loss resistant and robustly preserve EoR signal strength and turning points in the residuals. Finally, we demonstrate that in using a well-calibrated spectral radiometer and modeling foregrounds with MS functions, the global EoR signal can be detected with a Bayesian approach with 90\% confidence in 10 minutes' integration.
\end{abstract}
\section{Introduction}
\begin{figure}[ht]
\centering
\begin{minipage}[b]{.48\textwidth}
\includegraphics[trim={2cm 0 0 0},clip,height=2.5in,width=3.2in]{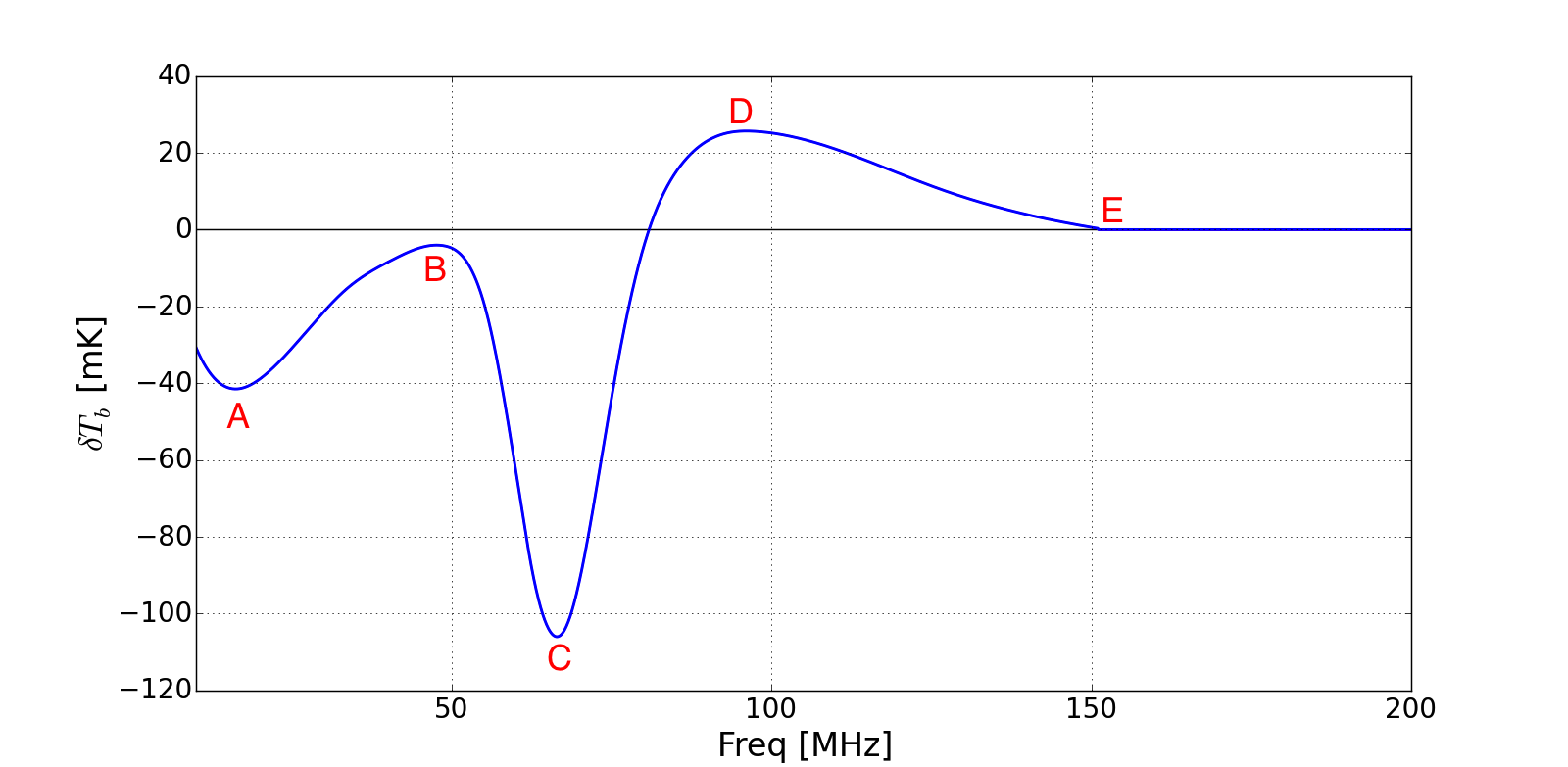}
\caption[Generic redshifted 21-cm EoR signature expected in the frequency range 10--200 MHz from thermal evolution in baryons at redshifts 150 to 6]{Generic redshifted 21-cm EoR signature expected in the frequency range 10--200 MHz from thermal evolution in baryons at redshifts 150 to 6. This is expected to appear as a global spectral distortion in the spectrum of the cosmic microwave background.}
\label{fig:eor_ideal_full}
\end{minipage}
\end{figure}

The epoch of reionization (EoR) represents one of the important transitions in the physical state of the universe, when the almost neutral baryonic matter from the dark ages transitioned to its mostly ionized form. Another important transition occurred earlier at even higher redshifts, when during the epoch of recombination the primordial plasma recombined and the universe became neutral, with atomic hydrogen, helium, and a small fraction of light elements. While the physics of the epoch of recombination is well understood from the theoretical point of view \citep{ah2011,ct2011,Glover2014} and constrained by observations of the cosmic microwave background (CMB) \citep{Calabrese2013,Farhang2013,Planck2015results1}, the EoR is relatively poorly understood. Among other aspects, the thermal evolution of baryons and nature of the first sources, the exact redshift and duration of reionization, and the dominant mechanisms that affect reionization are poorly constrained \citep{Planck2016}. The evolution of the strength of the redshifted 21-cm line of hydrogen against the radiation from CMB is expected to trace the thermal history of the gas across the EoR. Thus, the redshifted 21-cm signal is predicted to appear as a distortion of the CMB spectrum, encoding the physics of the EoR in the signal structure. A comprehensive review of EoR physics is presented in \citet{Furlanetto2006}.\\\\
There are a multitude of scenarios that predict different global redshifted 21-cm signatures \citep[see][]{Pritchard2010a}. The form of the generic signature from the EoR is shown in Fig.~\ref{fig:eor_ideal_full}.  The high-redshift absorption feature `\textit{A}' at frequencies below about 30 MHz arises from collisional coupling of the spin temperature to the relatively low gas kinetic temperature.  At relatively higher frequencies and below redshifts of about 40, a second absorption dip `\textit{C}'  arises from Wouthuysen-Field \citep[]{Wouthuysen1952, Field1958} driving of spin to the kinetic temperature by Ly-$\alpha$ from the first collapsed objects.  The subsequent X-ray and UV heating of the gas kinetic temperature, and consequently the spin temperature as well, by energetic radiation from the first stars and galaxies transforms the appearance of the gas from absorption to emission (at `\textit{D}'); the ionizing radiation then progressively results in the disappearance of the baryons in redshifted 21~cm and the gas is fully ionized and vanishes from this diagnostic by redshift 6 (at `\textit{E}').  The critical spectral features representing events in the thermal history, which appear as turning points and successive absorption and emission features, are in the frequency range 10--200~MHz corresponding to events at redshifts of about 150 to 6.  Identifying the true reionization signal amongst the multitude of plausible forms currently allowed by observations to date is a way to constrain the parameter space of the sources of reionization and their spatial-temporal distribution. 
At meter and decameter wavelengths, where the global EoR signal resides, foregrounds due to Galactic and extragalactic radio sources are relatively bright.  The reionization signals are of 10--100 mK brightness temperature, whereas the foregrounds are 100s to 10,000s of K; the spectral dynamic range required for the detection of the global EoR signature in any measurement of the absolute spectrum of the radio sky is about $10^3$--$10^6$. It is therefore not surprising that considerable effort has been placed on the development of methods for the separation of the faint 21-cm signals from the substantially brighter foreground component in measurement data in the context of detection of both the global and statistical EoR signals \citep[see, for example, ][]{Gnedin2004, Morales2006, Vedantham2012, LPTL2013, Thyagarajan2013, Harker2015, Bernardi2016, Chapman2016, Nhan2017}. 
\section{Motivation}\label{sec:motivation}
There are several ongoing and proposed experiments to detect the global EoR signal. Nevertheless, the methodology for the modeling, subtraction of foregrounds, and extraction of the much smaller EoR signal from spectral data, or for joint modeling of foregrounds and EoR signals, continues to be an open problem.\\\\
Most global EoR signal detection experiments employ a spectral radiometer connected to an elemental antenna. This provides as its response a measurement set that contains the signature arising from baryon thermal evolution and reionization of hydrogen gas along with the orders-of-magnitude larger foreground component. This radio foreground component appears as an average of sky spectra over the beam pattern of the telescope antenna.  Even though the foreground might be composed of emitting volume elements that individually have spectra of power-law form, the variation in the spectral indices of these sources along the line of sight and across the sky within the antenna beam results in an observed spectrum in which the foreground component has an unknown form and cannot be modeled as a simple power law anymore. As discussed below, experiments have employed high-order polynomials to model these foregrounds and also instrument systematics.\\\\  
The  EDGES experiment \citep{Bowman2010} observed the sky spectrum in the frequency range 100--200~MHz and excluded rapid reionization with timescales less than that corresponding to  $\Delta z <  0.06$. Polynomials of different orders in different frequency windows were used to jointly fit and model the foreground and instrumental systematics. The BIGHORNS experiment \citep{Sokolowski2015} used a ninth-order polynomial to model the receiver noise and the Global Sky Model (GSM) of de Oliveira-Costa et al. (2008), together with the simulated radiation pattern of their antenna, to estimate the foreground contribution in their data. The SCI-HI 21-cm all-sky experiment \citep{Voytek2014} also used the interpolated GSM to calibrate the foreground in their data. Experiments to date have indeed had to use high-order polynomials to fit out both foregrounds and uncalibrated bandpass; however, in this work, we focus entirely on complexity inherent in  foregrounds and assume a smooth bandpass of the observing instrument.\\\\
The order of the polynomial deemed necessary to fit foreground components in spectral data to $\sim$~mK level, lower than the expected signal, has varied in literature depending on the assumed sky model.  \cite{Pritchard2010a} argued that averaging sky spectra over large angular scales by wide-angle beams of telescopes designed for the global EoR detection would result in smooth frequency dependence for the foregrounds in spectral data and a third-order polynomial may suffice to model the foreground component and reduce their residuals to well below the expected EoR signal.  They note that these residuals are dominated by the numerical limitations of available sky models and also recognize that a simple polynomial approach to modeling the foreground---with order sufficient to describe the  foreground and systematics in spectral data---could substantially diminish the desired EoR signal in the residual. More recent results suggest the need to adopt polynomials of higher order ($> 4$) to model foregrounds to mK levels to be able to discern the global EoR signal \citep[see][]{Bernardi2015}.  While polynomials of lower order might not model the foreground components with sufficient precision, polynomials of higher orders risk over fitting the foregrounds and partially removing the EoR signal.  Adoption of sky models with increasing complexity  while generating synthetic spectra and mock observations in general leads to concluding that a higher order in the polynomial is required to fit to the foreground components in the mock observations so as to be able to discern the EoR signal.  This uncertainty in inferences of the degree of the polynomial required to model the foreground in measurements with sufficient precision arises both due to differences in the assumed spectral structure in the sky brightness and due to instrumental effects.  Since both the global EoR signal as well as the model for the low-frequency sky ($\lesssim 200~$MHz) are uncertain, appropriate choice of the functional form adopted to model the foreground in spectral data is critical to interpreting any global EoR signal detection experiment, failing which results in unknown biases in interpretation of data.\\\\ 
To summarize, the aforementioned approaches come with two problems. First, modeling with polynomials of arbitrary order, inclusive of GSM-based models, which also employ mathematical interpolation between measurements from publicly available all-sky radio surveys, can be unrealistic in their spectral shapes. Second, estimates of the percentage loss in EoR signal on subtracting a polynomial baseline and the level of foreground contamination that may remain in the residual, suggest that such a simple polynomial or GSM approach to modeling the foreground component in spectral data is far from ideal \citep{Pritchard2010a, Voytek2014, Harker2015}.\\\\ 
This paper represents an improvement on previous work in two ways. First, we use a new physically motivated model of the radio sky, GMOSS \citep{GMOSS2017}, which models the radio spectrum over the entire sky with parameters that describe continuum radiative processes. By beam-weighting such a physically motivated sky model, we generate synthetic spectra representing mock observations that have physical and plausible spectral forms.\\\\ 
Second, we suggest a new functional form for describing the foreground component, namely maximally smooth (MS) functions. We demonstrate that this is an improvement over polynomial or other interpolations between measured sky brightness at discrete and widely spaced frequencies, which would result in spectra with unphysical shapes and arbitrary spectral complexity.  This functional form for the foreground was recently applied to simulations of the detection of primordial recombination-line ripples from redshift $z\simeq 1000$ \citep{apsera2015}, where it was demonstrated that this approach is indeed powerful.  We demonstrate that for the more realistic physical GMOSS model of the foreground, the proposed modeling of foregrounds in measurement sets using MS functions is robust and superior to the more simplistic polynomial-form representation. We find that foregrounds in mock spectra generated using GMOSS, though spectrally more complex than previously assumed, are indeed smooth and describable by MS functions.\\\\
Most importantly, MS functions are loss resistant in the sense that when they are used to fit out foregrounds, they do not substantially attenuate the embedded EoR signal and also preserve the turning points in the EoR signature, unlike when high-order polynomials are used for foreground fitting.  We also examine detection likelihoods when mock observations are jointly modeled using theoretical EoR templates and MS functions to describe the foreground. 
\section{Toward a spectral model for the radio sky}
\label{sec:skymodel}
Previous efforts to simulate data analysis methods for global EoR signal detection experiments have assumed simple sky models that are,  in log-temperature versus log-frequency space (hereinafter referred to as log($T$) versus log($\nu$) space), either power laws or low-order polynomials \citep[see, for example,][]{Pritchard2010a, Harker2015}. The method adopted to separate the signal from foregrounds in mock observations is to fit the simulated spectra with polynomials of low order, possibly identical to those used to generate the input model sky.  Clearly, since the EoR signal contains higher order inflections than those used to describe the sky spectrum and model the foreground in data, this would leave behind a substantial part of the EoR signal in the residual. Such a simplistic input model for the sky with subsequent modeling of the foreground components in mock observations using polynomials of matching order belies the challenge when dealing with real measurements where the true functional form of the sky spectrum and that of the foreground component in data are unknown.\\\\ 
In this section, we first use a set of all-sky maps  to generate synthetic sky spectra, without any EoR component added.  We then combine spectra from individual pixels over wide beams to generate mock observations.  We estimate the degree of polynomial required to model the foreground in these mock data to mK precision in each case.   Combining spectra from individual pixels can result in a final beam-weighted mock spectrum with different spectral complexity.  Spectra with greater spectral complexity, in log($T$) versus log($\nu$) space, require a polynomial of higher degree to fit to the desired level.\\\\  
We investigate the consequences of adopting three separate sky models on the deduced order of the polynomial required to fit the final beam-weighted foregrounds to mK level.  The three different models we have investigated are (i) power laws or linear functions in log($T$) versus log($\nu$) space, (ii) polynomials in log($T$) versus log($\nu$) space and (iii) the Global MOdel for the radio Sky Spectrum (GMOSS; \citealt{GMOSS2017}).\\\\ 
\textbf{Inputs to the sky models.} A set of six all-sky maps are used to generate the sky models.  The maps used are at 150~MHz \citep{Landecker1970}, 408~MHz \citep{Haslam1982}, 1420~MHz \citep{Reich1982, Reich1986, Gorski2005} and  23 GHz (WMAP science data product\footnote{WMAP Science Team.}). The treatment of various maps to bring them to their final usable form are presented in \cite{GMOSS2017}. Maps at 22 and 45~MHz are generated from the GSM \citep[see][]{GSM2008}.  These are at frequencies at which input maps are available for GSM, hence they are close to the original maps used by the GSM.  In their final form, all input maps have a common resolution of $5^{\circ}$ and are represented in galactic coordinates in nested `R4'  scheme of HEALPix.\\ 
For each of the three sky models adopted here, spectra are derived via fits that are done separately at each sky pixel. The spectral fits are then averaged over wider beams to generate the final mock spectrum as might be observed with a telescope that has a wide field of view.\\\\ 
\textbf{Mock spectra and calibration.} Our code uses the aforementioned sky images to generate synthetic sky spectra at any location on Earth and at any local time, including appropriate corrections for effects including precession and atmospheric refraction. In this section we assume observing with an antenna having a frequency-independent cosine$^2$ beam with half-power beam width (HPBW) of $78^{\circ}$.  The antenna is assumed to be pointed toward zenith and observes the sky over Gauribidanur Observatory in southern India, which is at latitude $13\fdg01$~N and longitude $77\fdg58$~E.    Data are assumed to be recorded by a correlation spectrometer with in-built calibration for instrument bandpass with 1~MHz frequency resolution.  We focus on the frequency range 40--200 MHz, where signatures of the Cosmic Dawn and of Reionization are predicted to be present; these signatures are substantially more uncertain today than those from the Dark Ages, which lie below about 40 MHz.\\\\
\textbf{Assumptions.} We assume a Dicke-switching calibration scheme wherein differencing the spectra recorded separately with hot and cold loads (373.0 K and 273.0 K, respectively) on the antenna serves to calibrate the bandpass and absolute scale.  Admittedly, such schemes are limited in their accuracy owing to internal reflections within the signal path, which would be different for the sky signal and the injected calibration signal.   Here we focus on the spectral complexity of foregrounds  and ignore calibration errors, instrumental systematics, and mode coupling of spatial structures in the sky to the measured spectrum in frequency \citep[see][]{Mozdzen2016}.
\subsection{Case of power-law Form for Sky Spectra}
\label{sec:linear}
\begin{figure}[ht]
\centering
	\begin{minipage}{.48\textwidth}
	  \subfloat[]{
	\includegraphics[width=\columnwidth,height=2.0in]{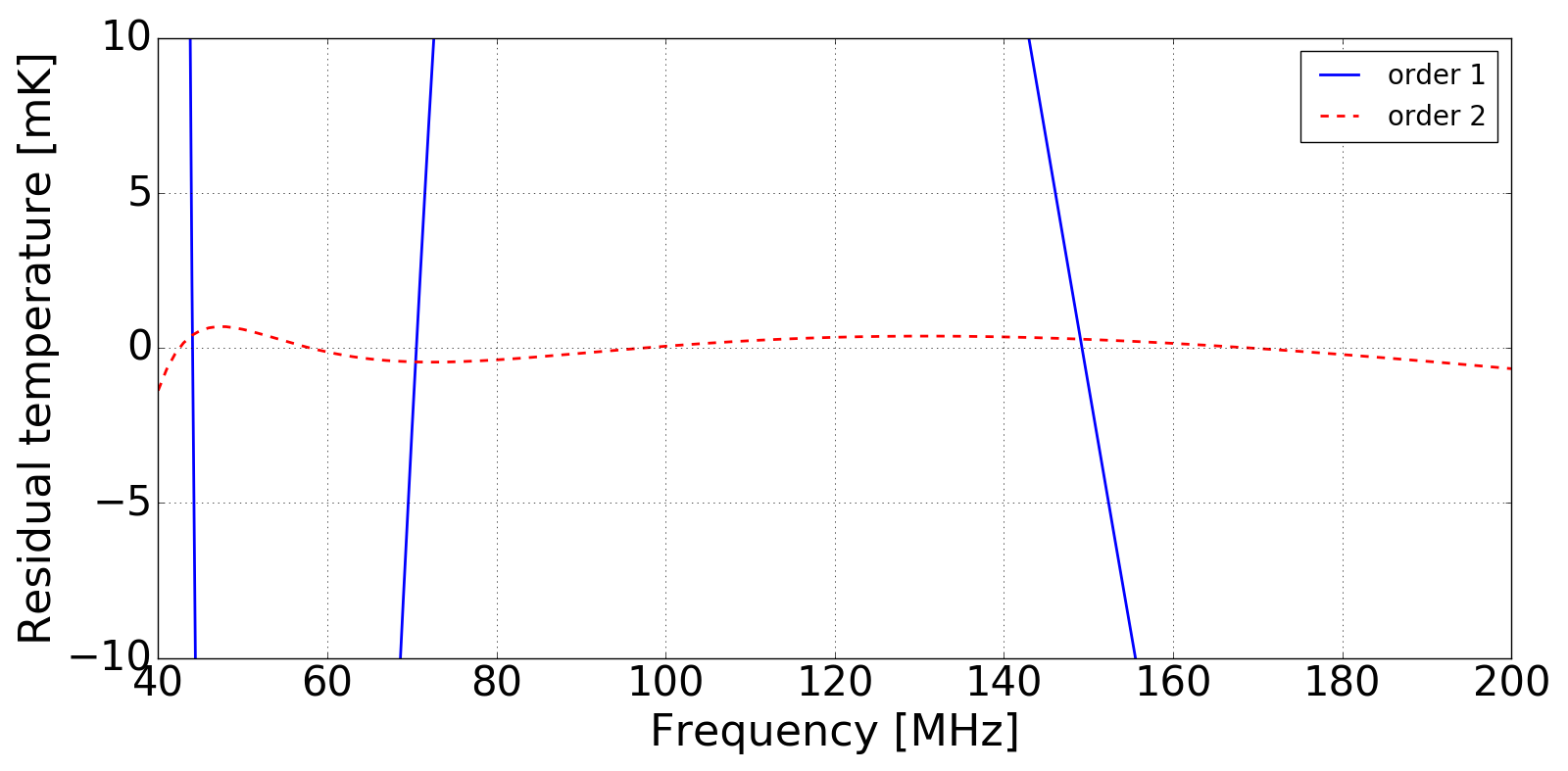}
	}
	\end{minipage}	
	\begin{minipage}{.48\textwidth}
	\subfloat[]{
	\includegraphics[width=\columnwidth,height=2.0in]{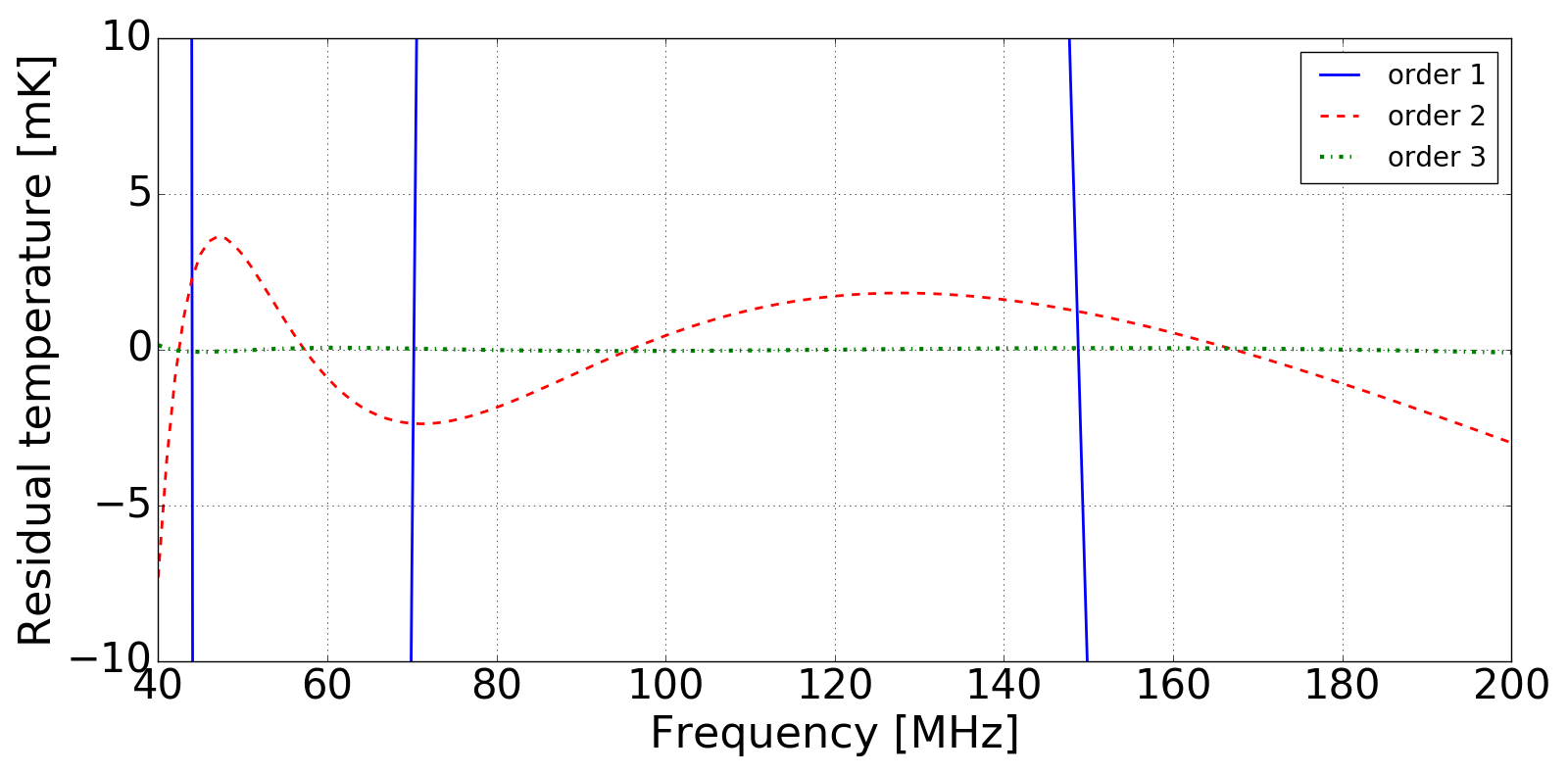}
	}
	\end{minipage}
\caption[Residuals obtained on fitting to mock observations of the foreground sky in log($T$)-log($\nu$) space with polynomials with orders between 1 and 3, assuming a power-law sky model]{Residuals obtained on fitting to mock observations of the foreground sky in log($T$)-log($\nu$) space with polynomials with orders between 1 and 3. The sky spectra were generated from a model sky that assumes a simple power-law form radio spectrum at each sky pixel. The panel on the left are residuals obtained for a mock observation that is away from the Galactic plane and the one on the right corresponds to a mock spectrum that includes the Galactic plane.}
\label{fig:resi_lin}
\end{figure}

Galactic synchrotron emission, which dominates the low-frequency radio sky (in our context $\nu \lesssim 1$ GHz), may in its simplest form be described as a power law:
\begin{equation}
T(\nu,n) = T_{\rm \nu_0,n} \times \Big(\frac{\nu}{\nu_0}\Big)^{\rm \alpha(n)},  \label{eq:simple_pl}\\
\end{equation}
which in log($T$) versus log($\nu$) space has a linear form:
\begin{equation}
{\rm log}(T(\nu,n)) = {\rm log}(T_{\nu_0,n}) + \alpha(n) \times {\rm log}\Big(\frac{\nu}{\nu_0}\Big). \label{eq:loglog_lin}
\end{equation}
Here $T(\nu,n)$ is the brightness temperature at frequency $\nu$ toward pixel $n$ and $\alpha(n)$ is the temperature spectral index of the power-law emission in that sky pixel. $T_{\rm \nu_0,n}$ is the temperature at frequency $\nu = \nu_0$ toward pixel $n$.\\\\
We realize an all-sky model by computing at each sky pixel such a power-law spectrum by a straight-line fit to the data points given by the six maps toward that pixel. Fits done in log($T$) versus log($\nu$) space, as given by Equation~(\ref{eq:loglog_lin}), weight the data to have uniform fractional errors.    Mock spectra, without any EoR signal added, were generated at times spaced 1~hr apart over the full LST range $0$--$24$~hr by averaging the pixel spectra with a weighting defined by the telescope beam at each sidereal time.   We show in Fig.~(\ref{fig:resi_lin}) the form of the residuals on fitting with polynomials of a few orders for representative spectra, which correspond to LSTs when the beam is on and off the Galactic plane.  The amplitudes of residuals on fitting these spectra with a polynomial of order two are a few mK, with residuals somewhat larger for mock observations toward the Galactic plane.
Fig.~(\ref{fig:resi_lin}) thus indicates that if we assume a power-law spectral form at every pixel that lies in a telescope beam, the beam-averaged foreground spectra might be fit with a quadratic polynomial with an accuracy that allows discerning the generic EoR signal.  However, it is not possible to fit a substantial number of sky pixels with such a simplistic power-law spectrum, and in sky regions having intrinsically curved spectra, the errors in the fit exceed measurement errors; therefore, the model considered above is implausible.
\subsection{Case of a polynomial sky model}
\label{sec:poly}
\begin{figure}[ht]
\centering
	\begin{minipage}{.48\textwidth}
	  \subfloat[]{
	\includegraphics[width=\columnwidth,height=2.0in]{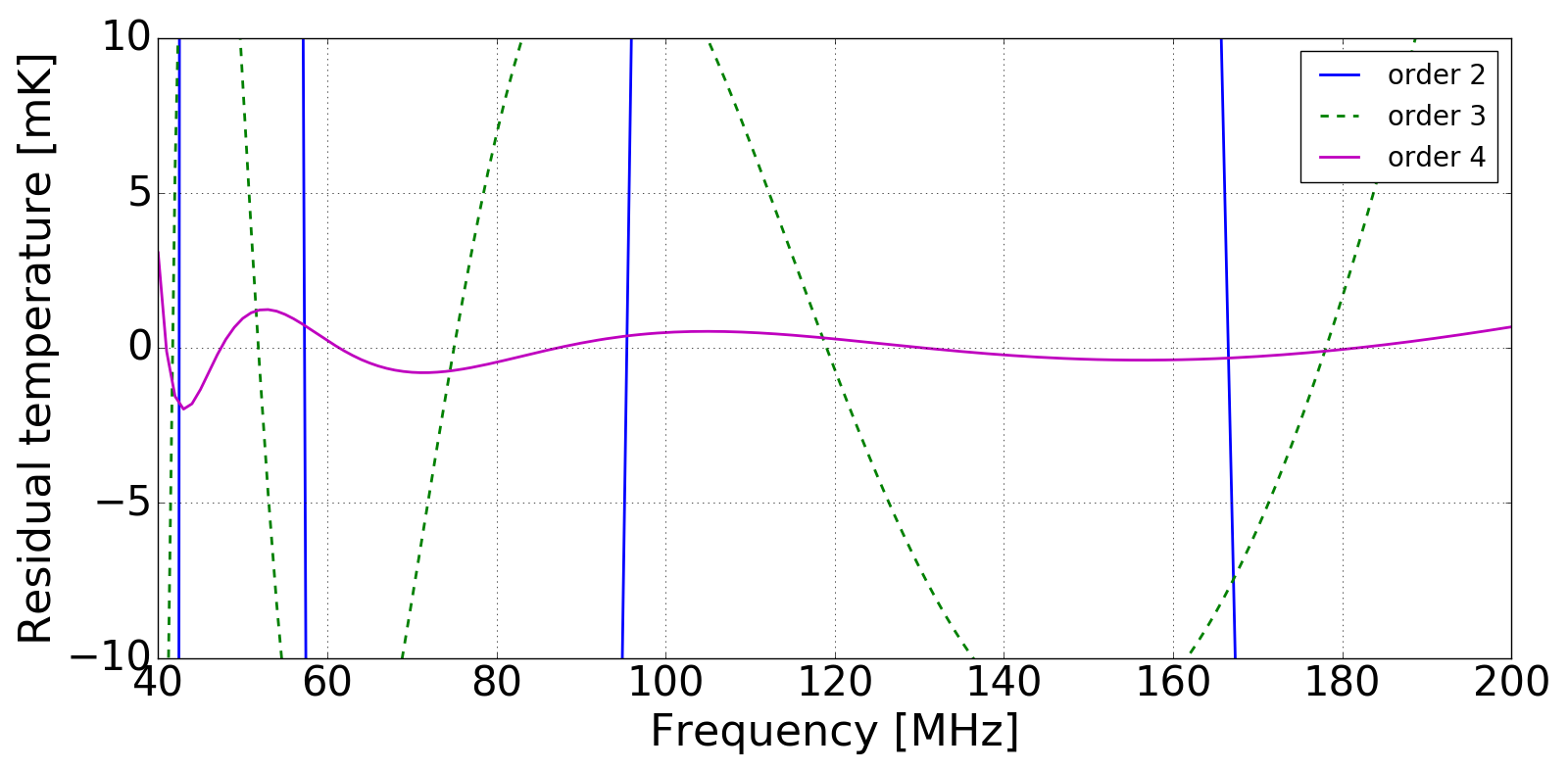}
	}
	\end{minipage}	
	\begin{minipage}{.48\textwidth}
	\subfloat[]{
	\includegraphics[width=\columnwidth,height=2.0in]{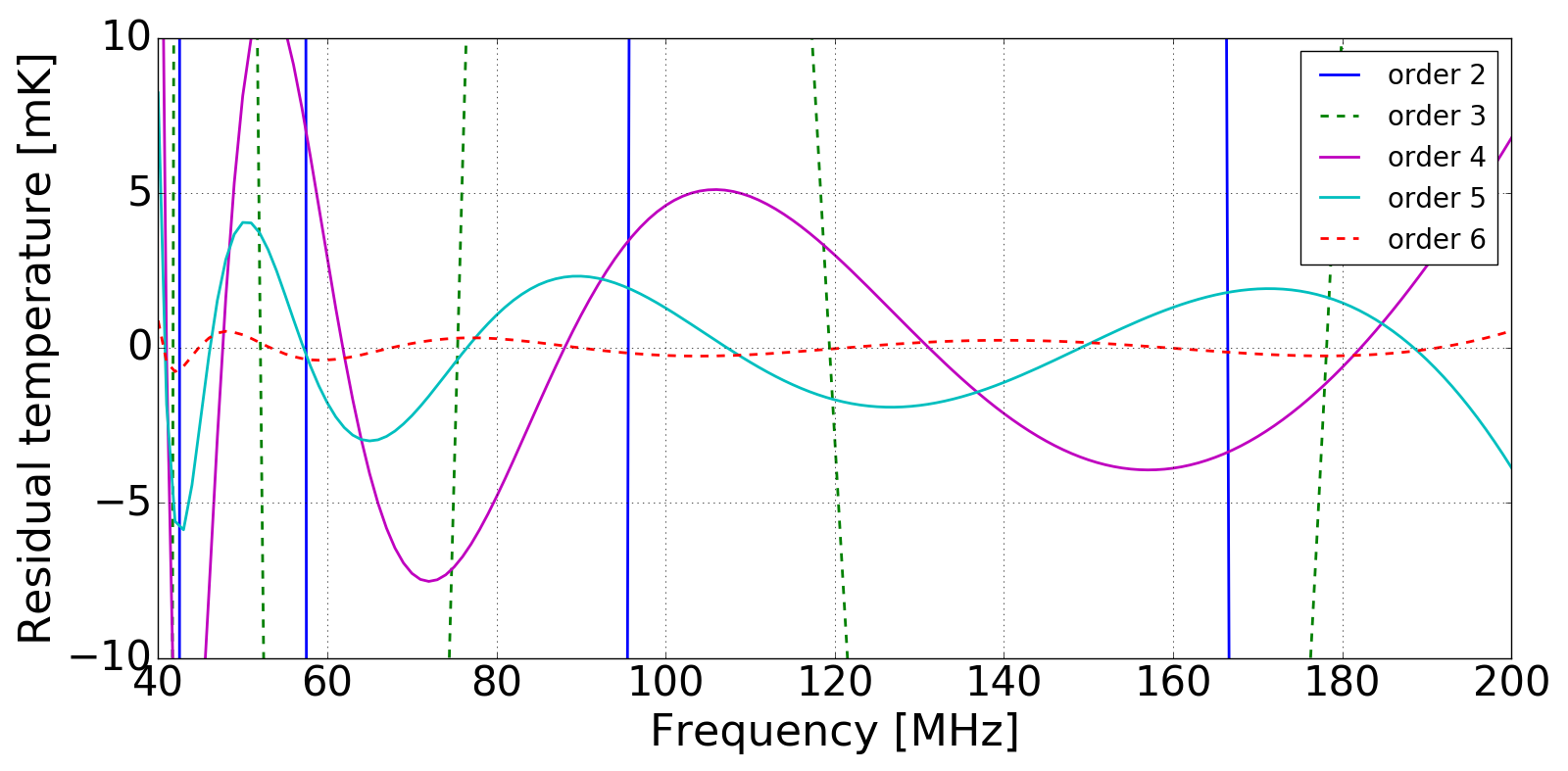}
	}
	\end{minipage}
\caption[Residuals on fitting mock spectra, generated from a model that assumes a quintic form for the spectrum at each pixel, with polynomials of increasing orders]{Residuals on fitting mock spectra, generated from a model that assumes a quintic form for the spectrum at each pixel, with polynomials of increasing orders. The panel on the left are residuals obtained for an observation that is away from the Galactic plane and the one on the right is for a mock spectrum that is toward the Galactic plane.}
\label{fig:resi_quad}
\end{figure}

As mentioned above, the six data points in the input sky maps do not lie on a straight line at every pixel in log($T$) versus log($\nu$) space. Clearly, the emission from the foreground, even in single pixels, is not precisely describable as a single power law. We allow for curvature in the sky spectrum by fitting the six data points at each pixel with a fifth-order polynomial in log($T$) versus log($\nu$) space:
\begin{equation}
\begin{split}
{\rm log}(T(\nu,n)) = {\rm log}(T_{\nu_0,n})  + \alpha(n)\times{\rm log}\Big(\frac{\nu}{\nu_0}\Big) + \beta_2(n)\times{\rm log}\Big(\frac{\nu}{\nu_0}\Big)^2 + \beta_3(n)\times{\rm log}\Big(\frac{\nu}{\nu_0}\Big)^3\\
+ \beta_4(n)\times{\rm log}\Big(\frac{\nu}{\nu_0}\Big)^4 + \beta_5(n)\times{\rm log}\Big(\frac{\nu}{\nu_0}\Big)^5 ,
\label{eq:loglog_quad}
\end{split}
\end{equation}
where the $\beta_i(n)$ coefficients provide for higher order curvature in the spectrum.\\\\
The optimized best-fit coefficients to such a quintic form at every pixel provides an input sky model.   We generate beam-convolved sky spectra for different LSTs over 0--24~hr and fit these with polynomials of varying order. The residuals so obtained for two such realizations of the sky for LSTs corresponding to off and on the Galactic plane are in the left and right panels of Fig.~(\ref{fig:resi_quad}), respectively. Clearly, if we adopt a quintic polynomial model for sky spectra and seek to model the foreground components in measurements to an accuracy so as to limit residuals to mK levels, polynomials of order four are required off the Galactic plane and orders about 5--6 are required for telescope pointings toward the Galactic plane.\\\\  
The above exercise demonstrates that if a higher level of complexity is allowed for the spectral structure in the model for the foreground, a higher order polynomial is necessary to fit the mock observations that simulate beam-averaged spectra. The spectral structure inherent in the sky model in simulations thus critically guides strategies developed to deal with foregrounds in measurements. Increasing the order of the interpolating polynomial to create sky models that are cubic, quartic, quintic, and so on require increasingly greater orders of polynomials for modeling the foreground component in mock observations.  Additionally, using higher order polynomials to model foregrounds risks generating unphysical sky models that would no longer usefully guide detection strategies.\\\\
It may be noted here that there do exist foreground models in the literature with greater sophistication than polynomial fits in log($T$) versus log($\nu$) space. Notably the Global Sky Model (GSM) \cite[][]{GSM2008,Zheng2016} uses a data-driven method to arrive at a foreground model using many data sets of large-area sky maps. Some studies in the literature that have examined foreground spectral complexity, as mentioned in Section \ref{sec:motivation}, use GSM in one of two ways. 
One approach is to first generate sky maps at widely spaced frequencies using GSM, construct beam-averaged temperature maps, then adopt polynomial interpolation between the image pixels to derive mock spectra. This is similar to the case of polynomial interpolation described above and potentially results in unphysical spectral shapes. 
A second approach is to use GSM to derive sky brightness temperature at every pixel over finely sampled frequencies and simulate pixel-by-pixel foreground spectra. In both approaches, GSM uses a cubic-spline interpolation to derive principal components at all frequencies other than those in the input data set. The spline interpolation is a mathematical fit without physical motivation, and there is no physics involved in deriving the principal components of GSM. This may lead to spectra with unphysical spectral shapes.\\\\
Since there are a limited number of maps (data points) available and the uncertainties in individual maps are orders of magnitude greater than the precision required for detection of the EoR, and with no physical rational to guide the order of the interpolating polynomial, we move toward the physically motivated sky model described below.
\subsection{GMOSS: Global MOdel for the radio Sky Spectrum}
\label{sec:GMOSS}
\begin{figure}[ht]
\centering
	\begin{minipage}{.48\textwidth}
	  \subfloat[]{
	\includegraphics[width=\columnwidth,height=2.0in]{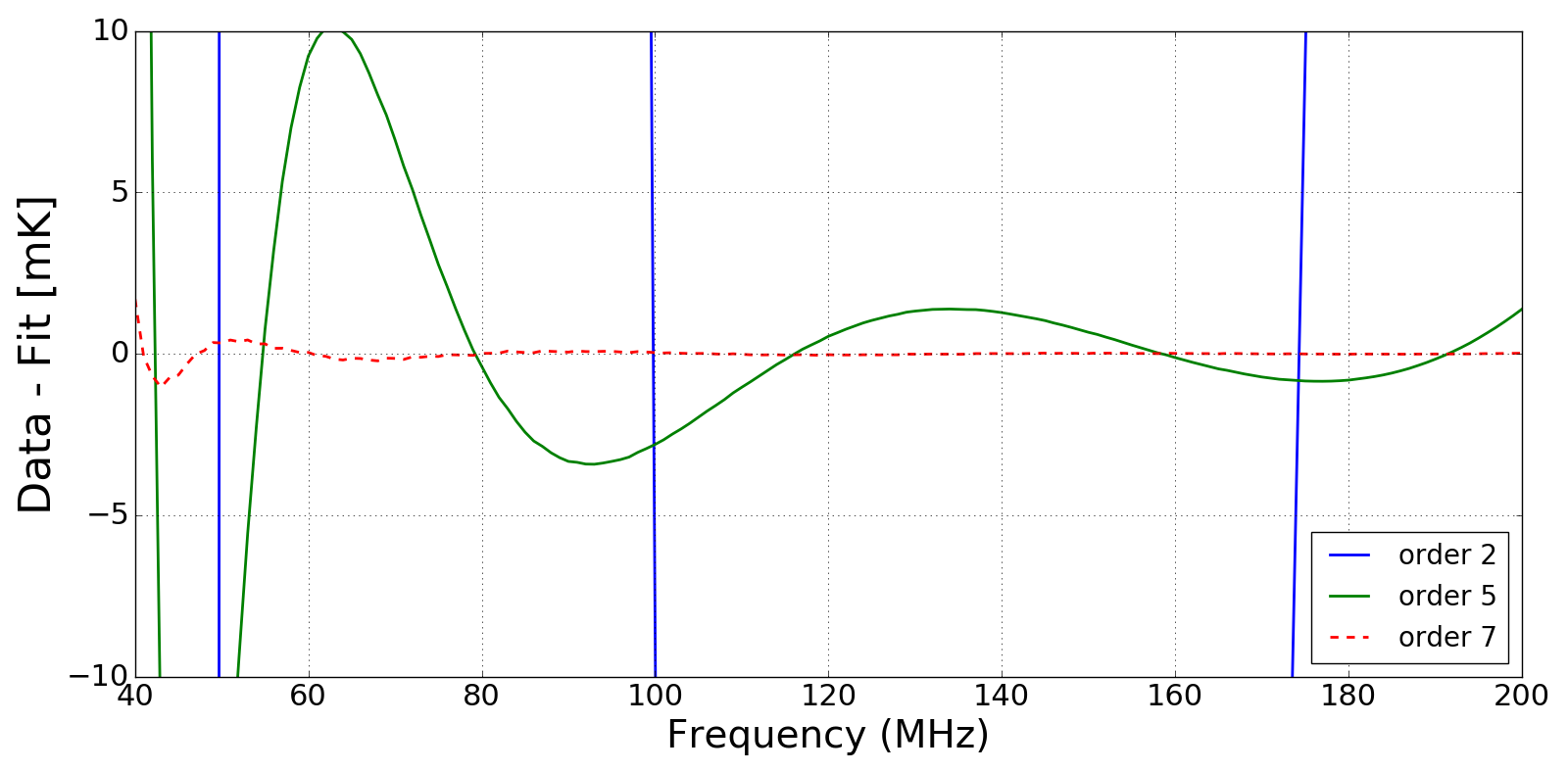}
	}
	\end{minipage}	
	\begin{minipage}{.48\textwidth}
	\subfloat[]{
	\includegraphics[width=\columnwidth,height=2.0in]{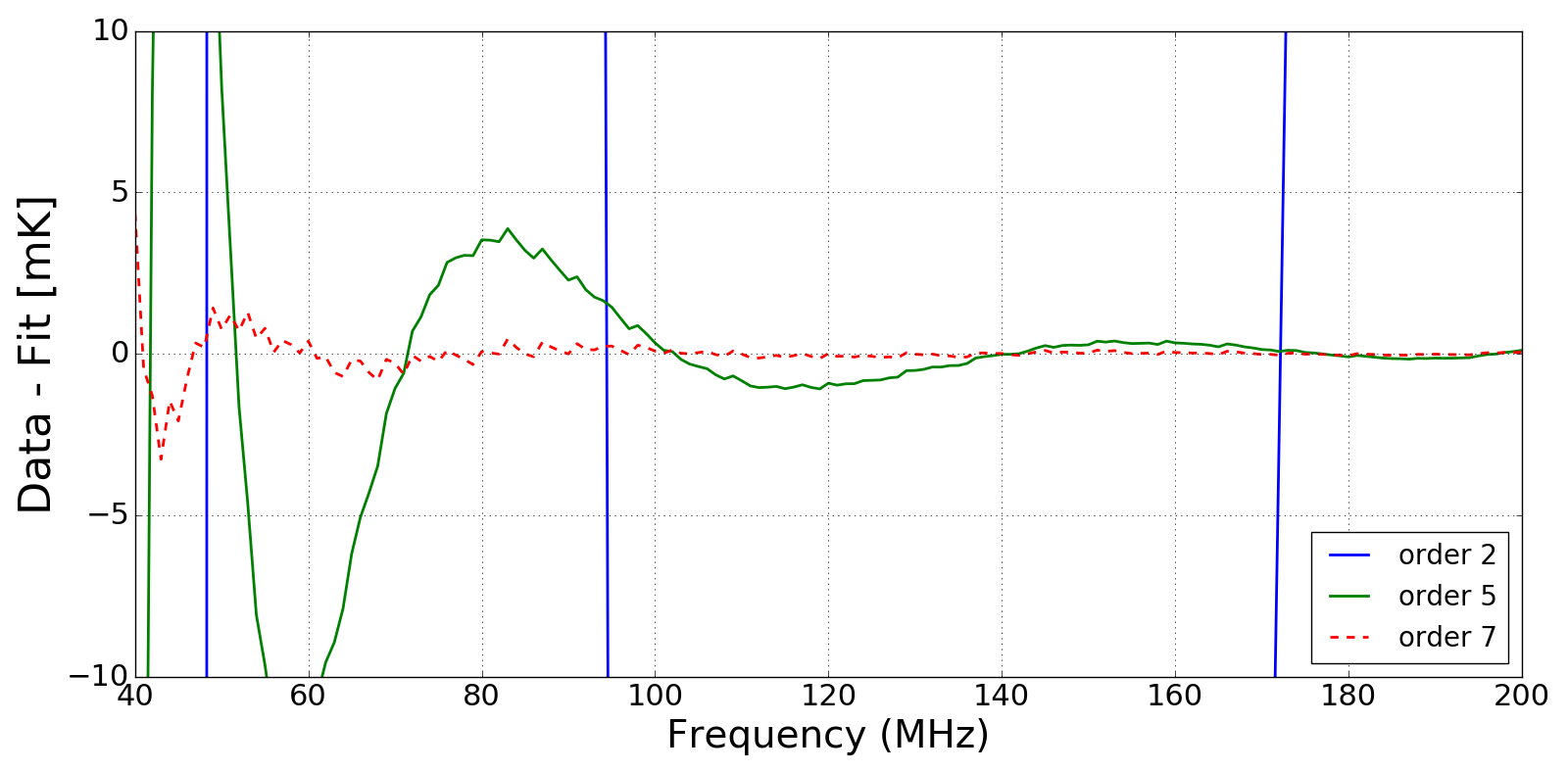}
	}
	\end{minipage}
\caption[Residuals on fitting mock spectra, generated from a model that assumes GMOSS as the sky model, with polynomials of increasing orders]{Residuals on fitting mock spectra, generated from a model that assumes GMOSS as the sky model, with polynomials of increasing orders. The panel on the left are residuals obtained for a mock observation that is away from the Galactic plane and the one on the right is for a mock spectrum that is toward the Galactic plane.}
\label{fig:resi_GMOSS}
\end{figure}

\begin{figure*}[t]
\centering
\begin{minipage}[b]{\textwidth}
\includegraphics[scale=0.4]{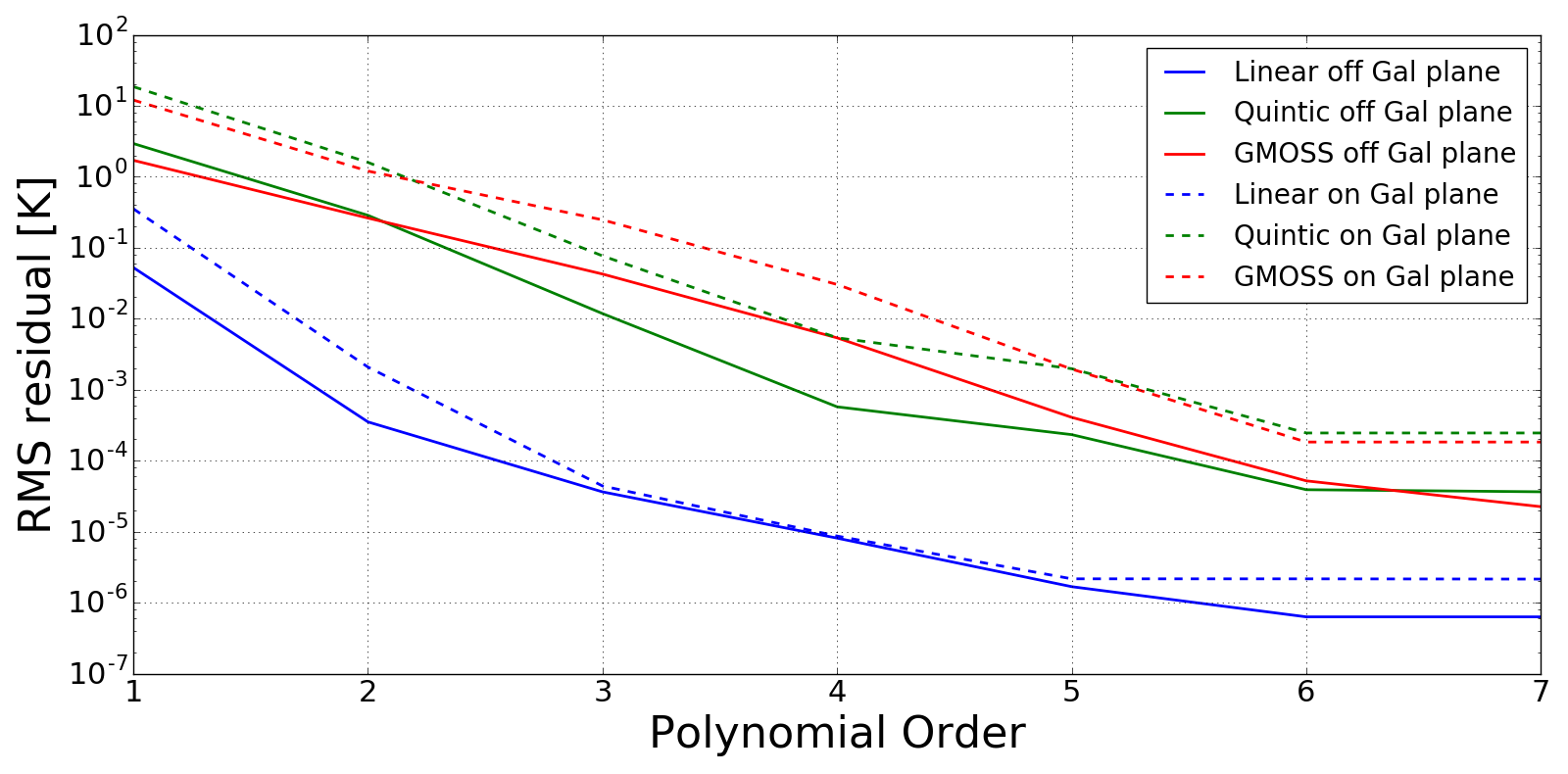}
\caption[rms of residuals on fitting mock observations of foregrounds, which are generated using different sky models, versus the order of the fitting polynomial]{rms of residuals on fitting mock observations of foregrounds, which are generated using different sky models, versus the order of the fitting polynomial.  Solid lines are for mock observations that are recorded away from the Galactic plane and the dashed lines are for those that are toward the Galactic plane.}
\label{fig:resi_rms_poly}
\end{minipage}
\end{figure*}

\begin{figure*}[t]
\centering
\begin{minipage}[b]{\textwidth}
\includegraphics[scale=0.35]{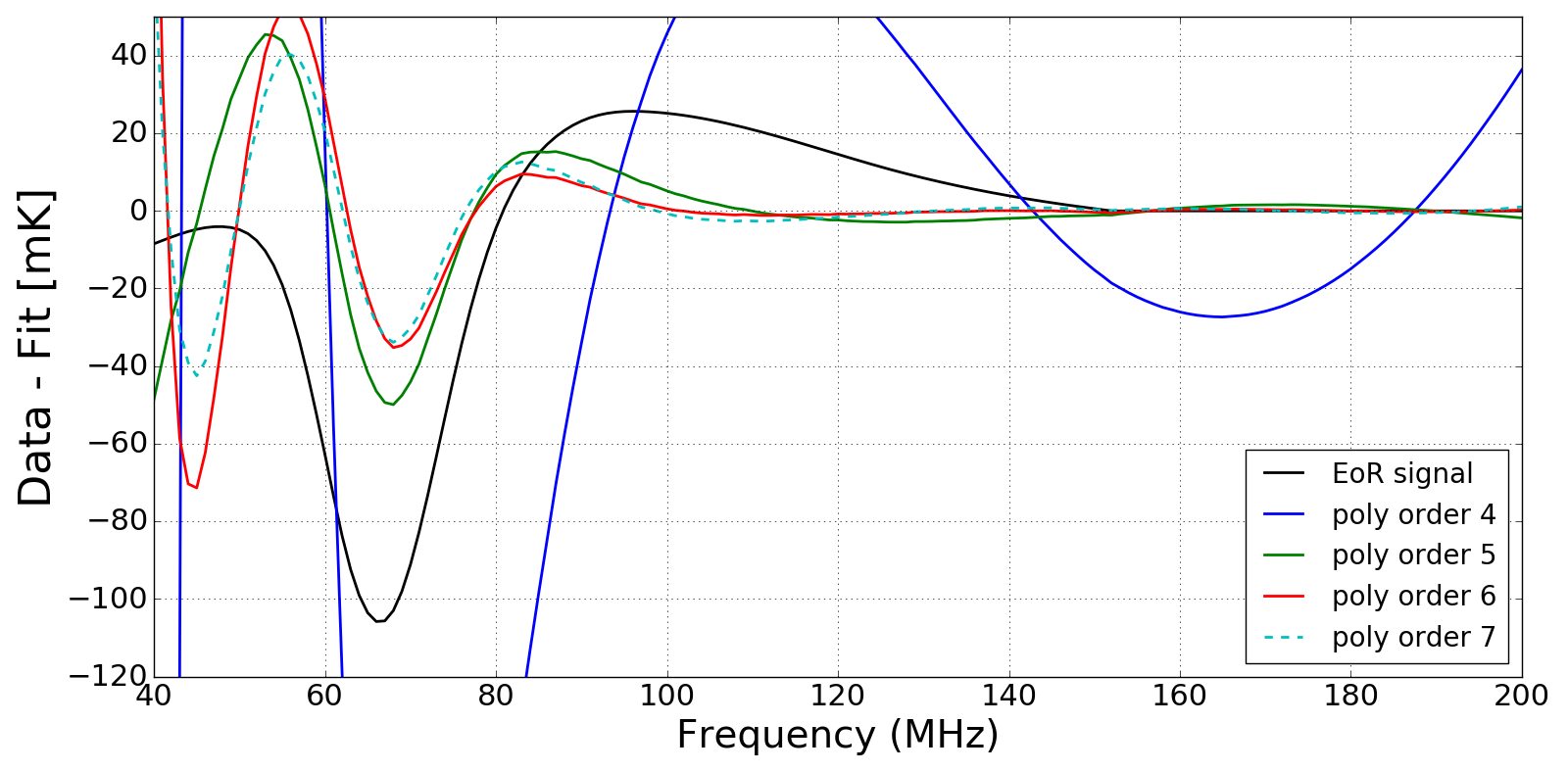}%EoR_residuals_polynomial_fit_to_GMOSS_spec.png 
\caption[Residuals obtained on fitting a mock observation of the sky spectrum, which uses GMOSS as the sky model and also contains the EoR signal, with polynomials of orders 4, 5, 6 and 7]{Residuals obtained on fitting a mock observation of the sky spectrum, which uses GMOSS as the sky model and also contains the EoR signal, with polynomials of orders 4--7. Overlaid as a solid black line is the generic EoR signal. Increasing the order of the polynomial used, in log($T$) vs. log($\nu$) space, to model and remove the foreground component of the spectrum results in a residual that progressively departs from the generic EoR signal. Not only does the amplitude of the EoR signal reduce, additional turning points are also introduced.}
\label{fig:EoR_res_poly}
\end{minipage}
\end{figure*}

GMOSS \citep{GMOSS2017} is a physically motivated model of the radio sky spectrum in which seven physical parameters define the radiative processes that generate the spectrum of the brightness at each sky pixel. GMOSS incorporates plausible physics including (i) synchrotron emission that is assumed to arise from a power-law form electron energy spectrum that may have a break as well, (ii) composite emission from flat and steep spectrum synchrotron sources, (iii) thermal absorption at low frequencies, and (iv) optically thin thermal free-free emission at high frequencies. GMOSS provides synthetic spectra representative of the intensity distribution toward HEALPix pixels of $5^{\circ}$ over the whole sky. GMOSS parameters are optimized by fits of the spectra to six all-sky maps at frequencies distributed between 22~MHz and 23~GHz, same as for the case of sky models described in Sections \ref{sec:linear} and \ref{sec:poly}. Details of GMOSS and the goodness of the model are described in detail in \citet{GMOSS2017}.  GMOSS describes spectra of individual $5^{\circ}$ pixels toward different sky directions using one of two forms, described below, depending on whether the pixel spectra are convex or concave.\\\\
For the case when the spectral index toward high frequencies is steeper than that toward low frequencies, GMOSS adopts an underlying model of synchrotron spectral steepening in which the frequency dependence of brightness temperature $T(\nu)$ is described by 
\begin{multline}
%\MoveEqLeft
T(\nu) = C_{\rm 1}\Bigg(\nu^{-2}\Big\{\gamma_{\rm break}^{2\alpha_{\rm 1} -3}\int\limits_{\gamma_{\rm min}}^{\gamma_{\rm break}} F(x)~x~\gamma^{-(2\alpha_{\rm 1} -3)}\,d\gamma~+~\gamma_{\rm break}^{2\alpha_{\rm 2} -3}\int\limits_{\gamma_{\rm break}}^{\gamma_{\rm max}} F(x)~x~\gamma^{-(2\alpha_{\rm 2} -3)}\,d\gamma\Big\}~+ \\
I_{\rm x}\nu^{-2.1}\Bigg)~e^{-(\frac{\nu_{\rm t}}{\nu})^{2.1}} + T_{\rm e}\Bigg(1 - e^{-(\frac{\nu_{\rm t}}{\nu})^{2.1}}\Bigg).
\label{eq:GMOSS_steep} 
\end{multline}
Here $\alpha_1$ and $\alpha_2$ denote the spectral indices of electrons over Lorentz-factor distributions in the range $\gamma_{\textrm{min}}$ and $\gamma_{\textrm{max}}$ with a break at $\gamma_{\textrm{break}}$. $\alpha_1$ and $\alpha_2$ are both constrained to lie in the typically observed range $2\le\alpha_1,\alpha_2\le3$ and are related by the change in spectral index $\delta_{\alpha} = \alpha_2 - \alpha_1$,  with $\delta_{\alpha} \ge 0$. $T_{\textrm{e}}$ parameterizes the electron temperature and is constrained to be no greater than $10^4$~K. $\nu_{\textrm{t}}$ represents the low-frequency absorption turnover where the medium becomes optically thick. $I_{\textrm{x}}$ represents the optically thin thermal component that dominates toward higher frequencies. $C_{\rm 1}$ represents the normalization. $x$ and $F(x)$ follow standard notation in synchrotron radiation literature for which we refer the reader to \cite{rl1986}.\\\\ 
For the case of pixels with high-frequency spectral flattening, the model adopted is a composite of emission from steep and flat spectrum sources in which the spectrum is described by
\begin{equation}
\begin{split}
%\MoveEqLeft
T(\nu) = C_1\Bigg(\nu^{-\alpha_{\rm 1}} + \frac{C_{\rm 2}}{C_{\rm 1}}\nu^{-\alpha_{\rm 2}} + I_{\rm x}~\nu^{-2.1}\Bigg){\rm e}^{-(\frac{\nu_{\rm t}}{\nu})^{2.1}}+ T_{\rm e}\Bigg(1 - e^{-(\frac{\nu_{\rm t}}{\nu})^{2.1}}\Bigg).
\end{split}
\label{eq:GMOSS_flat}
\end{equation}
In this model, $\frac{C_1}{C_2}$ denotes the relative contributions from steep and flat spectrum sources with spectral indices $\alpha_1$ and $\alpha_2 = \alpha_1+\delta_{\alpha};~\delta_{\alpha}\le0$.  Other notations and constraints are the same as those for Equation~\ref{eq:GMOSS_steep} above.\\\\
There are seven parameters describing the physical processes to fit six data points at each sky pixel, which may seem an over fit. However, the total spectra that incorporate thermal free-free and synchrotron processes simply cannot fit any arbitrary set of six measurement points for two reasons.  First, the overall shape of the spectrum is strongly constrained to be of the form corresponding to the sum of free-free and synchrotron emission. Second, physically motivated constraints have been placed on the electron temperature of the thermal foreground medium (Te) and the two spectral indices describing the synchrotron emission. Additionally, there exists interdependence between these two spectral indices: the second spectral index is parameterized as an offset from the first, while still being bound by the aforementioned constraints. 
Thus, the seven parameters that describe the physical models of GMOSS do not over fit the six data points at each pixel. Consistent with this expectation, GMOSS spectra do not exactly pass through all six data points and the residuals to the fit are in the ballpark of the measurement errors in the data.\\\\
The synthetic spectra from GMOSS over the 40-200~MHz band and over HEALPix pixels of $5^{\circ}$ over the whole sky are averaged over telescope beams to yield mock observations.  Residuals on fitting polynomials of varying order to such mock observations are given in Fig.~(\ref{fig:resi_GMOSS}a) and Fig.~(\ref{fig:resi_GMOSS}b) for two LSTs where the telescope beam points off and on the Galactic plane, respectively. The complexity in spectral structure due to the underlying physical processes necessitates a polynomial of order seven in log($T$) versus log($\nu$) space to fit the mock observations of the foreground to an accuracy sufficient to be able to limit residuals to a few mK and well below that of the EoR signal.
We infer that increasing the complexity of the model that describes the foreground demands, unsurprisingly, higher order polynomials to model the mock observations to sufficient accuracy for detection of the EoR. This trend is shown in Fig.~(\ref{fig:resi_rms_poly}), which plots the root mean square (RMS) of residuals against the polynomial order; these residuals were obtained on fitting mock observations generated assuming the three different sky models.  When the order of the fitting polynomial is increased to a sufficiently large value, all curves will eventually reach an asymptotic RMS level corresponding to the numerical noise inherent in the mock spectra. The solid lines represent residuals corresponding to spectra away from the Galactic plane and the dashed lines are for those looking at the Galactic plane. 
Although at first glance a low RMS residual might be encouraging, it may be noted that increasing the order of the fitting polynomial would also fit to the EoR signal, thus subsuming the EoR signal into the estimate of the foreground and hence compromising the detection. This is demonstrated by fitting polynomials of varying order to mock observations that also contain the generic vanilla model for the EoR signal. GMOSS is adopted for describing the foreground sky spectrum since it is physically motivated and most realistic.  The residuals in this case are shown in Fig.~(\ref{fig:EoR_res_poly}).  Increasing the order of the polynomial changes the form of the residual: the peak amplitude of the residual progressively diminishes and the number of turning points in the residual increases, and thus the residual progressively departs from the form of the generic vanilla model for the EoR signal.
To fit and subtract foregrounds or for joint modeling of the foreground with the EoR signal, without compromising the information contained in the EoR signal and preserving the turning points and amplitude of the signal, we propose below that foreground components of spectral radiometer measurements be modeled not with polynomials but with Maximally Smooth (MS) functions. 
\section{Modeling the foreground using MS functions}
\label{sec:fit}
\begin{figure*}[t]
\begin{minipage}{0.5\textwidth}
\subfloat[cos$^2$ beam]{\includegraphics[clip,width=\columnwidth]{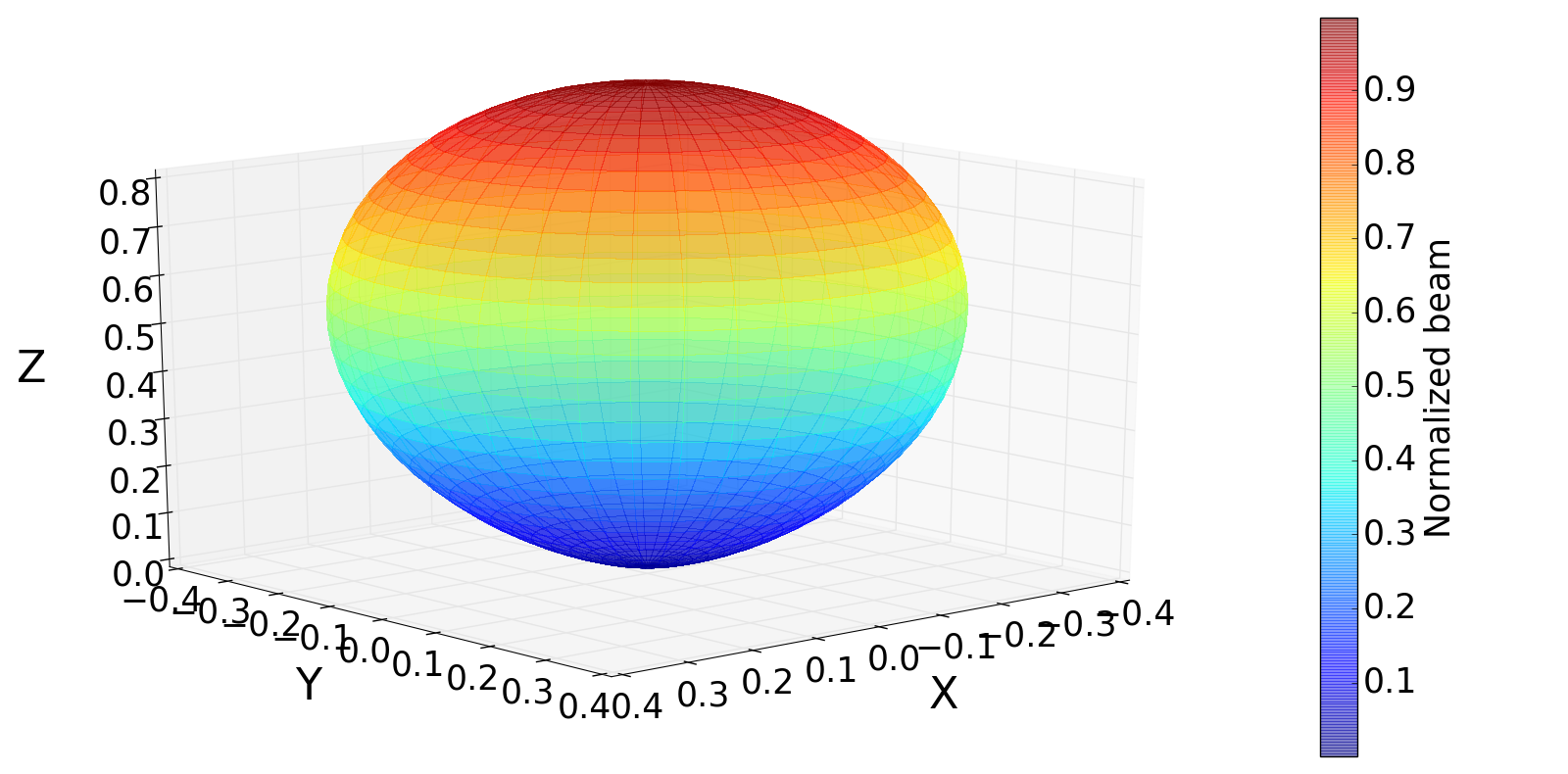}\label{fig:cos2_beam}
}
\subfloat[SARAS 2 beam]{\includegraphics[clip,width=\columnwidth]{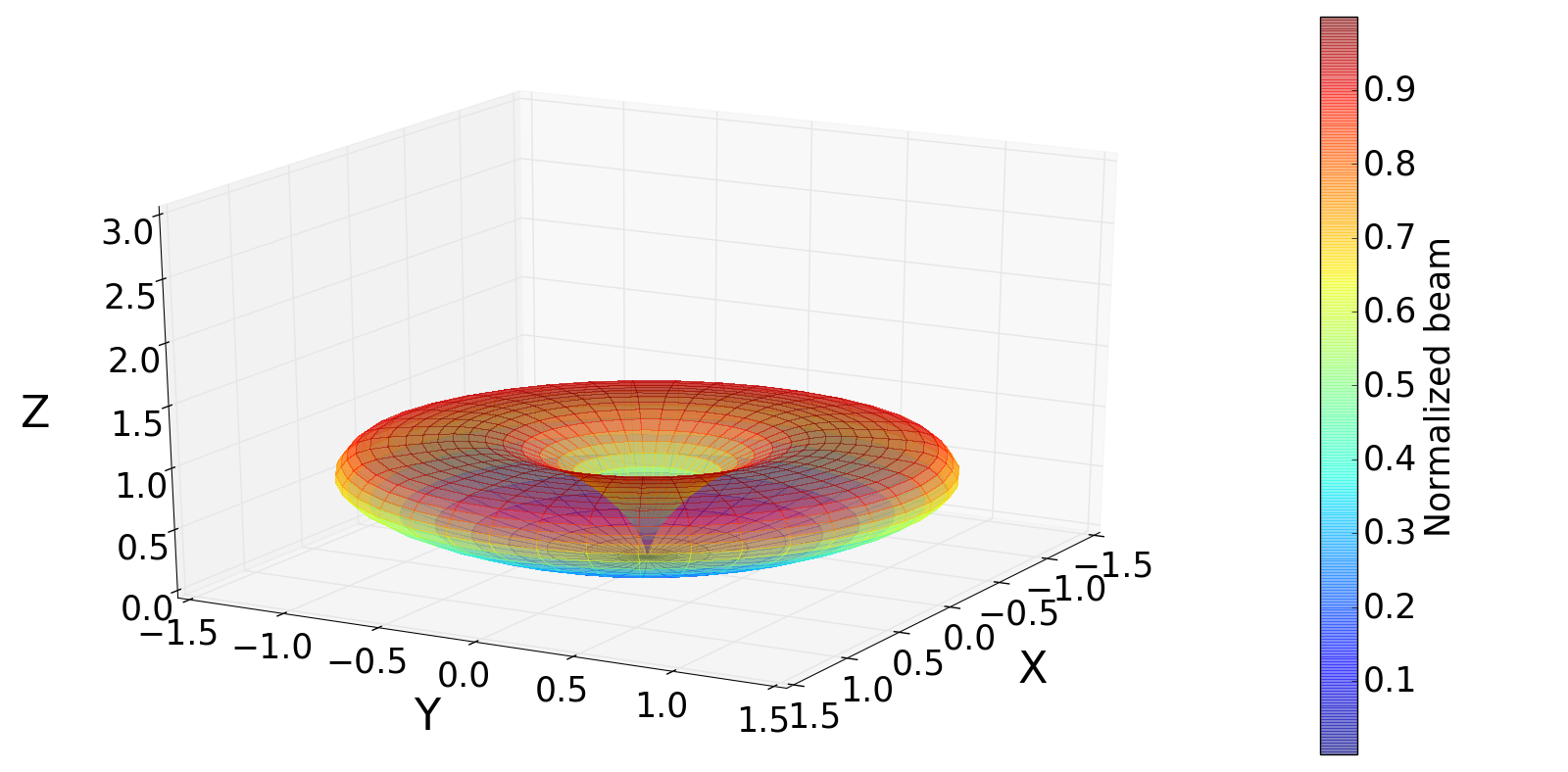}\label{fig:saras2_beam}
}
\end{minipage}
\caption[Frequency-independent antenna beam patterns adopted in simulations of mock sky spectra]{Frequency-independent antenna beam patterns adopted in simulations of mock sky spectra. The cos$^2$ beam pattern in panel (a) represents a beam with a peak toward the zenith; this is the radiation pattern of a conical spiral antenna. The SARAS 2 beam in panel (b), on the other hand, has a null toward the zenith and a peak toward an elevation of 30$^{\circ}$; this is the radiation pattern of a short monopole. Both beams have a null toward the horizon. The beam shapes have been chosen to be complementary to one another and provide different beam-weighted samplings of the sky.}
\label{fig:beams}
\end{figure*}

We hereinafter adopt GMOSS as the sky model due to its physically motivated nature. As described above and in \cite{GMOSS2017}, GMOSS describes individual pixels ($5^{\circ}$ wide) to have varying spectral shapes such as concave, convex, or of more complex form. We now examine whether beam-averaged mock spectra generated using GMOSS are smooth in the sense that they do not have embedded small-amplitude ripples that may confuse a detection of fine-scale cosmological signals from reionization.\\\\ 
We attempt to fit the mock spectra using MS functions \citep[see][]{apsera2015}.  An MS function $f(x)$ is a polynomial of degree $n$
\begin{equation}
\begin{split}
%\MoveEqLeft
f(x) = p_0 + p_1 (x -x_0) + p_2 (x - x_0)^2 + p_3 (x - x_0)^3
&+ p_4 (x - x_0)^4 + ... + p_\textrm{n} (x - x_0)^\textrm{n}
\end{split}
\end{equation}
in which the polynomial coefficients $p_j$ are constrained so that there are no zero crossings within the domain for any derivative of order $m \ge 2$.  The order-$m$ derivative of the polynomial is 
\begin{equation}\label{eq:MS_def}
\frac{\text{d}^mf(x)}{\text{d}x^m} = \sum_{i=0}^{n-m} \{(m+i)! / i! \} p_{m+i} (x - x_0)^i 
\end{equation}
and the coefficients $p_j$ are constrained so that for all $m$ in the range  2, 3, 4, .... $(n-1)$ the above derivative functions are never zero within the domain of interest. 
Thus the smooth component in a data set is described by an MS function with its coefficients optimized by minimizing the chi-square ($\chi^2$) between data and the MS functional form:
\begin{equation}
\chi^2 = \frac{(10^{ \sum_{i=0}^{n} \left[\log_{10}(x/x_0)\right]^i \,p_{i}} - \textrm{data})^2}{\textrm{number of data points}}.
%sqrt(((func(p,x)-y)**2).sum()/float(len(x)))
\end{equation}
Coefficients $p_i$ ($i\ge2; ~i\subset \mathbb{Z}$) are disallowed from changing signs within the domain of the spectrum and hence the fitting polynomial is consistent with the condition for smoothness.\\\\
Critical to the modeling is the formulation of a robust method for solving for the parameters: the MS function is defined in log-temperature versus log-frequency space.  In this space, the function is written as an expansion about log($x0$), where $x0$ is a pivot frequency.  The coefficients of the expansion are constrained so that there are no zero crossings in the second derivative and higher derivatives. Beginning with a low-order polynomial $f(x)$, the order $n$ of the MS function is incremented by unity in each iteration and the parameters returned in the previous step are used as initial guess in the next.\\\\ 
Constraining the polynomial in this manner while fitting to the measured sky spectrum in log($T$)-log($\nu$) space allows the function to fit to the mean spectral index, a constant spectral curvature and higher order curvatures without allowing the polynomial to follow any ripple or multiple turning points in embedded spectral components.
\begin{figure*}[t]
\begin{minipage}{0.5\textwidth}
\subfloat[]{\includegraphics[clip,width=\columnwidth]{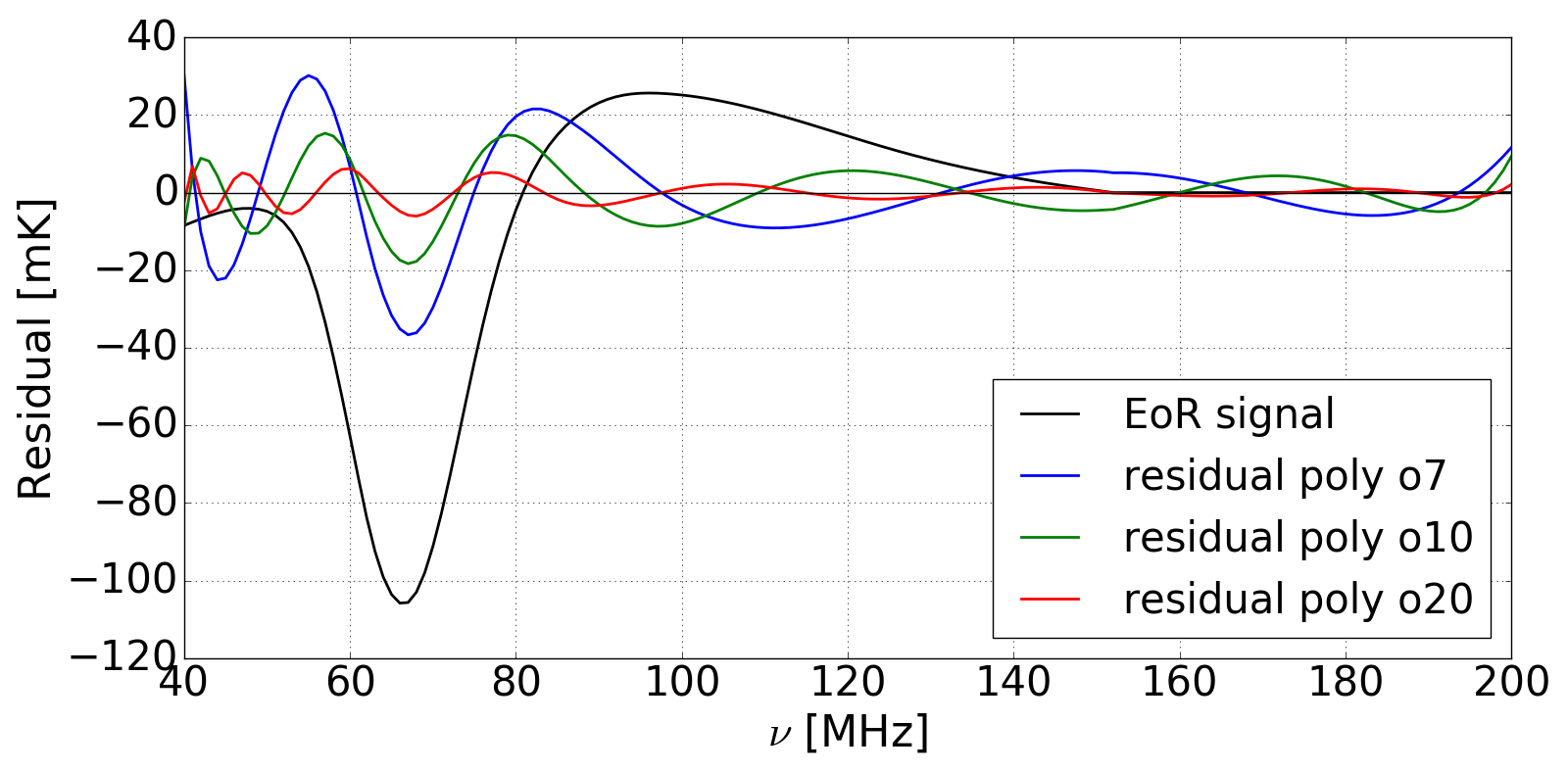}\label{fig:EoR_MS_polya}
}
\subfloat[]{\includegraphics[clip,width=\columnwidth]{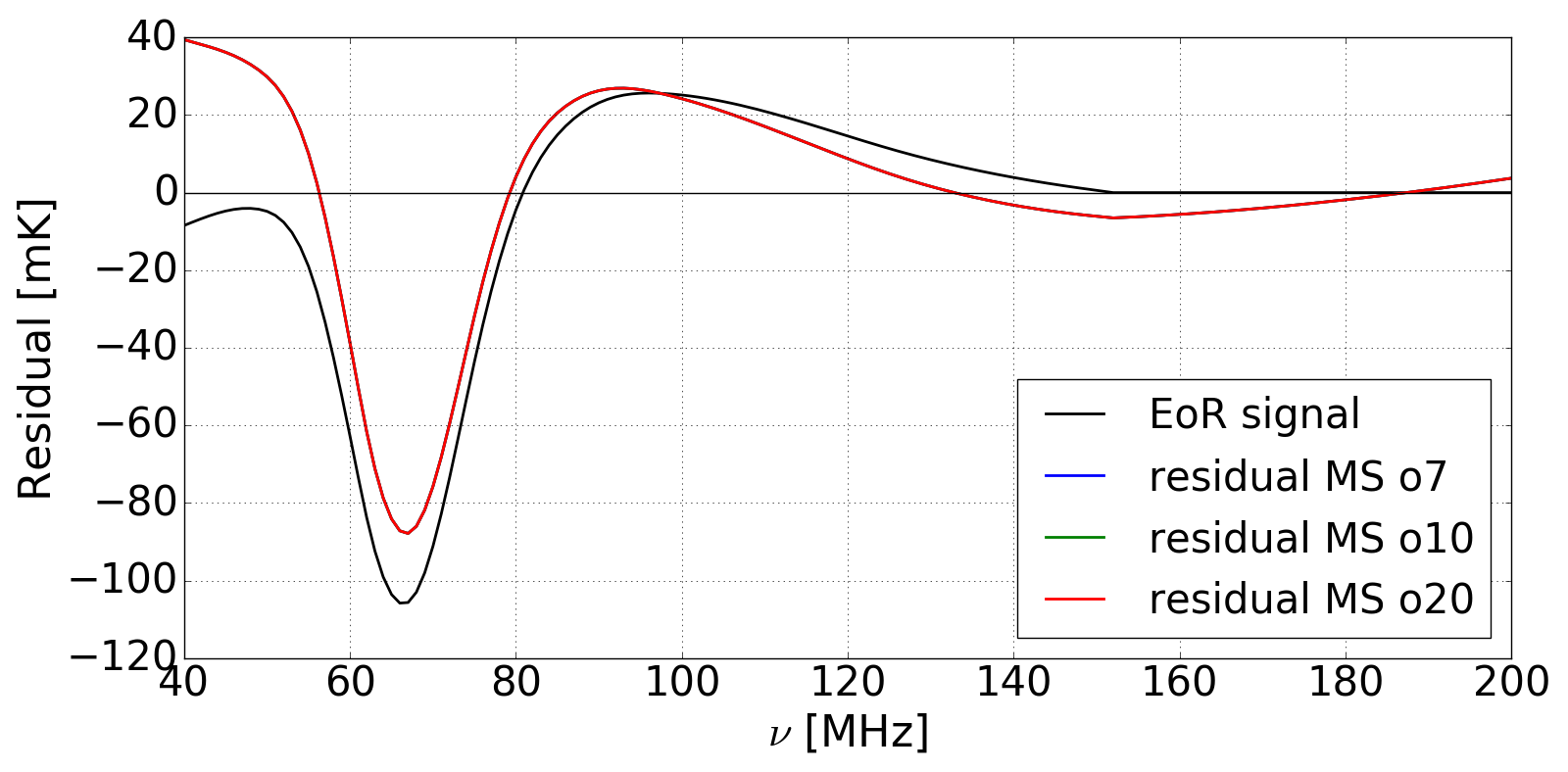}\label{fig:EoR_MS_polyb}
}
\end{minipage}
\caption[Residuals on fitting the global EoR signal with varying orders of polynomials and MS functions]{Residuals on fitting the global EoR signal with varying orders of (a) polynomials and (b) MS functions; fits are made in log($T$)-log($\nu$) space to the assumed global EoR signal plus CMB with temperature $T_{\circ} = 2.73$~K.  Residuals to polynomial fits substantially differ from the EoR signal assumed, displaying reduced amplitude as well as multiple turning points. On the other hand, fitting the EoR signal with an MS function of arbitrary order preserves turning points and after the smooth component of the signal is entirely removed, the residual saturates and does not change any further on increasing the order of the MS function as evidenced by the curves of all residuals lying perfectly one above the other. Shown as a black solid line in both panels is the global EoR signal that is being fit to. A polynomial of order at least 7 is adopted as is necessitated by GMOSS, as described in Section \ref{sec:GMOSS}.}
\label{fig:EoR_MS_poly}
\end{figure*}
As a first step, to demonstrate that there is no loss in the EoR signal on modeling the foreground using an MS function, we examine the residuals on fitting the adopted generic form of the EoR signal with MS functions and compare with residuals on fitting them with polynomials that have no constraints on coefficients.  We present in Fig.~(\ref{fig:EoR_MS_poly}) residuals on fitting the global EoR signal, which has  multiple turning points over the 40--200~MHz band and is hence not MS, with polynomial and MS functions of varying orders.  We add the CMB monopole temperature to the global EoR signal and then fit the resulting spectrum in log($T$)-log($\nu$) space. Since foregrounds generated using GMOSS as the sky model  require at least a seventh order polynomial in log($T$)-log($\nu$) space to fit to the accuracy needed for global EoR detection, we fit foregrounds with polynomials of orders 7, 10 and 20. As shown in Panel (a) of Fig.~(\ref{fig:EoR_MS_poly}), modeling the EoR signal with polynomials results in turning points being introduced in the residual, which were not present in the EoR signal itself.  As a consequence, the shape of the residual is no longer qualitatively similar to the cosmological signal. Further, the amplitude of the residual progressively reduces and if fit with a sufficiently high-order polynomial the signal risks being subsumed in any polynomial model for the foreground and hence being entirely fitted out.  As a comparison, residuals on modeling the same spectrum with MS functions of orders 7, 10 and 20 are shown in Panel (b) of Fig.~(\ref{fig:EoR_MS_poly}). Clearly, the residual obtained on fitting MS functions retains all the turning points of the EoR signal. The residual is similar but not identical to the EoR signal itself and represents a smooth baseline subtracted version of the signal. Additionally, increasing the order of the MS function does not deteriorate the amplitude of the signal in the residual once all the smooth components in the data have been removed. This suggests that if the only non-smooth component in an EoR detection experiment is the signal itself, fitting an MS function of arbitrarily high order should leave behind in the residual a distinctive signature of the EoR signal with all turning points intact and with minimal degradation of signal strength.\\\\ 
As a second step, we examine whether beam-averaged mock sky spectra generated using GMOSS as the sky model are indeed maximally smooth, at least to the precision needed to detect an embedded generic EoR signal. We generate mock sky spectra at different observing sites to include different parts of the sky, and hence investigate the smoothness of spectra over different samplings of sky coverage and spectral complexity. At any observing site, we generate 24 mock sky spectra one hour apart and with a uniform sampling over LST. The observing sites adopted for the simulations were (a) Hanle in North India where the Indian Astronomical Observatory is situated (Latitude $= 32.79^{\circ}$N) and (b) the Murchison Radio Observatory (MRO) site in Western Australia (Latitude $= 26.68^{\circ}$S). With large antenna beams as are typically employed by most global EoR detection experiments \citep[see][]{fatdipole2013}, the Galactic plane will come into the beam at some time in the day. Mock sky spectra were generated by simulating observations with (a) an ideal frequency-independent cos$^2$ beam pattern that has maximum toward zenith (see Figure \ref{fig:cos2_beam}) and (b) adopting the short-monopole type beam pattern of the SARAS\footnote{see {\tt http://www.rri.res.in/DISTORTION/saras.html}}~2 antenna that has null toward zenith, null toward horizon and maximum toward elevation of about $30^{\circ}$ (see Figure \ref{fig:saras2_beam}). The choice of observing sites, close to $+30^{\circ}$ and $-30^{\circ}$ of latitude, and beams that are contrasting in shapes, was made to have a wide variety in plausible spectra. The final calibrated mock spectra in each case are generated over the 40--200~MHz band and have, in addition to the foreground spectrum, contribution from a Planck form CMB with no spectral distortions.

\begin{sidewaysfigure}
\begin{minipage}{0.3\textwidth}

\subfloat[cos$^2$ beam Hanle off Galactic plane]{%
  \includegraphics[clip,width=\columnwidth,height=2.0in]{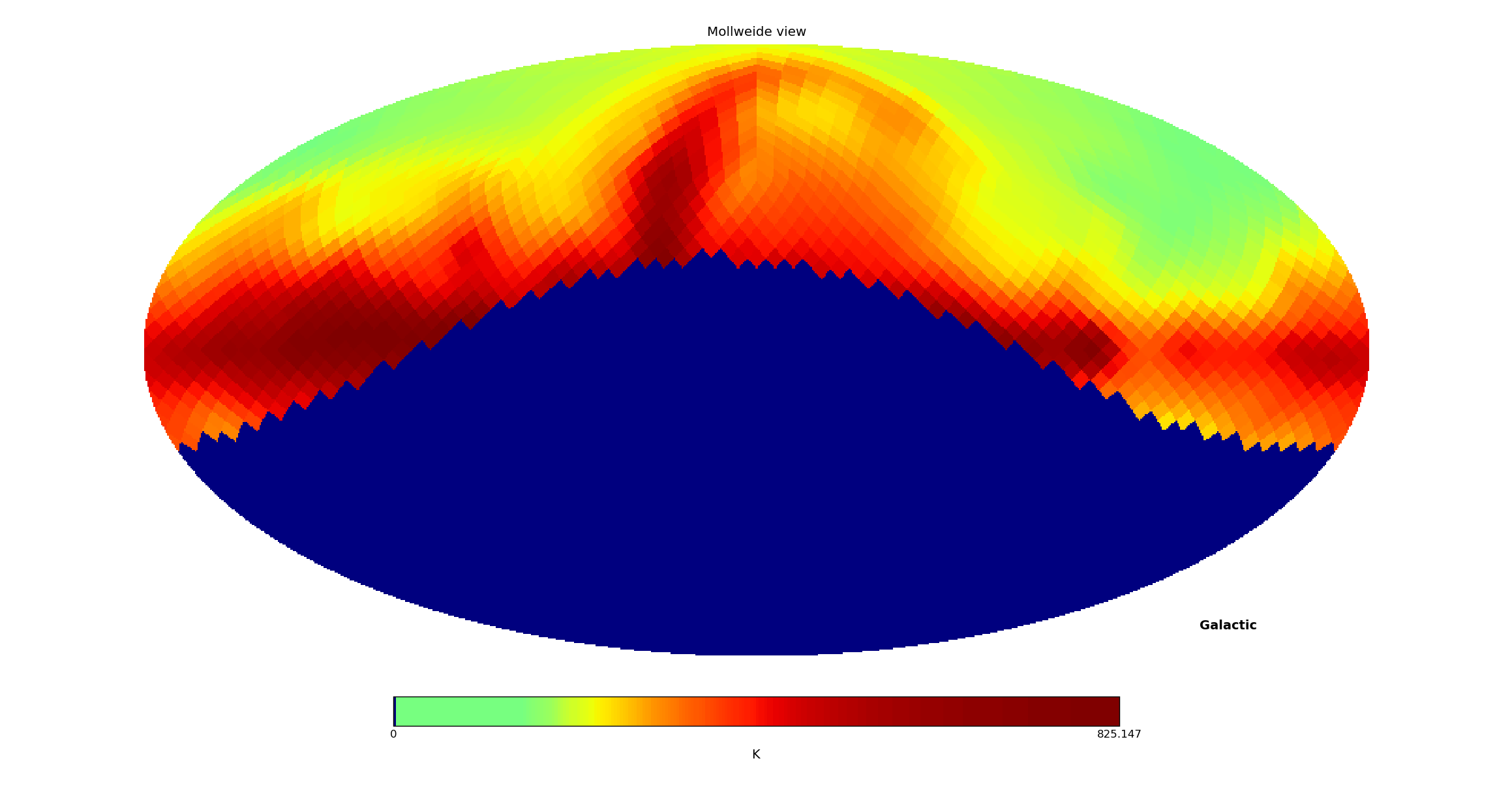}%
}

\subfloat[cos$^2$ beam Hanle on Galactic plane]{%
  \includegraphics[clip,width=\columnwidth,height=2.0in]{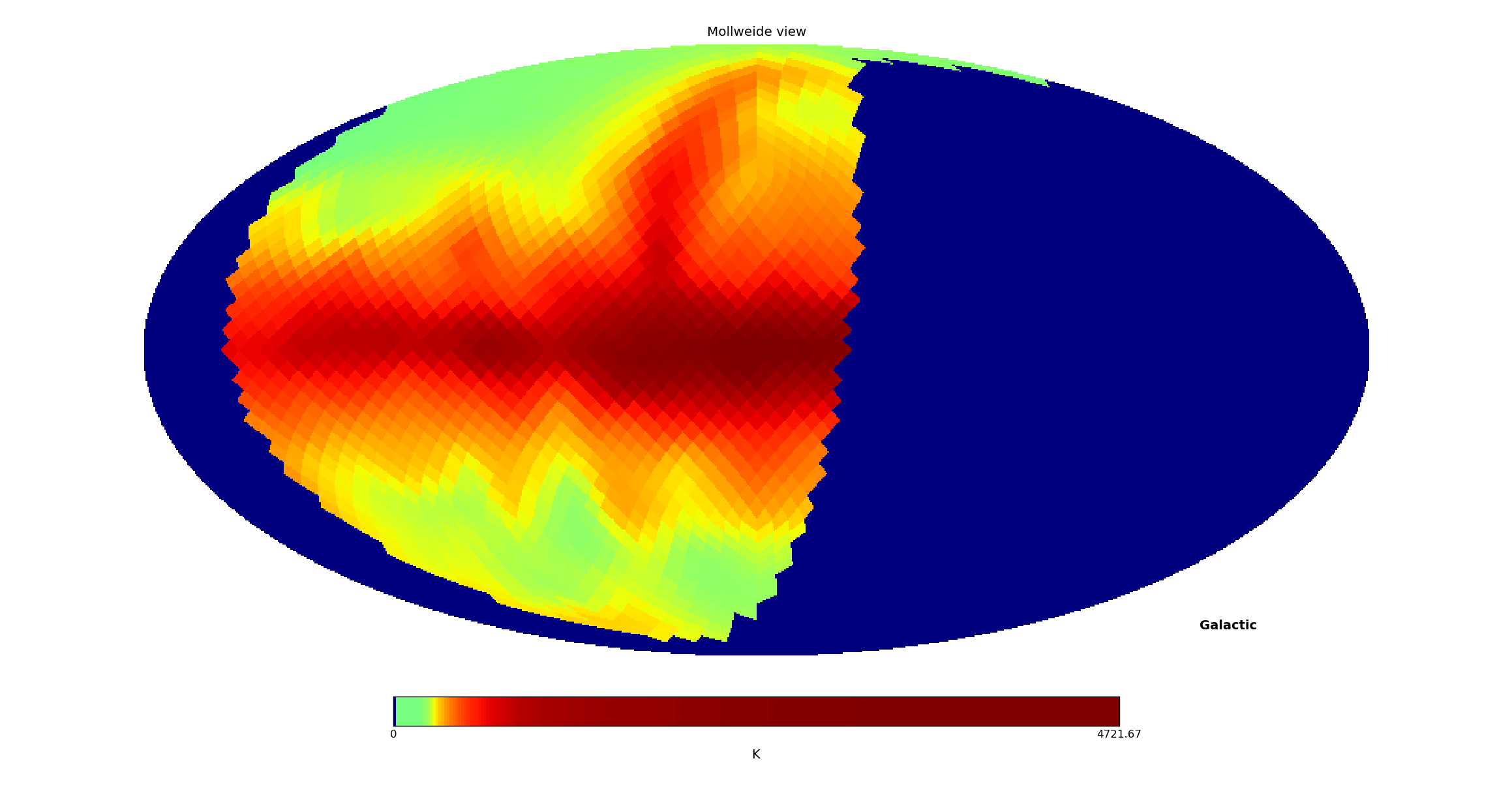}%
}

\end{minipage}
\begin{minipage}{0.3\textwidth}

\subfloat[Spectrum off Galactic Plane]{%
  \includegraphics[clip,width=\columnwidth,height=2.0in]{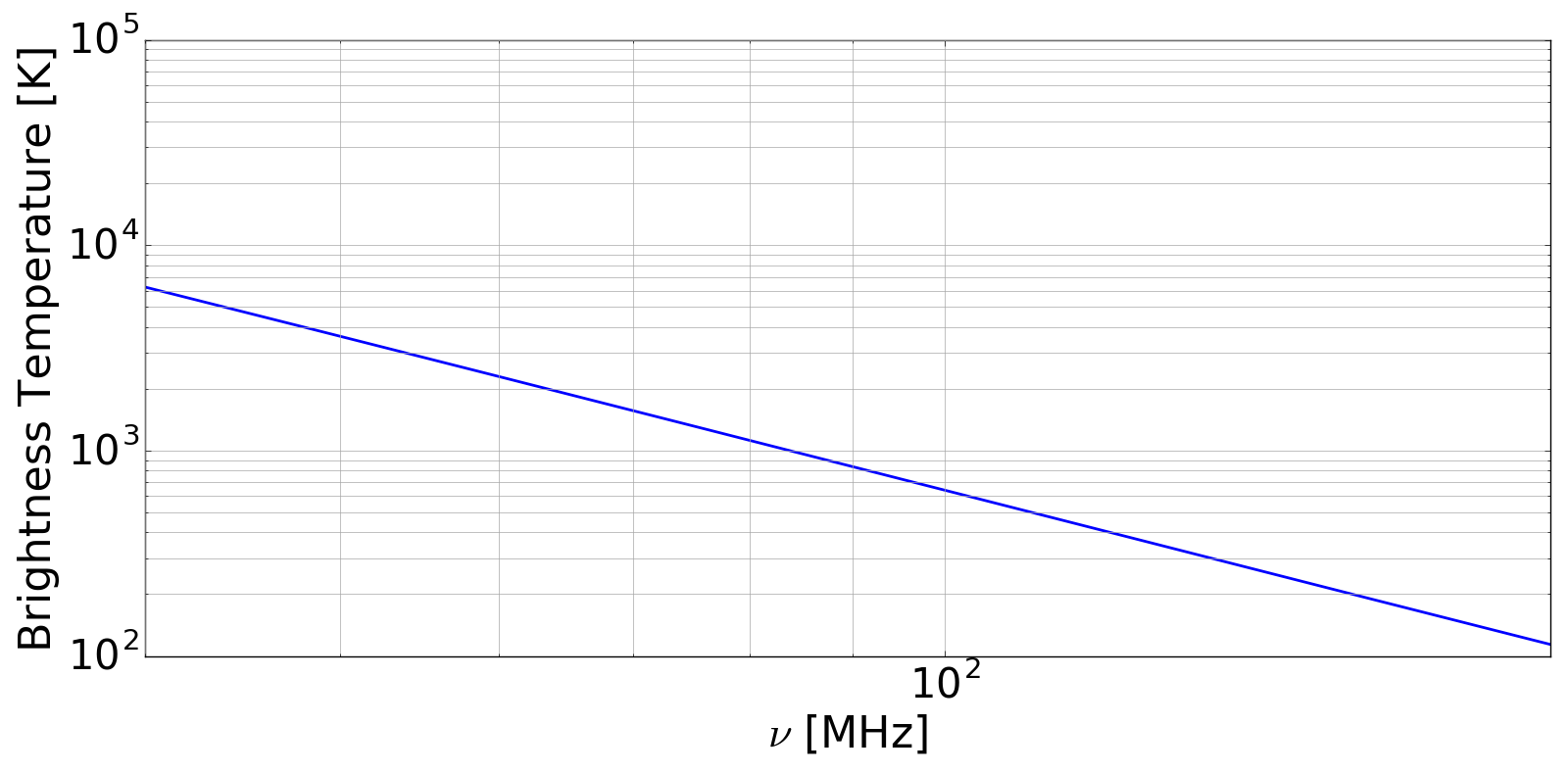}%
}

\subfloat[Spectrum on Galactic Plane]{%
  \includegraphics[clip,width=\columnwidth,height=2.0in]{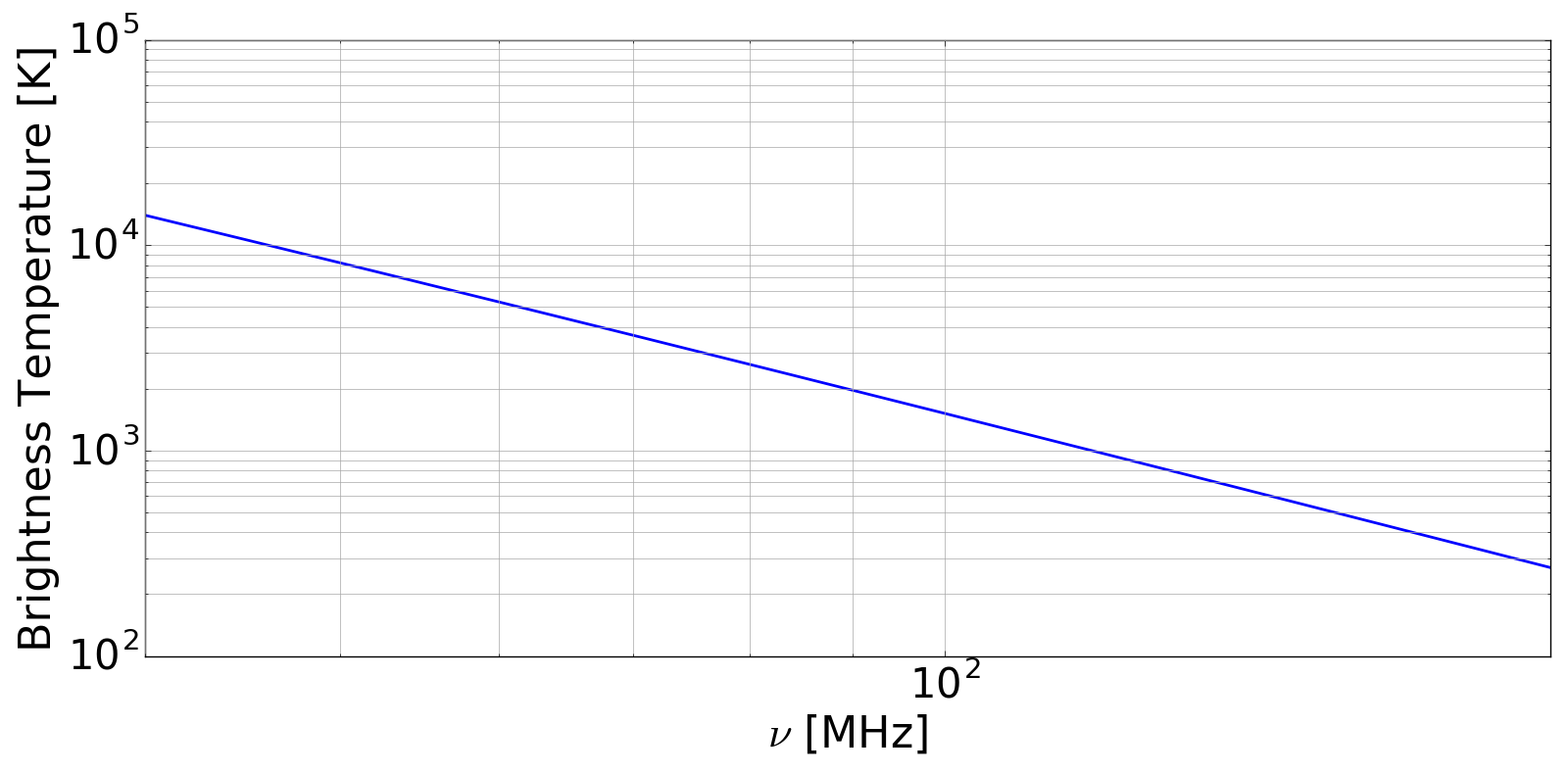}%
}
\end{minipage}
\begin{minipage}{0.3\textwidth}

\subfloat[Residual]{%
  \includegraphics[clip,width=\columnwidth,height=2.0in]{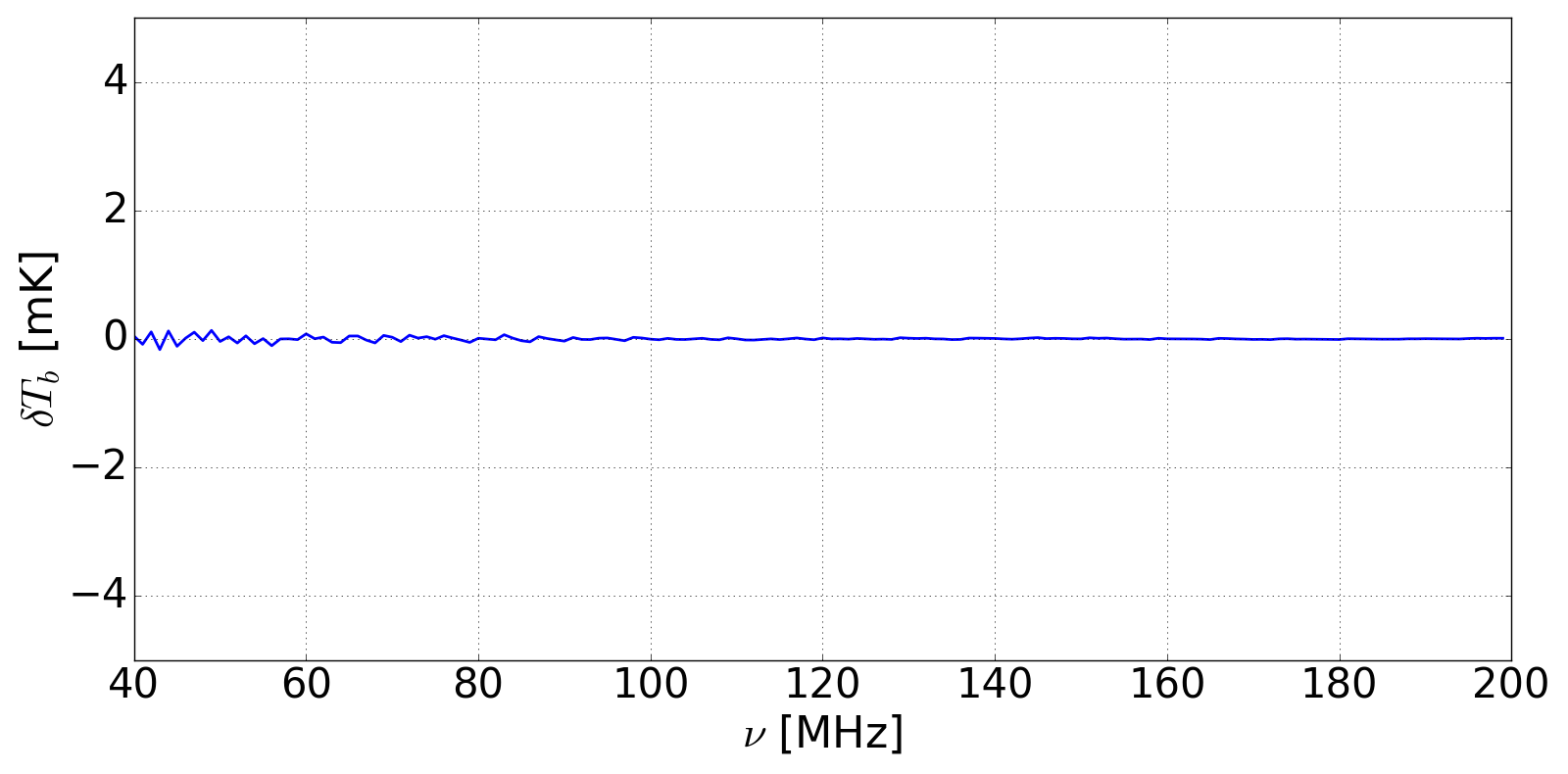}%
}

\subfloat[Residual]{%
  \includegraphics[clip,width=\columnwidth,height=2.0in]{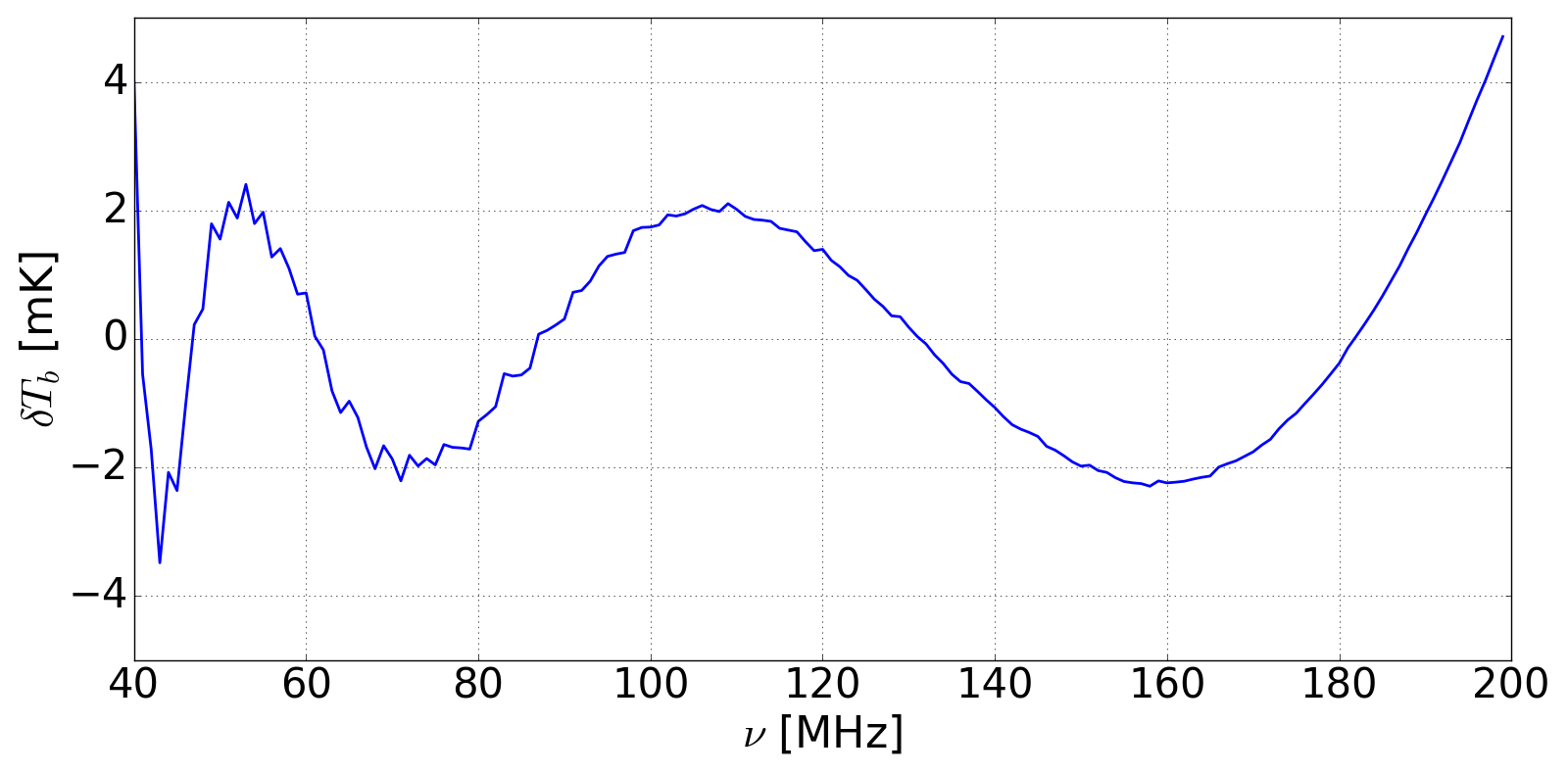}%
}
\end{minipage}
\caption[Testing the smoothness of foregrounds in mock sky spectra that are simulated using GMOSS as the sky model for the sky over Hanle (Latitude $= 32.79^{\circ}$N), as observed by an antenna beam that has a cosine$^2$ profile]{Testing the smoothness of foregrounds in mock sky spectra that are simulated using GMOSS as the sky model for the sky over Hanle (Latitude = $32.79^{\circ}$N). The first column shows the pixels that lie in the antenna beam used to generate spectra; these are displayed on an all-sky map at 150 MHz with a resolution of 5$^{\circ}$.  Pixels that lie outside the beam are blanked. The spectra themselves are shown in the second column. The third column shows the residual on fitting these spectra with the form given by Equation~\ref{eq:fit_func}, which represents our adopted Maximally Smooth model for the foreground.  Spectra are simulated for a frequency-independent beam with a cosine$^2$ pattern. Two spectra at the LSTs in which they have maximum and minimum foreground brightness are shown as examples. Three figures that follow show corresponding plots for the sky at other locations and for different assumed telescope beams.}
\label{fig:cos2_Hanle}
\end{sidewaysfigure}

\begin{sidewaysfigure}
\begin{minipage}{0.3\textwidth}

\subfloat[cos$^2$ beam MRO off Galactic plane]{%
  \includegraphics[clip,width=\columnwidth,height=2.0in]{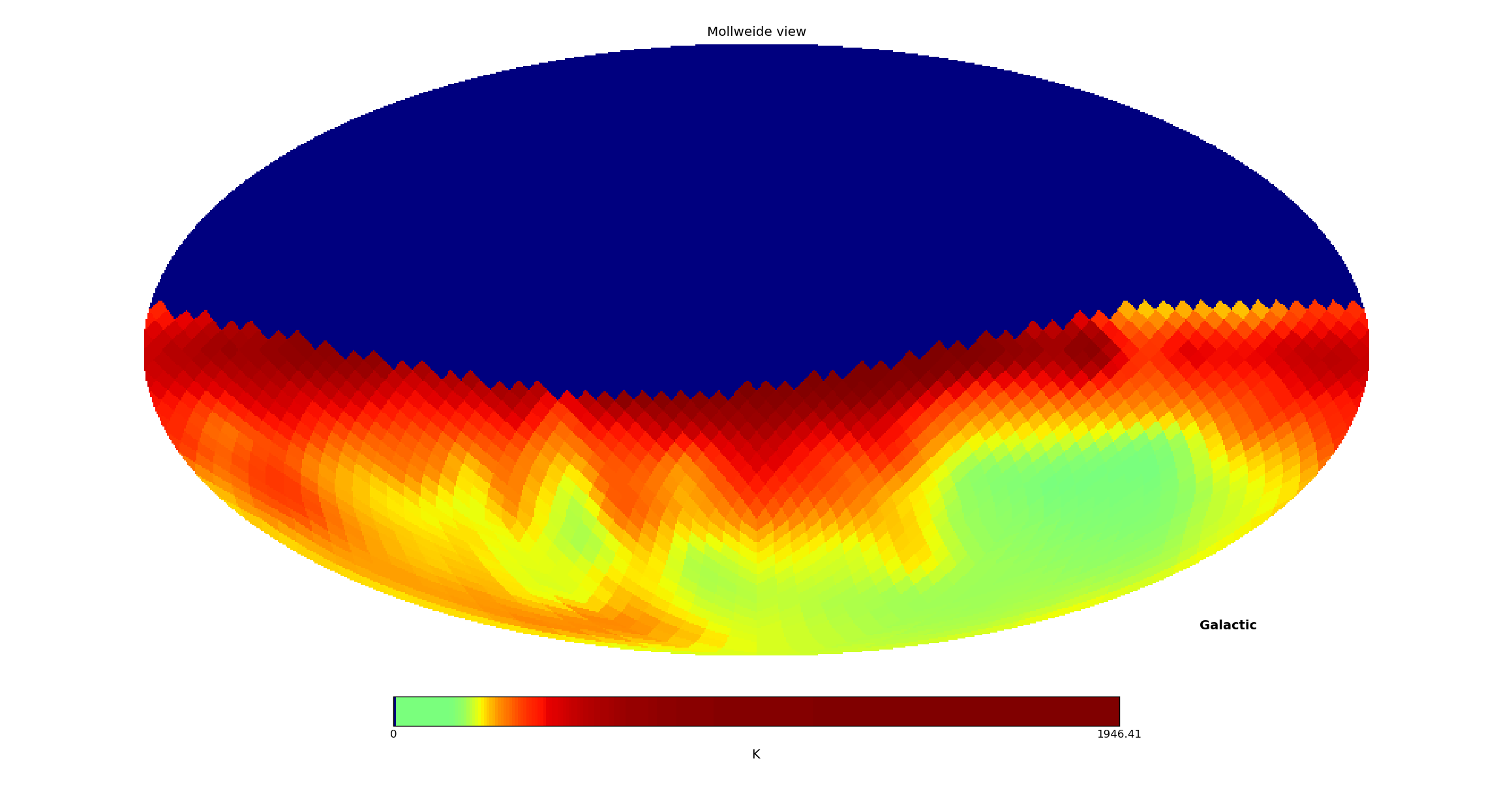}%
}

\subfloat[cos$^2$ beam MRO on Galactic plane]{%
  \includegraphics[clip,width=\columnwidth,height=2.0in]{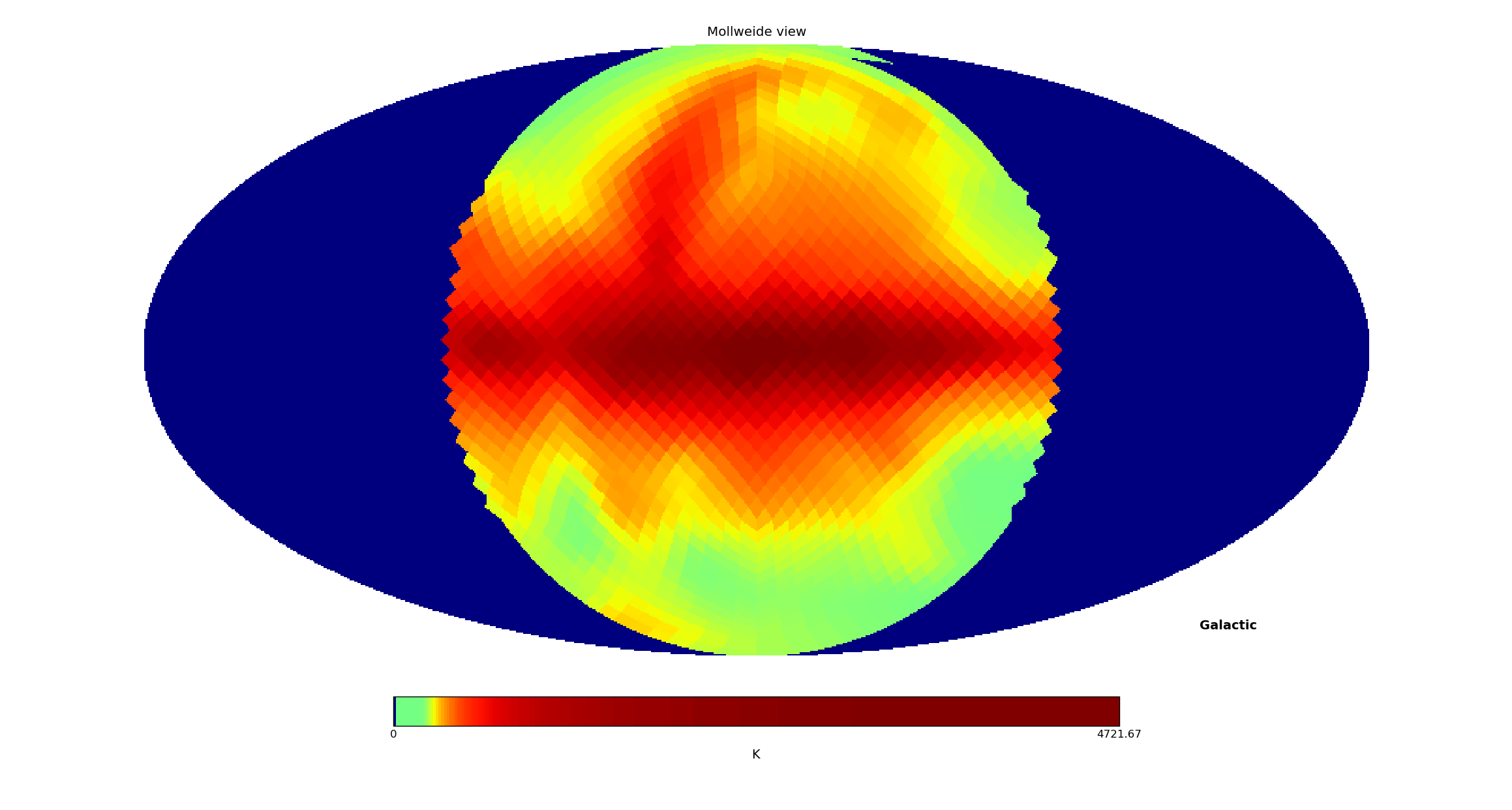}%
}
\end{minipage}
\begin{minipage}{0.3\textwidth}

\subfloat[Spectrum off Galactic Plane]{%
  \includegraphics[clip,width=\columnwidth,height=2.0in]{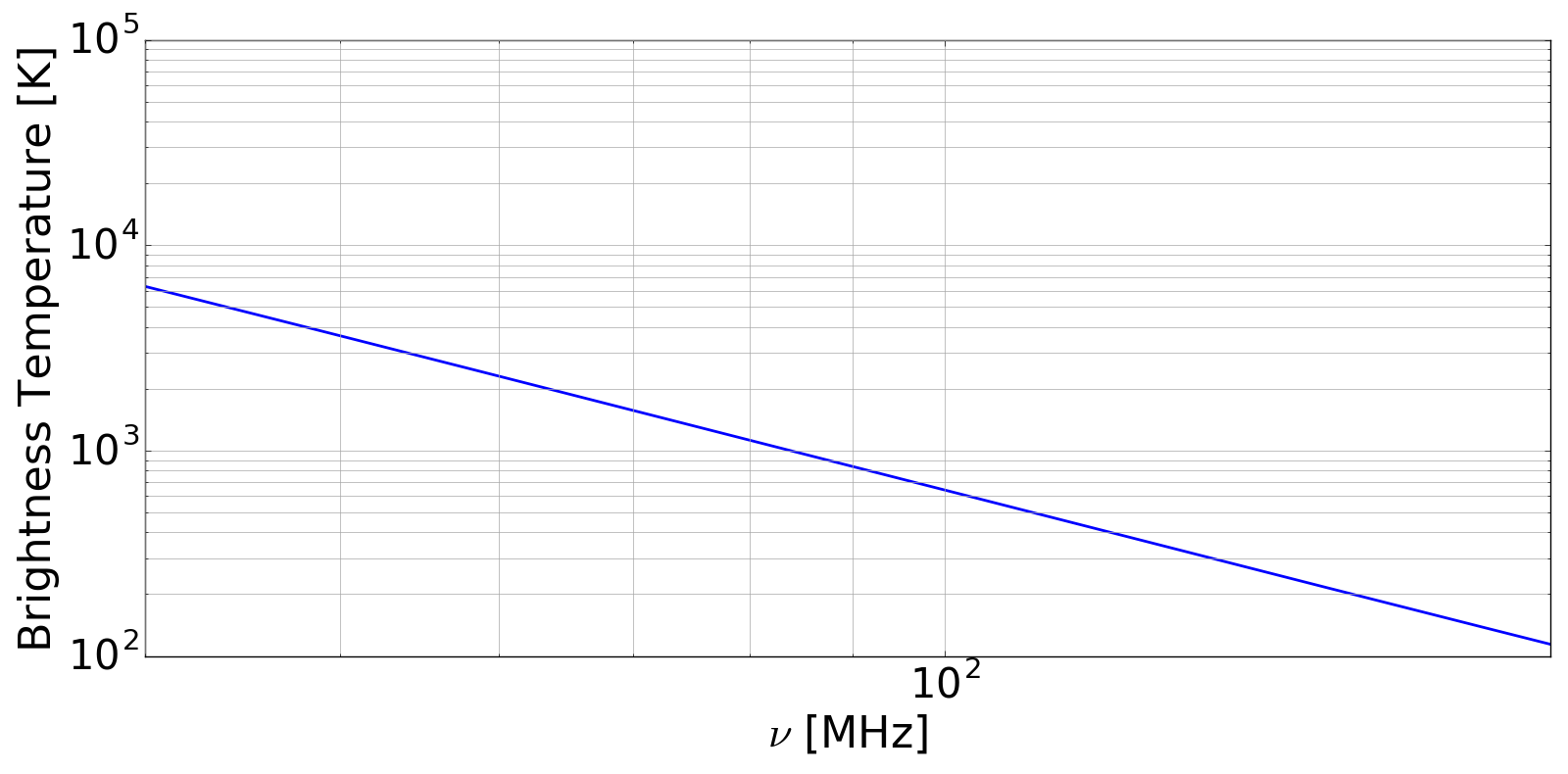}%
}

\subfloat[Spectrum on Galactic Plane]{%
  \includegraphics[clip,width=\columnwidth,height=2.0in]{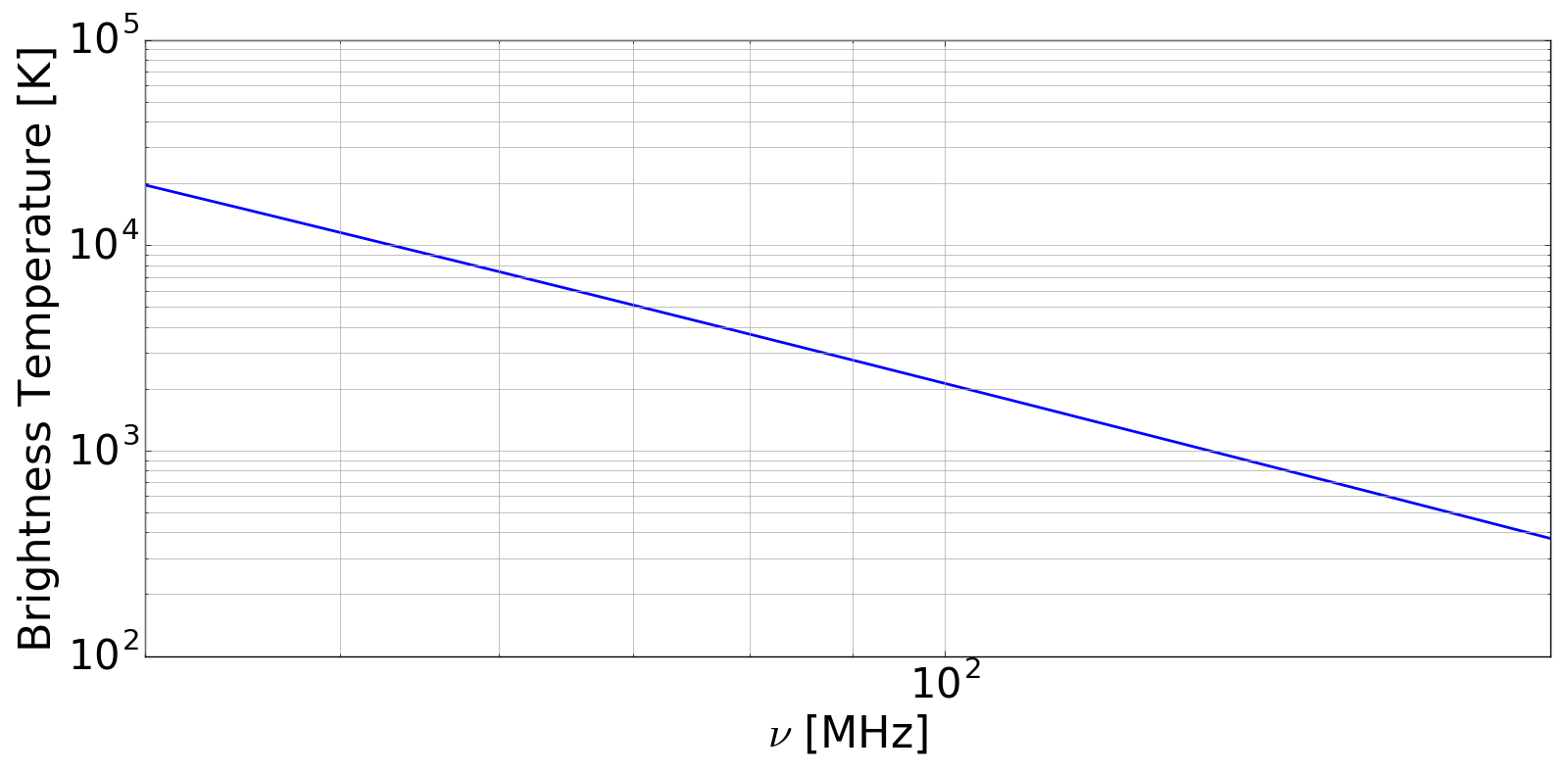}%
}

\end{minipage}
\begin{minipage}{0.3\textwidth}

\subfloat[Residual]{%
  \includegraphics[clip,width=\columnwidth,height=2.0in]{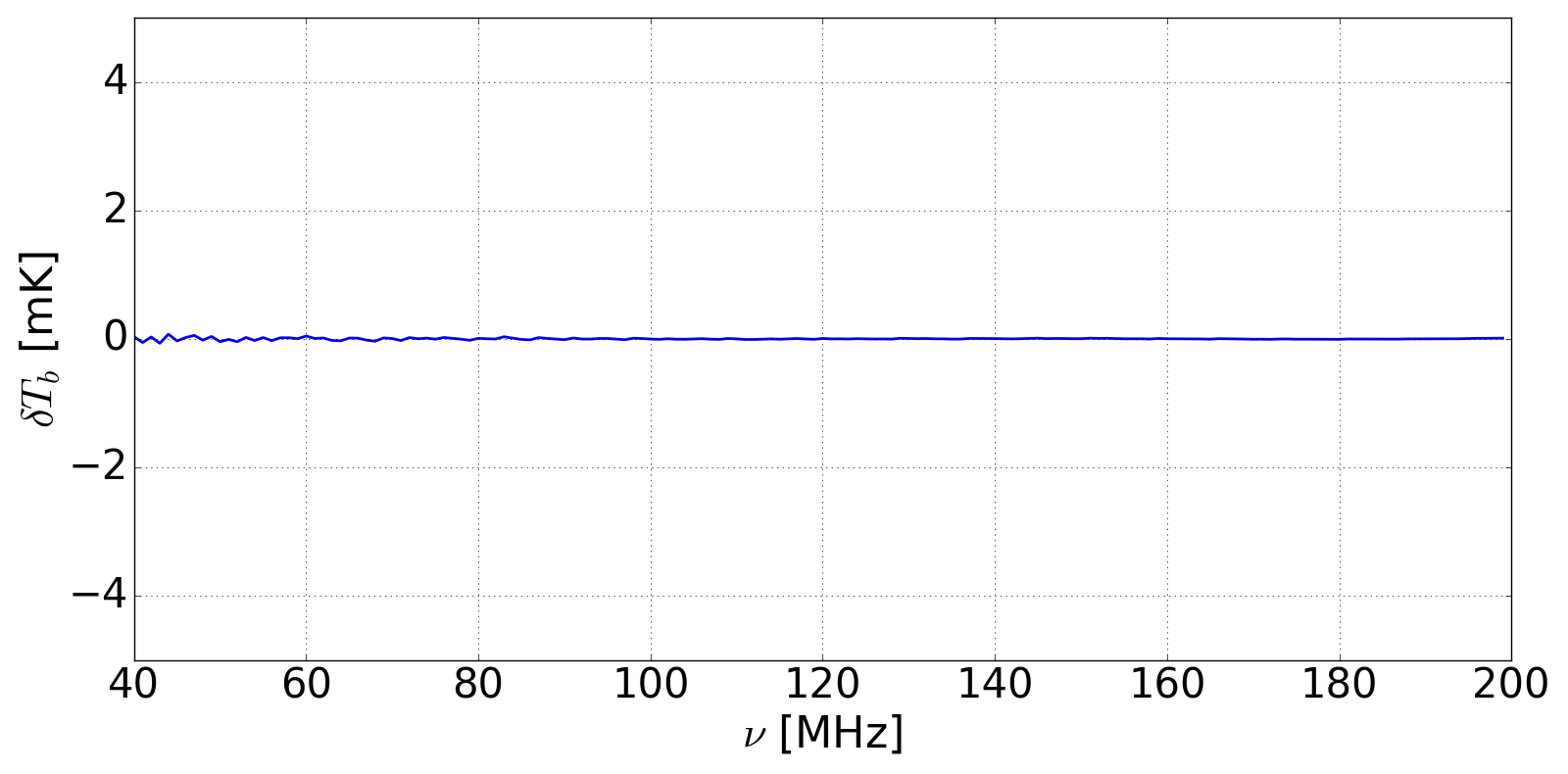}%
}

\subfloat[Residual]{%
  \includegraphics[clip,width=\columnwidth,height=2.0in]{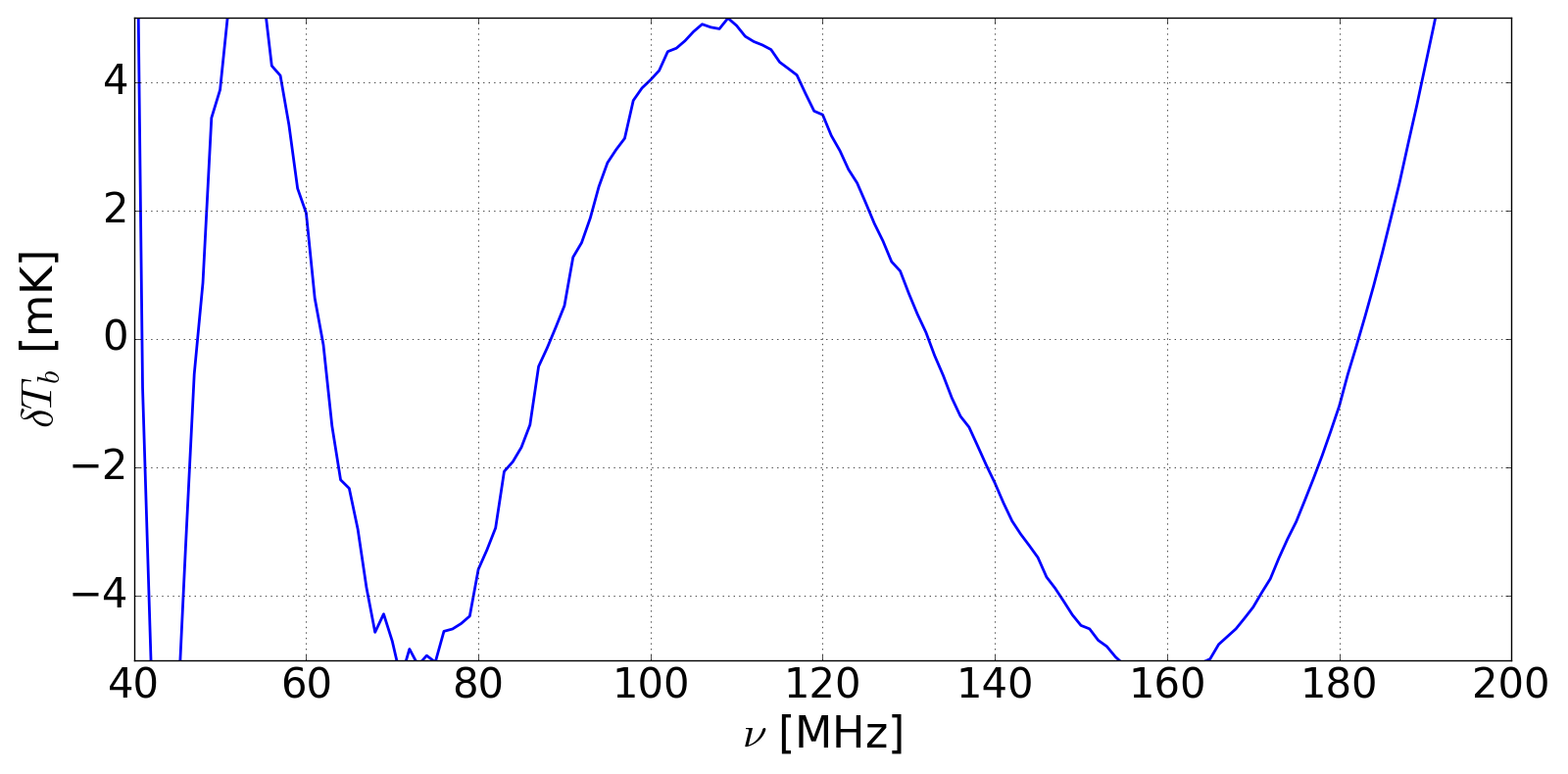}%
}
\end{minipage}
\caption[Testing the smoothness of foregrounds in mock sky spectra that are simulated using GMOSS as the sky model for the sky over MRO (Latitude = 26.68$^{\circ}$S), as observed by an antenna beam that has a cosine$^2$ profile]{
Same as Fig.~\ref{fig:cos2_Hanle}, but with mock sky spectra simulated using GMOSS for the sky over MRO (Latitude $= 26.68^{\circ}$S). 
uency-independent beam with cosine$^2$ pattern. Two spectra at the LSTs in which they have maximum and minimum foreground brightness are shown, as examples.}
\label{fig:cos2_MRO}
\end{sidewaysfigure}

\begin{sidewaysfigure}
%%\ContinuedFloat
\begin{minipage}{0.3\textwidth}

\subfloat[SARAS2 beam Hanle off Galactic plane]{%
  \includegraphics[clip,width=\columnwidth,height=2.0in]{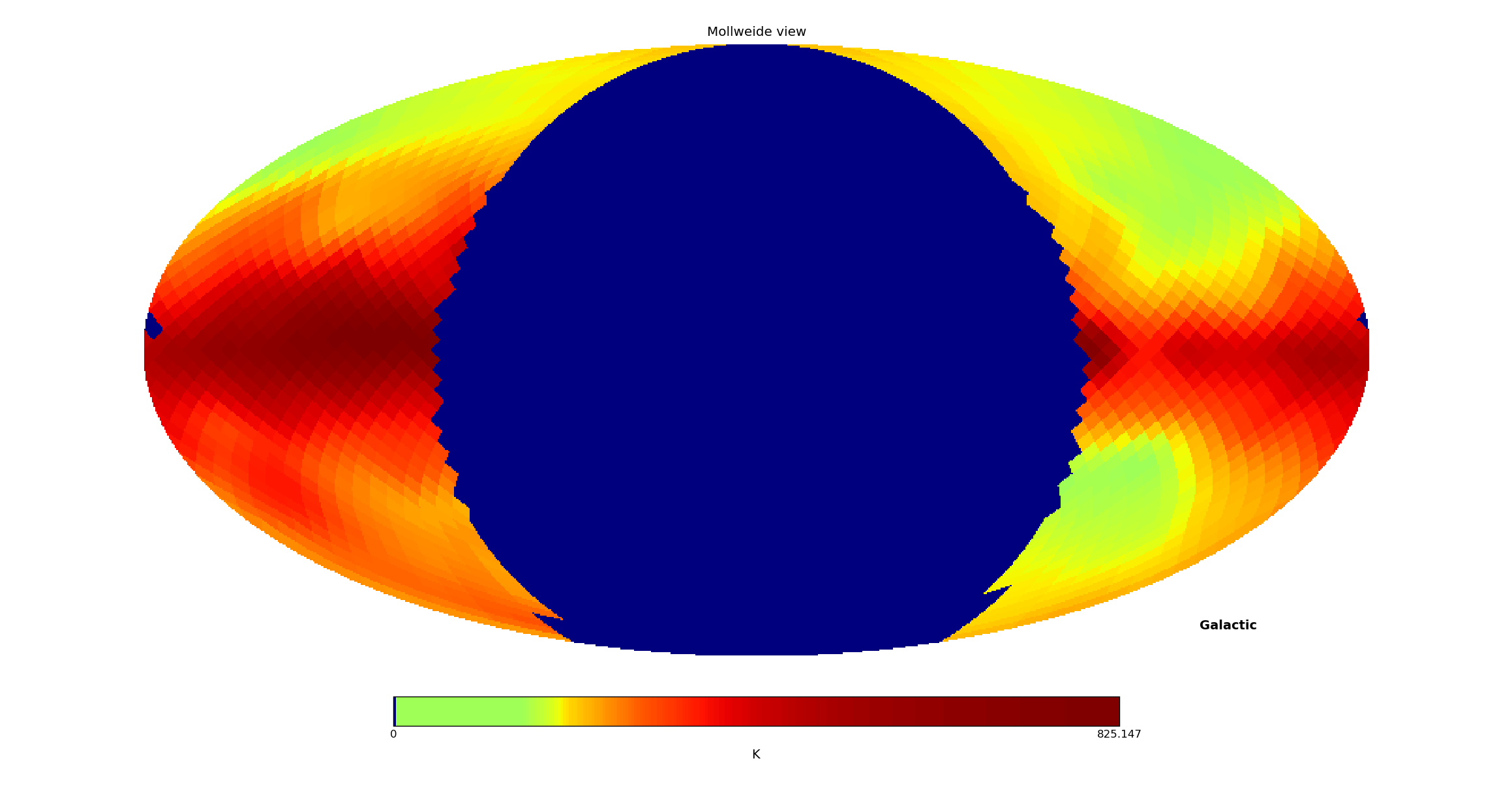}%
}

\subfloat[SARAS2 beam Hanle on Galactic plane]{%
  \includegraphics[clip,width=\columnwidth,height=2.0in]{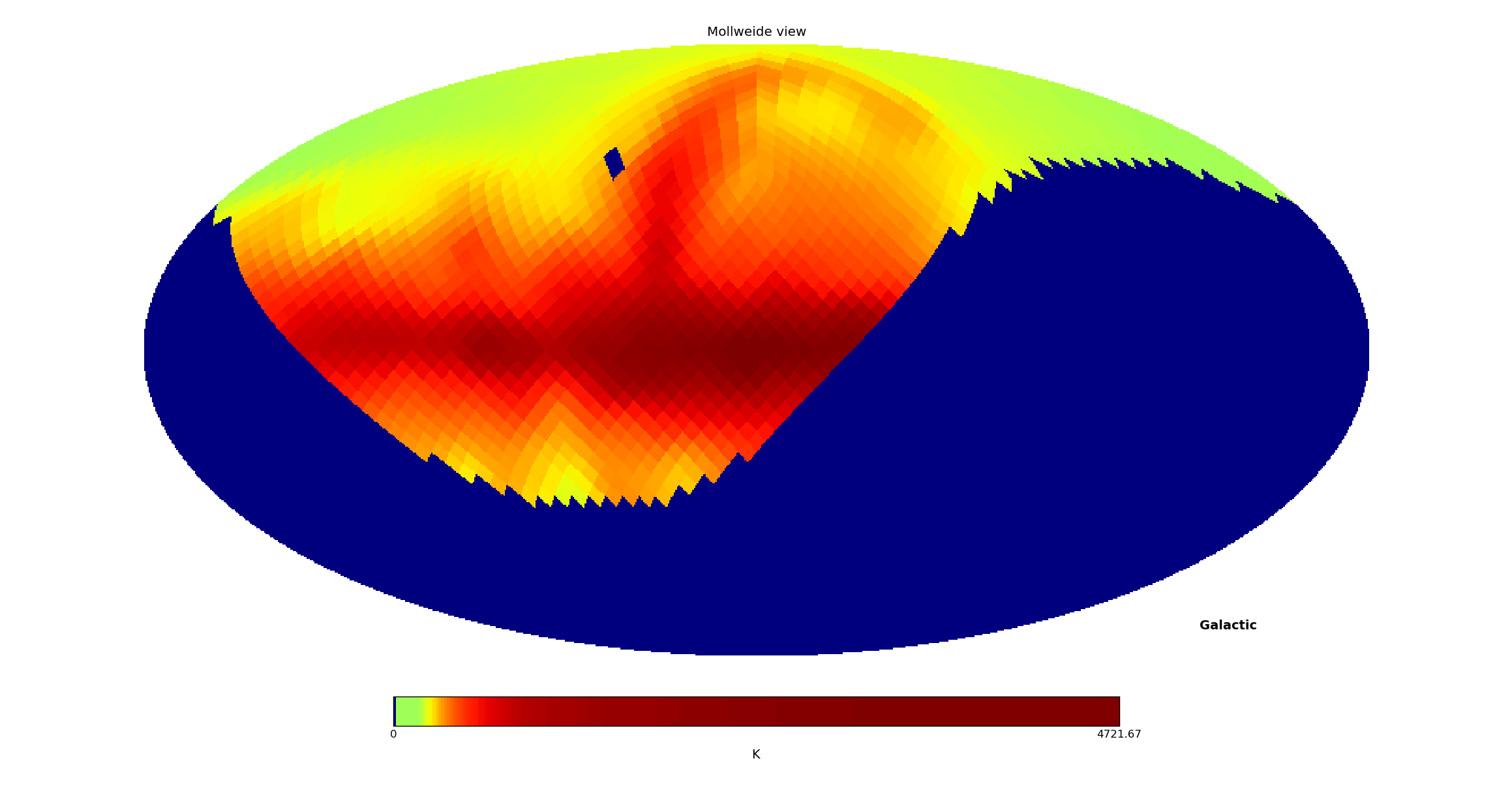}%
}
\end{minipage}
\begin{minipage}{0.3\textwidth}

\subfloat[Spectrum off Galactic plane]{%
  \includegraphics[clip,width=\columnwidth,height=2.0in]{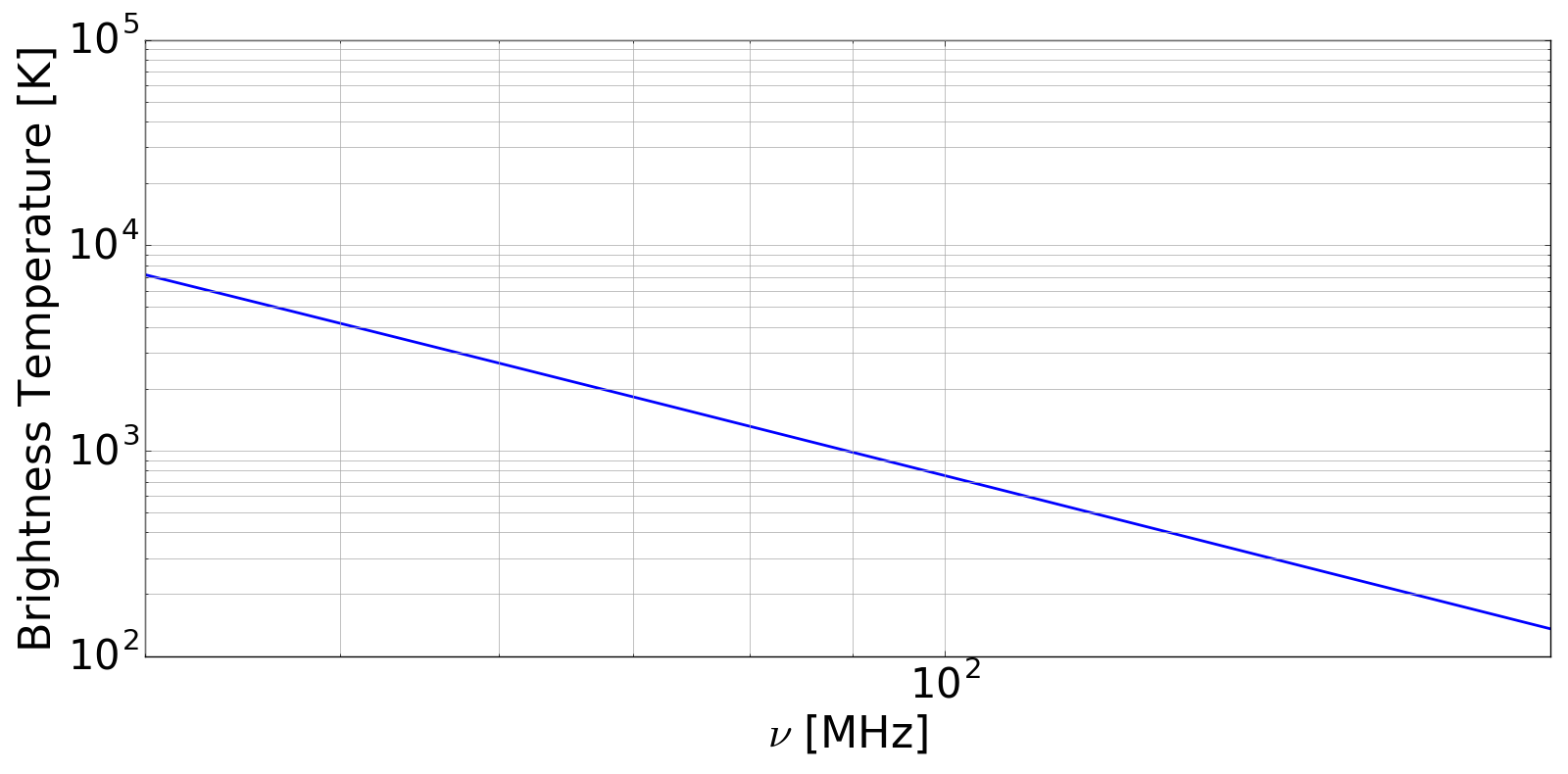}%
}

\subfloat[Spectrum on Galactic plane]{%
  \includegraphics[clip,width=\columnwidth,height=2.0in]{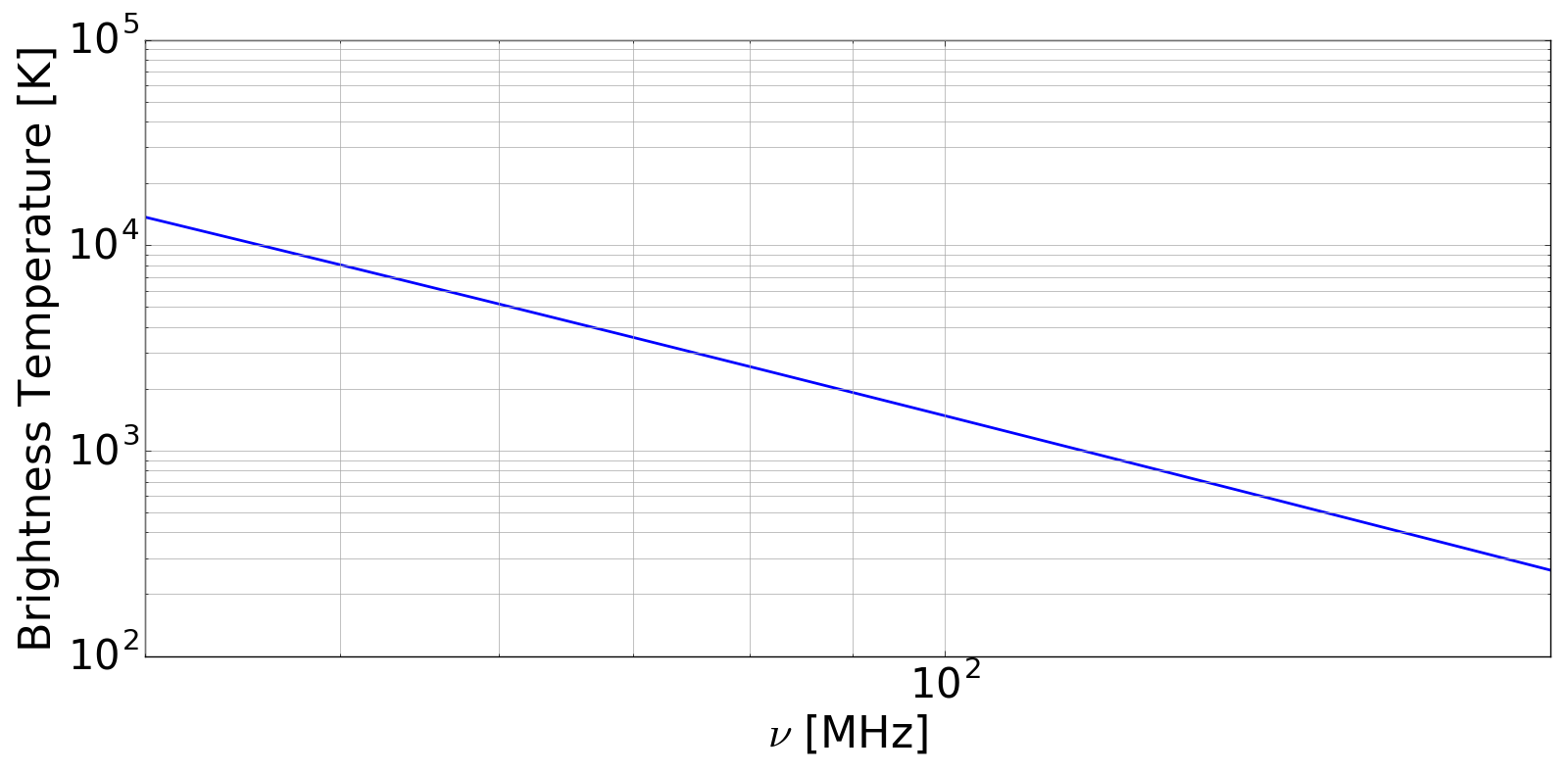}%
}

\end{minipage}
\begin{minipage}{0.3\textwidth}

\subfloat[Residual]{%
  \includegraphics[clip,width=\columnwidth,height=2.0in]{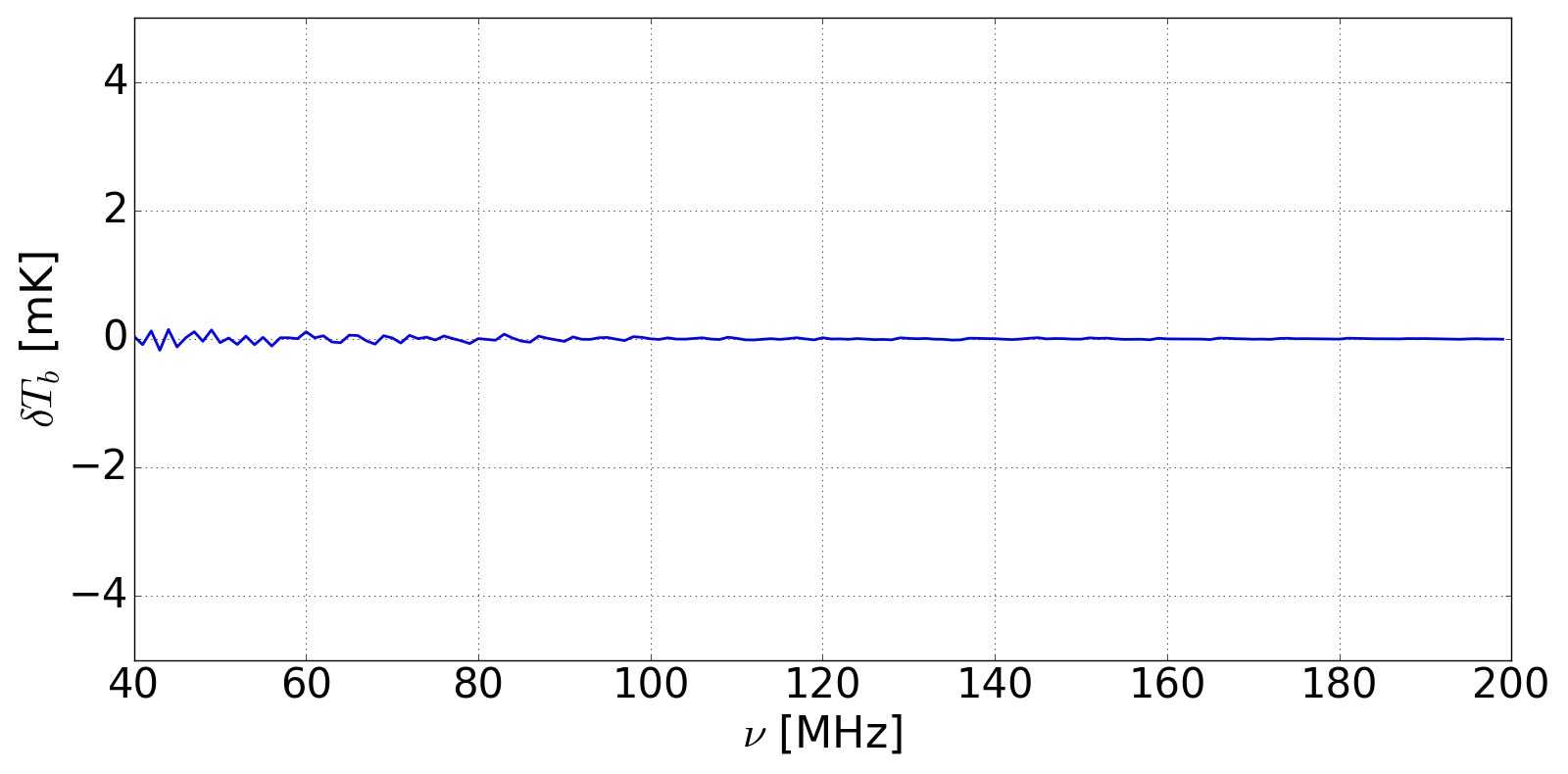}%
}

\subfloat[Residual]{%
  \includegraphics[clip,width=\columnwidth,height=2.0in]{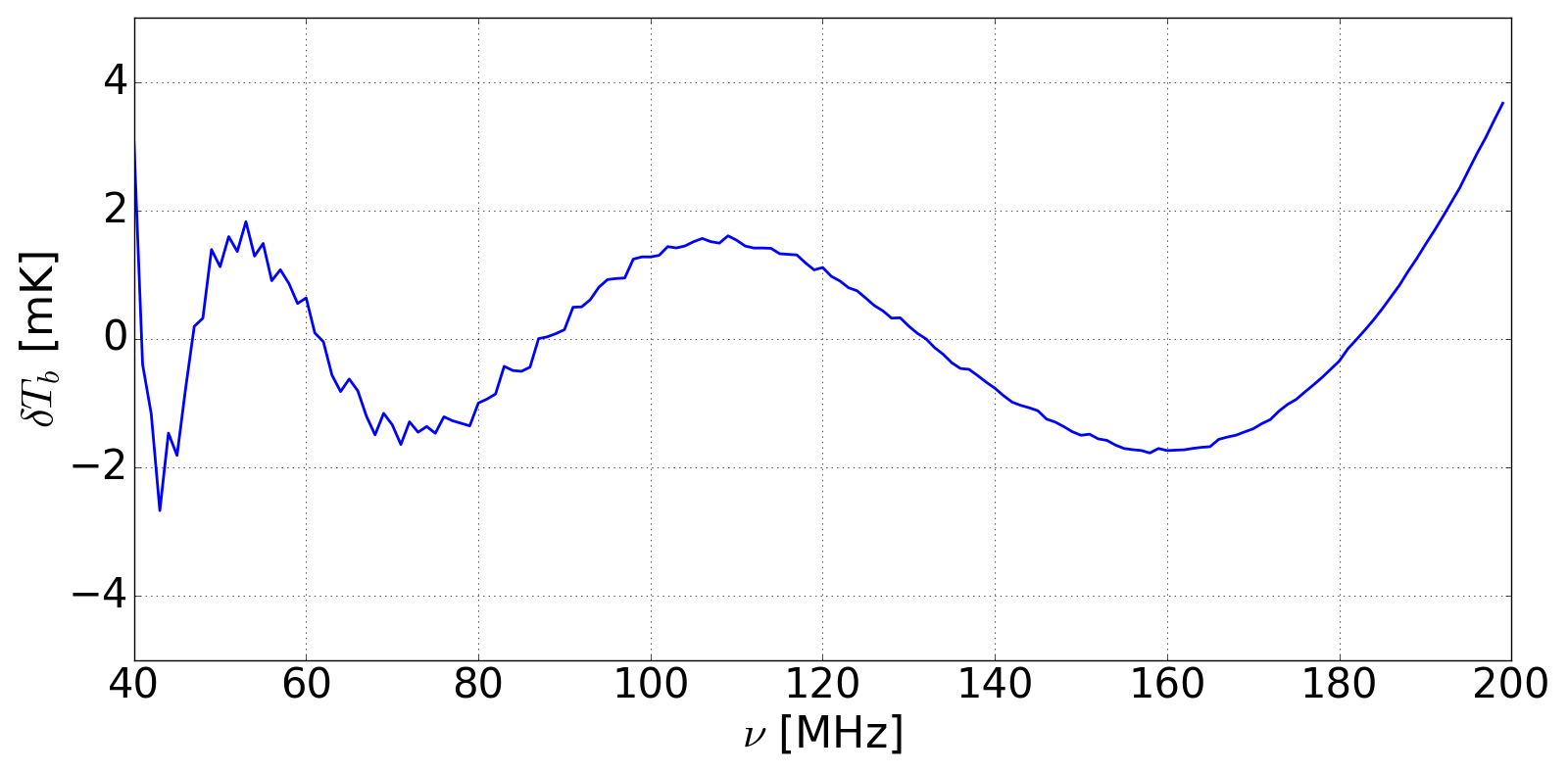}%
}
\end{minipage}
\caption[Testing the smoothness of foregrounds in mock sky spectra that are simulated using GMOSS as the sky model for the sky over Hanle (Latitude = 32.79$^{\circ}$N), as observed by the SARAS~2 short monopole antenna]{Same as Fig.~\ref{fig:cos2_Hanle}, but with beam adopted to be that of the SARAS~2 short monopole antenna.}
\label{fig:SARAS2_Hanle}
\end{sidewaysfigure}

\begin{sidewaysfigure}
\begin{minipage}{0.3\textwidth}

\subfloat[SARAS2 beam MRO off Galactic plane]{%
  \includegraphics[clip,width=\columnwidth,height=2.0in]{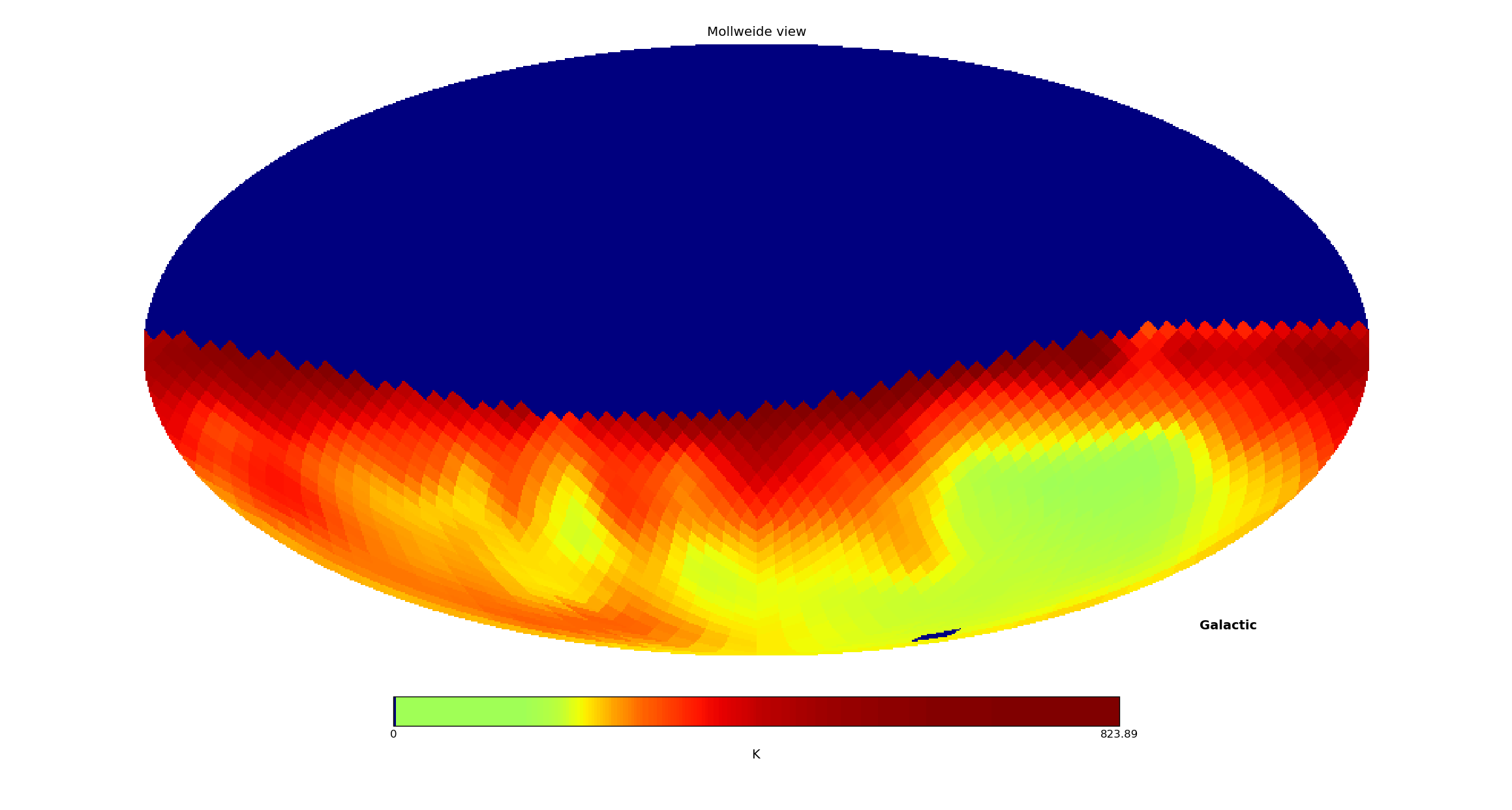}%
}

\subfloat[SARAS2 beam MRO on Galactic plane]{%
  \includegraphics[clip,width=\columnwidth,height=2.0in]{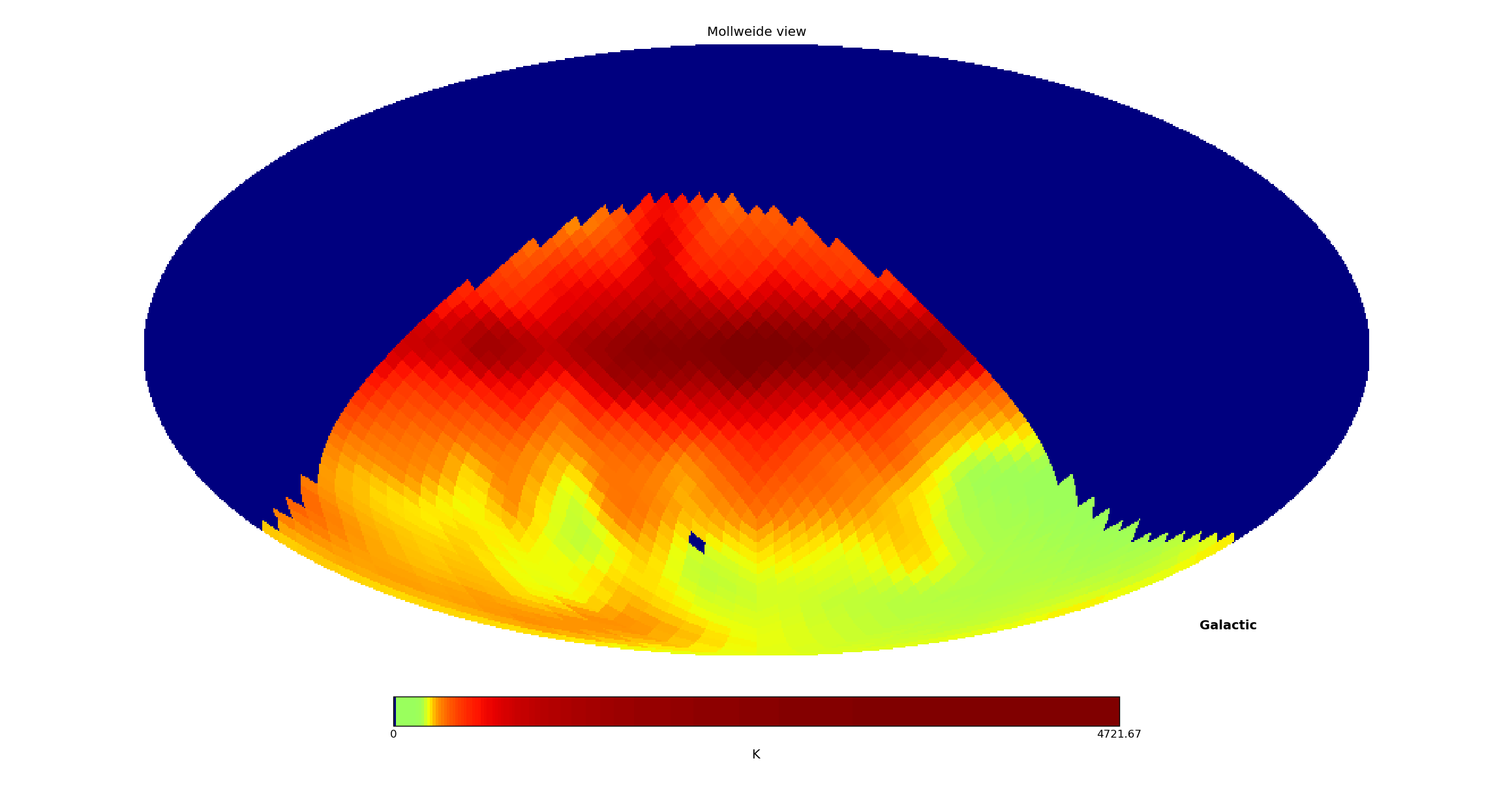}%
}
\end{minipage}
\begin{minipage}{0.3\textwidth}

\subfloat[Spectrum off Galactic plane]{%
  \includegraphics[clip,width=\columnwidth,height=2.0in]{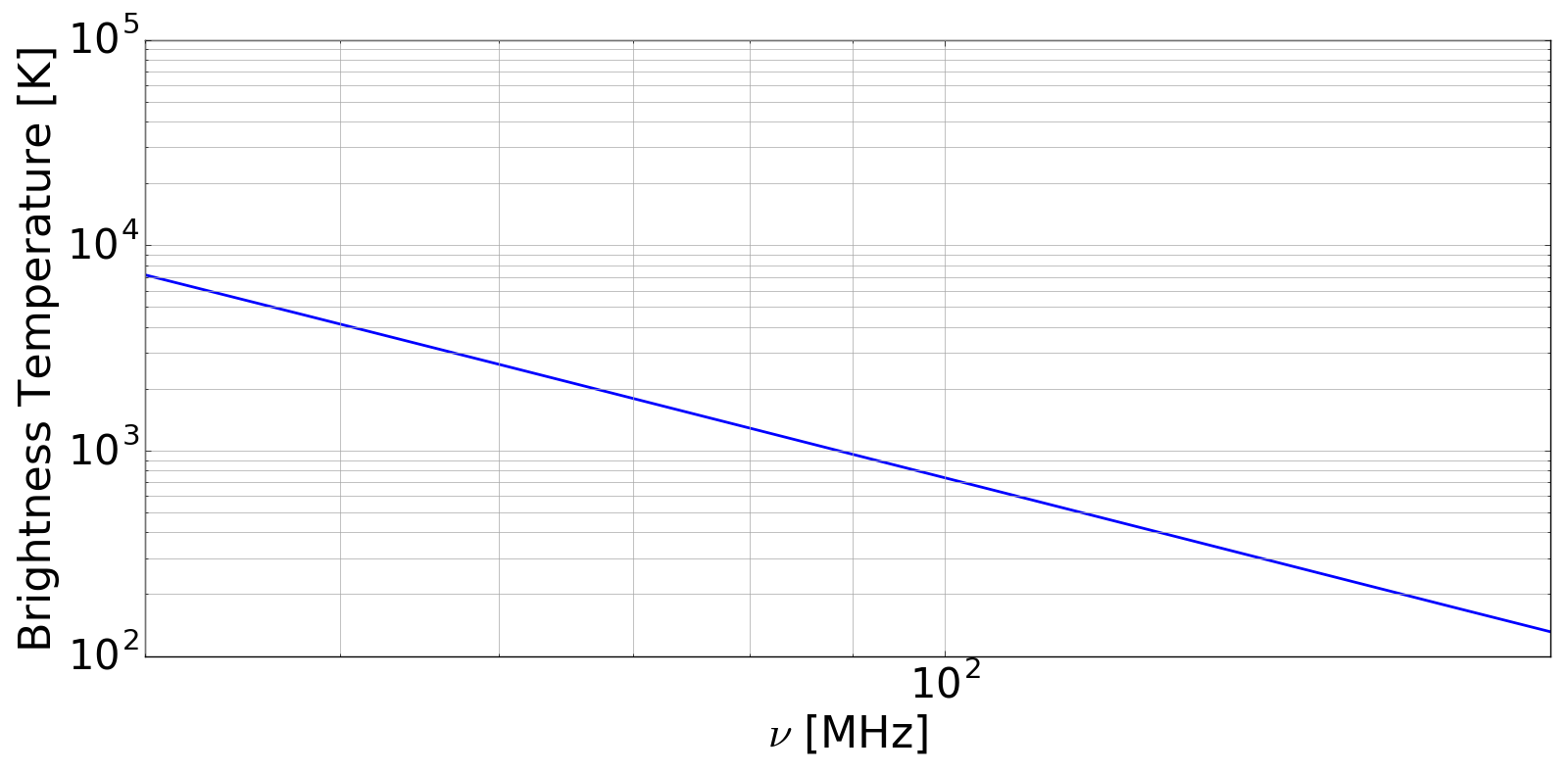}%
}

\subfloat[Spectrum on Galactic plane]{%
  \includegraphics[clip,width=\columnwidth,height=2.0in]{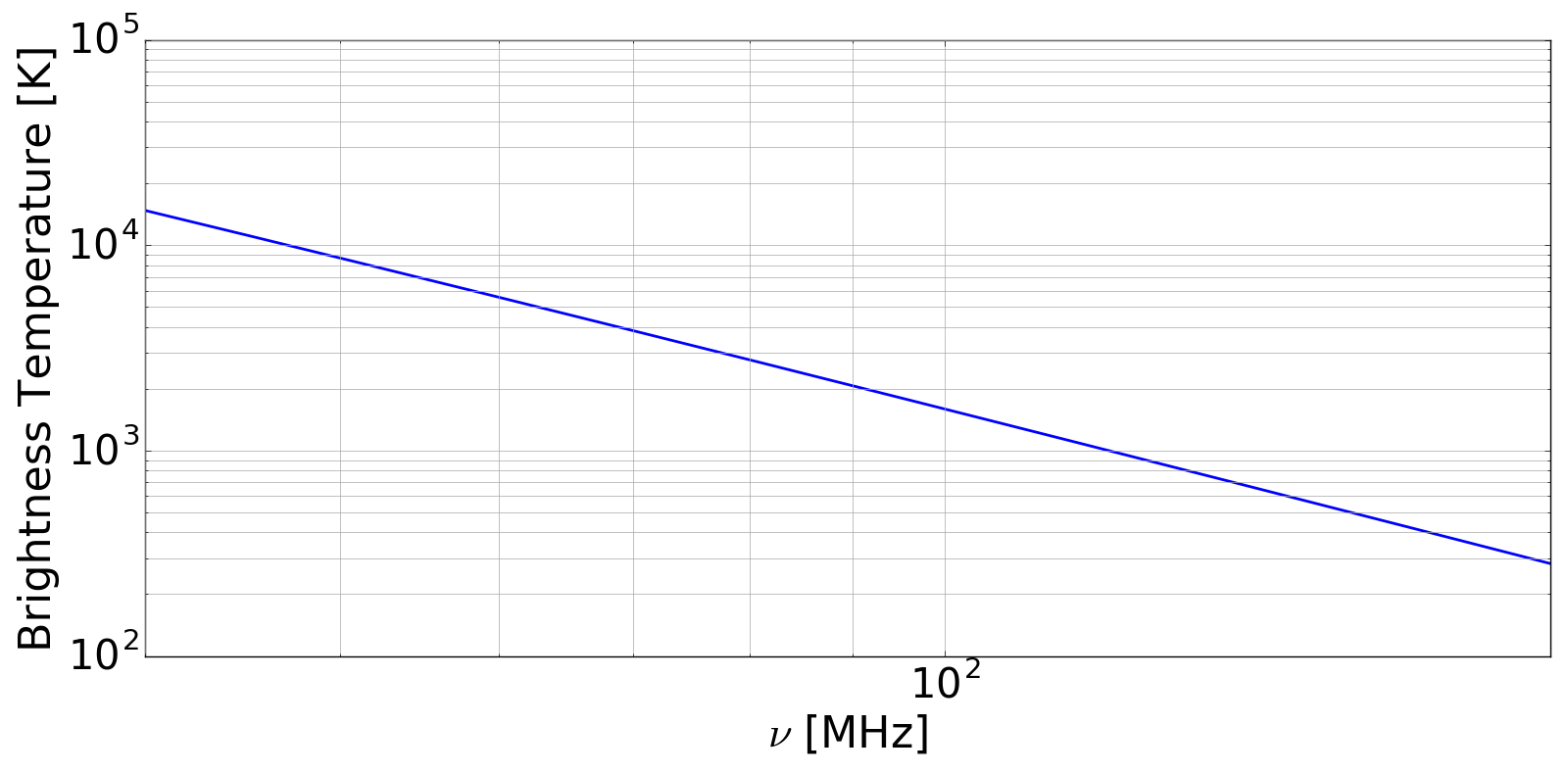}%
}

\end{minipage}
\begin{minipage}{0.3\textwidth}

\subfloat[Residual]{%
  \includegraphics[clip,width=\columnwidth,height=2.0in]{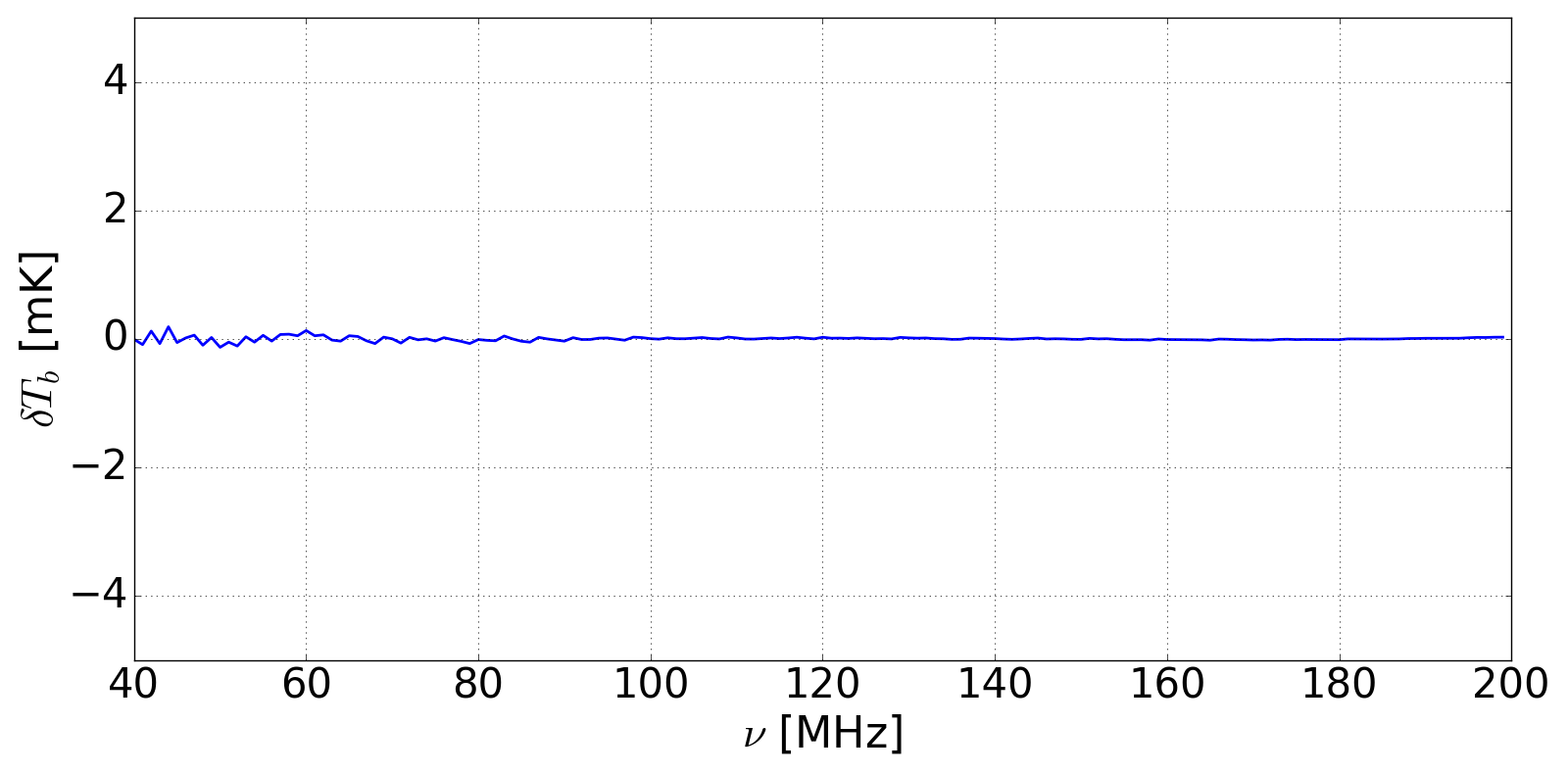}%
}\label{fig:norm2}

\subfloat[Residual]{%
  \includegraphics[clip,width=\columnwidth,height=2.0in]{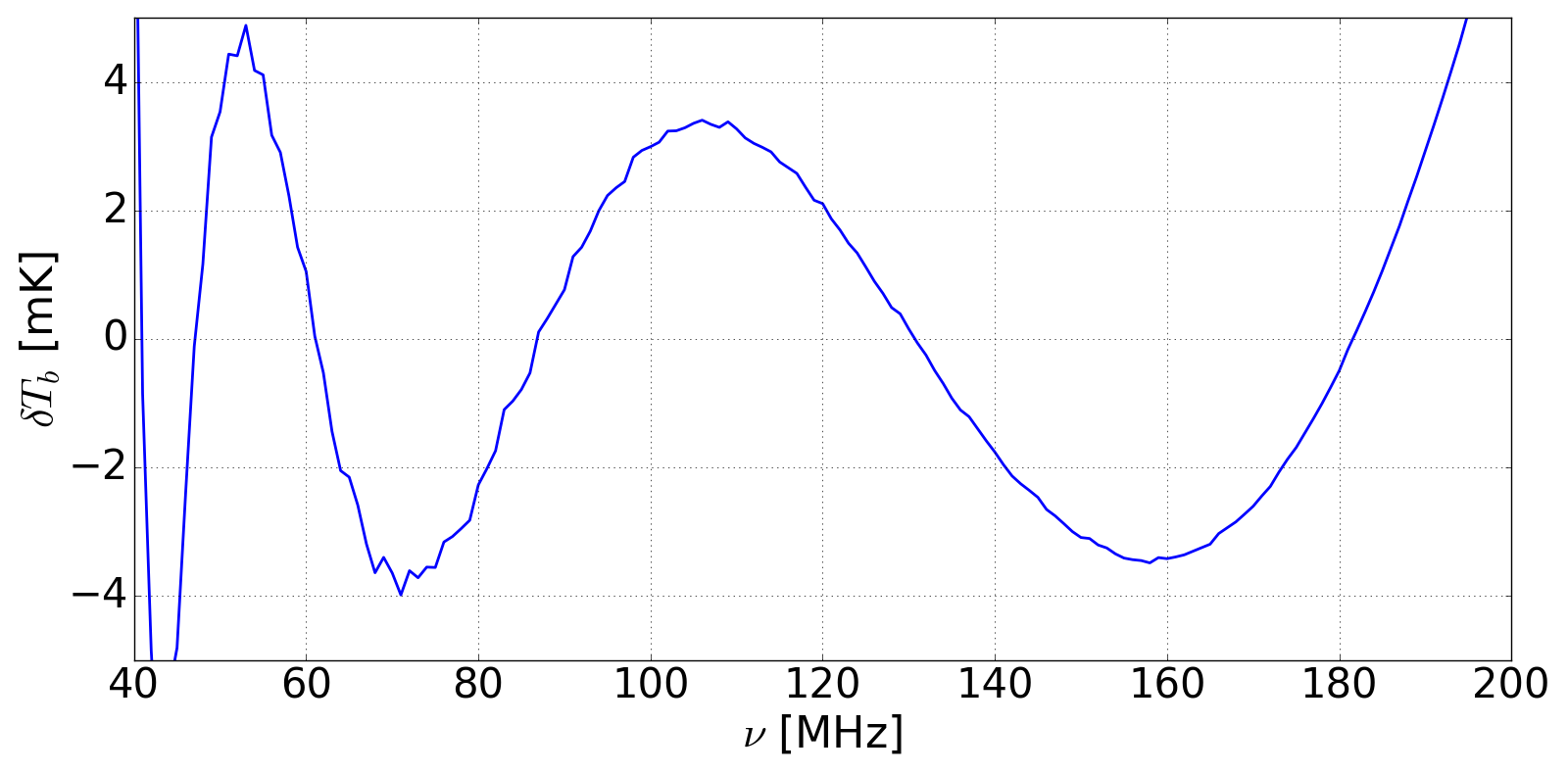}%
}
\end{minipage}
\caption[Testing the smoothness of foregrounds in mock sky spectra that are simulated using GMOSS as the sky model for the sky over MRO (Latitude = 26.68$^{\circ}$S), as observed by the SARAS~2 short monopole antenna]{Same as Fig.~\ref{fig:cos2_MRO}, but with beam adopted to be that of the SARAS~2 short monopole antenna.}
\label{fig:SARAS2_MRO}
\end{sidewaysfigure}

Though in the frequency range of interest (40--200~MHz) the difference between the Planck form and its Rayleigh-Jeans approximation is about 1 Jy~sr$^{-1}$ in specific intensity, we still explicitly fit to the CMB using its Planckian form to avoid any spectral shapes that may arise from approximately modeling the CMB as a constant in brightness temperature.\\\\ %However, CMB spectral distortions of the $y$ and $\mu$ type \citep{Hu1993, Chluba2011therm} as well as the cosmological radiation \citep[e.g.,][]{Chluba2006, Sunyaev2009} may be neglected in the frequency--sensitivity space being considered here for the detection of global EoR.
Thus the functional form used to describe the mock spectrum is 
\begin{equation}\label{eq:fit_func}
T(\nu) = \left( {\frac{h\nu}{k}} \right) / \left( {e^{\frac{h\nu}{kp_{0}}}-1} \right) + 10^{ \sum\limits_{i=0}^{N} \left[\log_{10}(\nu/p_1)\right]^i \,p_{i+2}}.
\end{equation}
Here, $p_0$ is the CMB temperature, $p_1$ is the frequency about which the expansion is centered, and $p_2$ through $p_{2+N}$ are the coefficients of the terms of the N$^{\rm th}$-order MS function that models the foreground. The remaining terms follow standard notation.\\\\
The mock spectra representing the foreground and undistorted CMB at different LSTs and for different sites and telescope beams were all uniformly fit with MS functions of order 10. Sample spectra corresponding to LSTs when the antenna temperatures are minimum and maximum, the corresponding sky coverage by the telescope beam for each of these spectra, and the residuals on fitting these spectra with the function given by above Equation~\ref{eq:fit_func}, are shown in Figures~\ref{fig:cos2_Hanle}, \ref{fig:cos2_MRO}, \ref{fig:SARAS2_Hanle} and \ref{fig:SARAS2_MRO}. The sky-coverage plots give the all-sky map at 150 MHz with a resolution of 5$^{\circ}$ such that pixels that lie outside the beam are blanked. To examine for just the smoothness of foregrounds, these mock spectra have not had the EoR signal added.\\\\ 
If the residual on fitting these spectra as the sum of a Planckian and MS function is at the level of a few mK, the foregrounds may be considered to be smooth at a level required to discern the EoR signal. This would provide impetus to use the MS function to separate foregrounds from the more complex cosmological signal. However, if the foreground spectra are themselves not smooth, the residuals in this exercise would be large, possibly larger than the EoR signal, which would indicate that there may potentially be spectral structure in the foregrounds that may limit or prohibit detection of the EoR signal, particularly if adopting signal extraction strategies that assume smooth foregrounds.
\begin{figure}[t]\centering
\subfloat[]{
\includegraphics[scale=0.3]{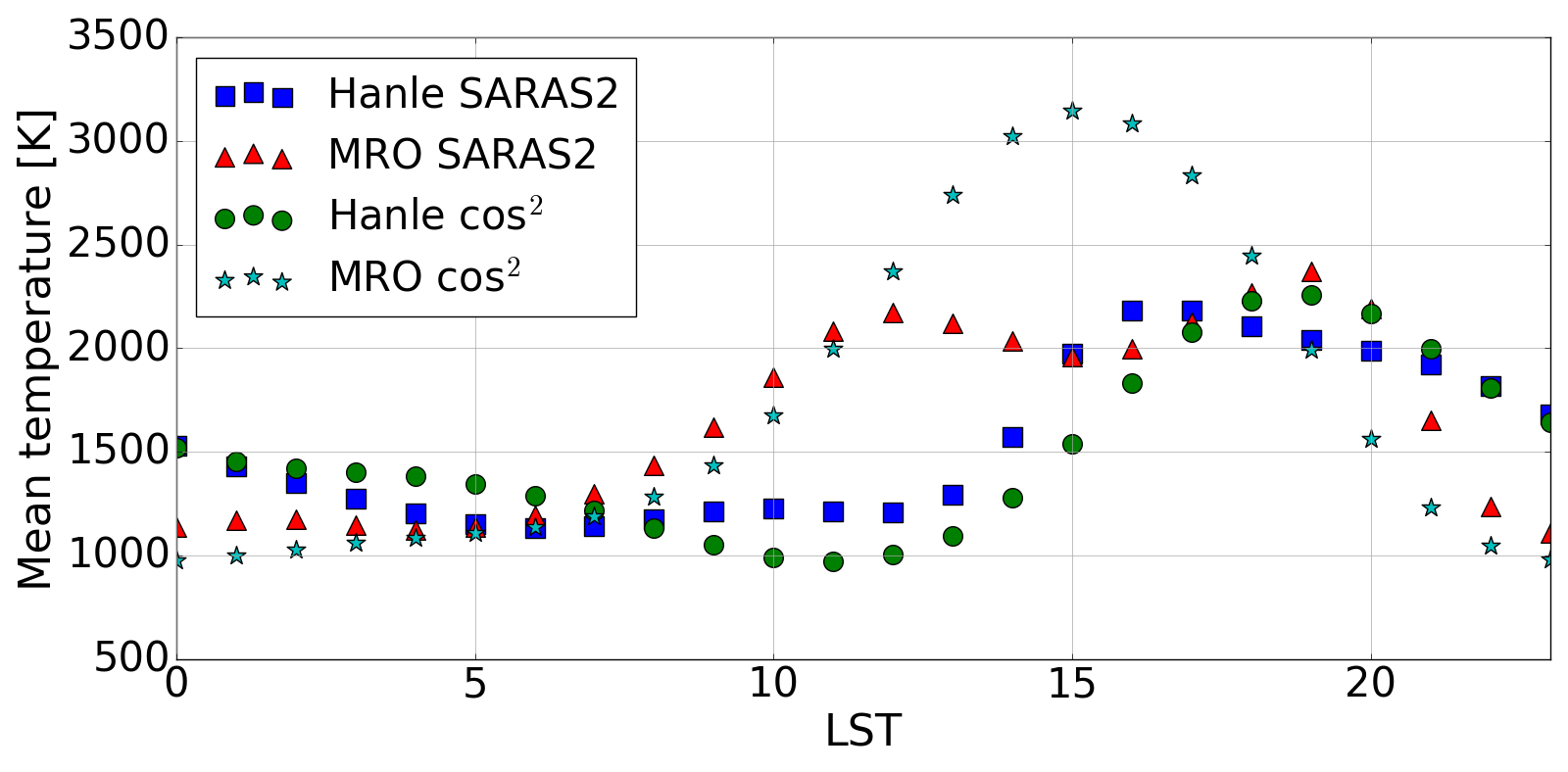}
}

\subfloat[]{
\includegraphics[scale=0.3]{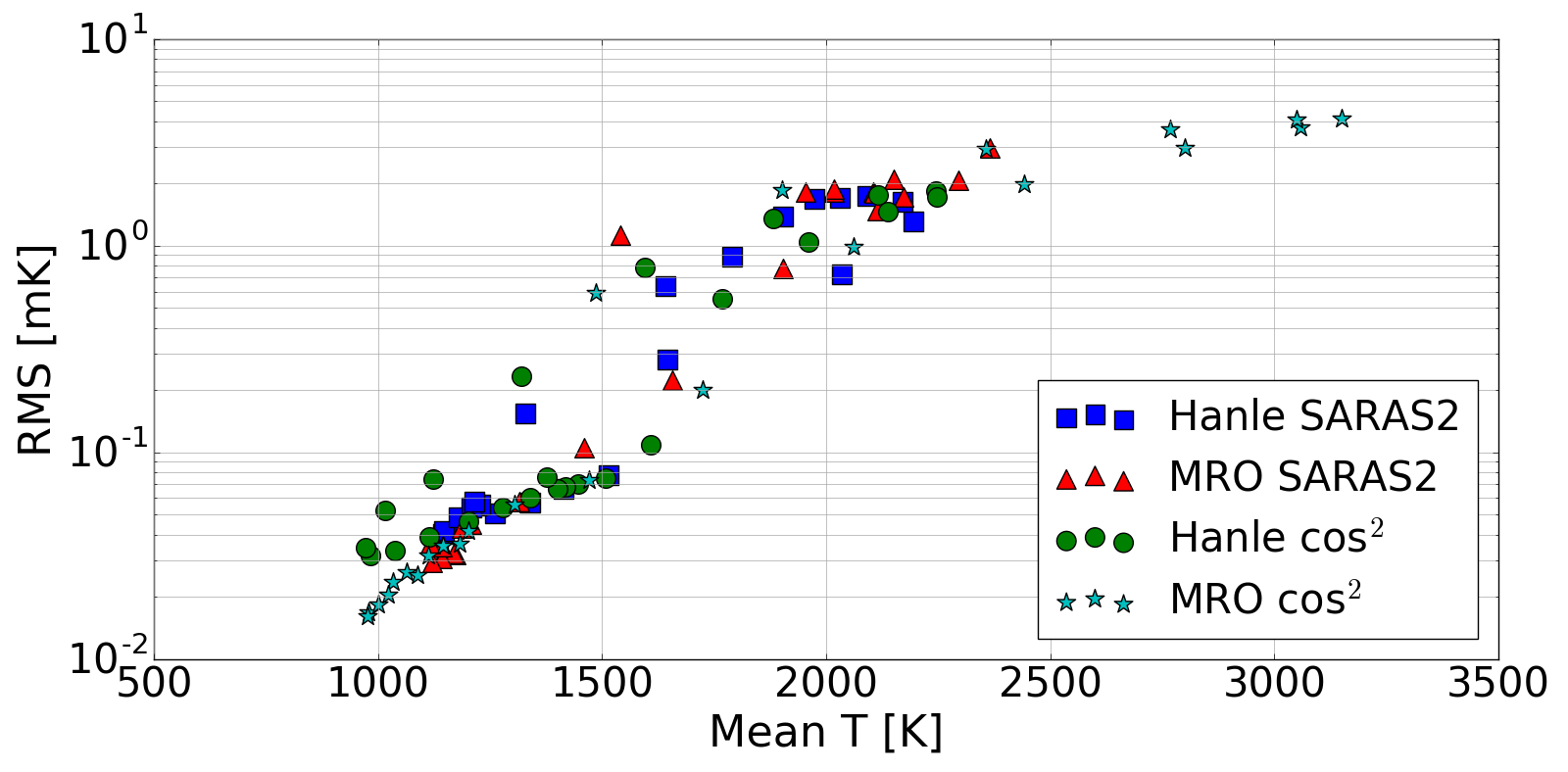}
}
\caption[Mean temperature of mock spectra versus LST and the RMS of residuals on fitting MS functions to these spectra]{Mean temperature of mock spectra versus LST is shown in panel (a).   The RMS of residuals on fitting MS functions to these spectra is shown in panel (b). The RMS residuals for spectra simulating observations with a cos$^2$ beam at Hanle and MRO are shown as filled circles and filled stars, respectively. The RMS of residuals for spectra simulating observations with the beam of the SARAS 2 antenna, at Hanle and MRO, are given by filled squares and triangles, respectively. Clearly, the RMS residuals are larger for spectra that are brighter, which are those having relatively greater contribution from the Galactic plane. A spectrum recorded at MRO by an antenna with cos$^2$ beam  would have a maximum temperature at the LST when the Galactic center is at zenith. This also corresponds to the spectrum having the largest RMS residual amongst the various mock spectra considered here.}
\label{fig:T_vs_rms}
\end{figure}
Inferences from the above exercise, whose results are presented in Figures~\ref{fig:cos2_Hanle}, \ref{fig:cos2_MRO}, \ref{fig:SARAS2_Hanle} and \ref{fig:SARAS2_MRO}, are summarized below
\begin{enumerate}
\item When mock observations of sky spectra are generated using GMOSS and made with wide frequency-independent beams and over the entire LST range, MS functions are capable of accurately modeling the shape of the foreground component to leave residuals less than 5~mK, demonstrating that the foreground spectrum is indeed smooth to such a precision.  It follows that if the mock spectrum contains the generic EoR signal, which is inherently not smooth over the 40--200~MHz wide band, the residual on fitting such a spectrum with an MS function would be expected to be dominated by the EoR signal, with a smooth baseline subtracted. The smooth foreground would be largely removed and thus separated from the EoR signal. This expectation, which is based on the above analysis, is demonstrated below in Section~\ref{sec:detection}.
\item Spectra that have relatively lower contamination from the Galactic plane and Galactic center are fit by MS functions to greater absolute precision and yield residuals that have RMS of $\sim$ 0.01~mK, which is substantially lower than the amplitude of the expected EoR signal. However, spectra at LSTs at which there is relatively large contribution from the Galactic plane and Galactic center yield residuals with RMS in the range 2--5~mK when fit with MS functions.  As discussed in \cite{GMOSS2017}, spectra toward the Galactic plane indeed have more complex spectral shapes.\\   
Figure~\ref{fig:T_vs_rms} shows the RMS of residuals obtained on fitting MS functions to model foregrounds versus the mean temperature of the corresponding spectrum. This is shown for the four sets of data corresponding to the two sites and two beams described above. The mock spectra were synthesized at separations of one hour and distributed over the entire 24~hr in LST. Clearly there exists a correlation between the mean temperature of the spectrum and the RMS of the residual. 
\item In the case of spectral measurements that are made with beams well off the Galactic center, with minimal contribution from the Galactic plane, the mean temperatures are low and the corresponding residuals have an RMS lesser than 1~mK. We find that an eighth-order MS function is sufficient to fit the foreground to a level that is lower than most predicted EoR signals. Therefore, provided observations are made at LSTs where Galactic contamination is relatively low, the foreground will be removed to relatively high precision. Toward these sky directions,  any global EoR signal in a class of models similar to the vanilla model, with multiple inflections in the band, is expected to be detectable in the residual. 
\item On the other hand, for spectra toward regions of the sky that are intrinsically brighter, such as those toward the Galactic center, the RMS of the residual on fitting MS functions is higher. The highest RMS is expected for spectra recorded at MRO with a cos$^2$ beam, when the Galactic center transits nearly overhead. This is despite fitting with an MS function of order 10 and there is no significant change in the residual on increasing the order of the MS function to even 20. However, the RMS is at worst a few mK, which implies that even toward these directions that have significantly larger brightness, generic models of the EoR signal can nevertheless be usefully distinguished from the foreground.   On the other hand, since these regions do have relatively brighter emission with more complex spectral shapes, and the fitting process is computationally challenging, it is best to avoid the Galactic plane ($\pm10^{\circ}$) and the Galactic center while attempting to detect the weak global EoR cosmological signal. 
\end{enumerate}
For various large-area samplings of the sky, we have thus demonstrated that the foreground component in a beam-averaged spectrum is adequately describable by an MS function.  Thus for observations with a frequency-independent beam, MS functions can be used to separate foregrounds from the global EoR signal.  We next discuss mathematical statistical detection based on MS-form modeling of foregrounds.   
\section{On the detection of the global reionization signal}
\label{sec:detection}
\begin{figure}[t]
\centering
\begin{minipage}[b]{0.8\textwidth}
\subfloat[]{
\includegraphics[width=\columnwidth]{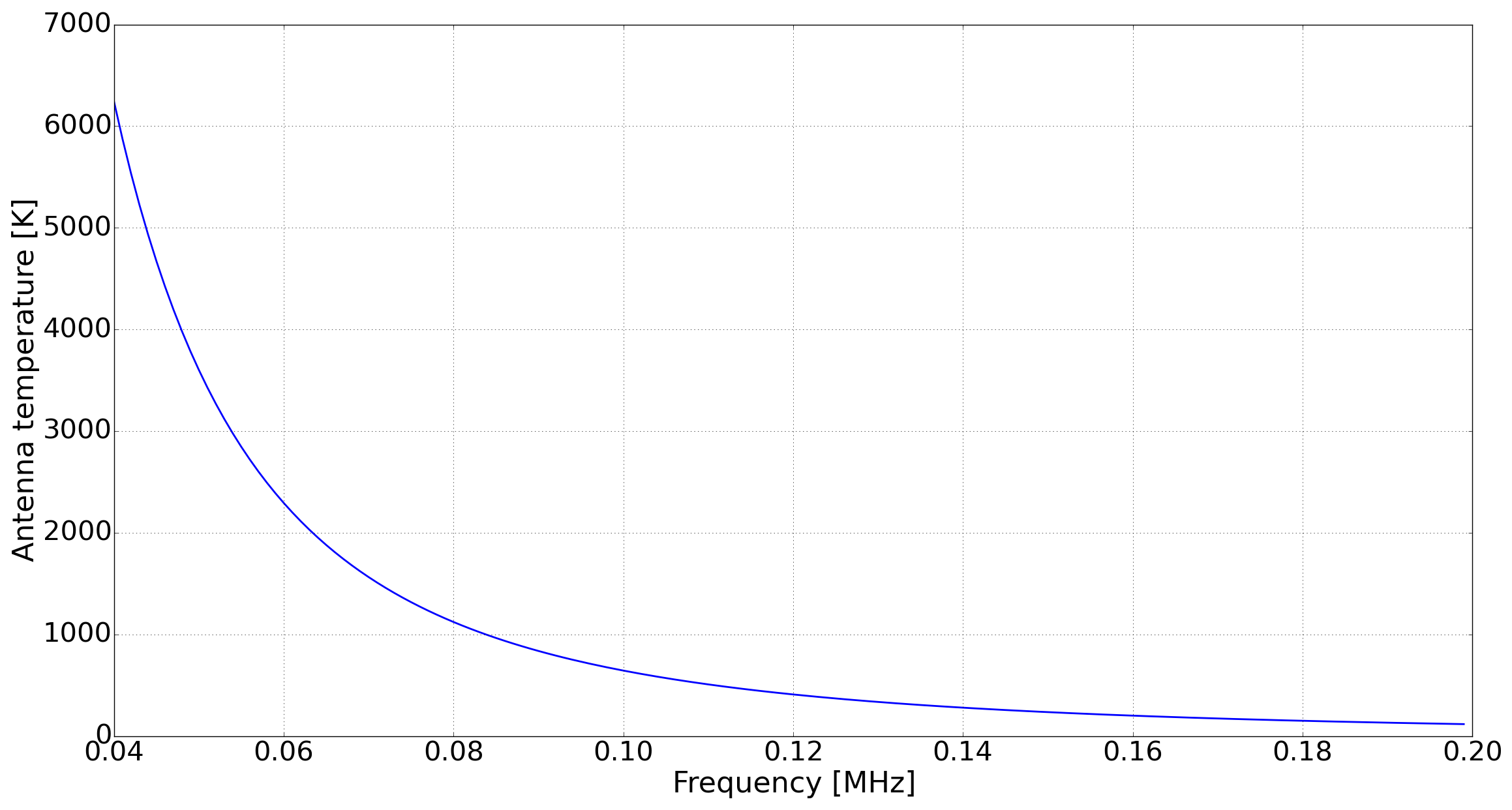} 
}
\end{minipage}

\begin{minipage}{0.8\textwidth}
\subfloat[]{
\includegraphics[width=\columnwidth]{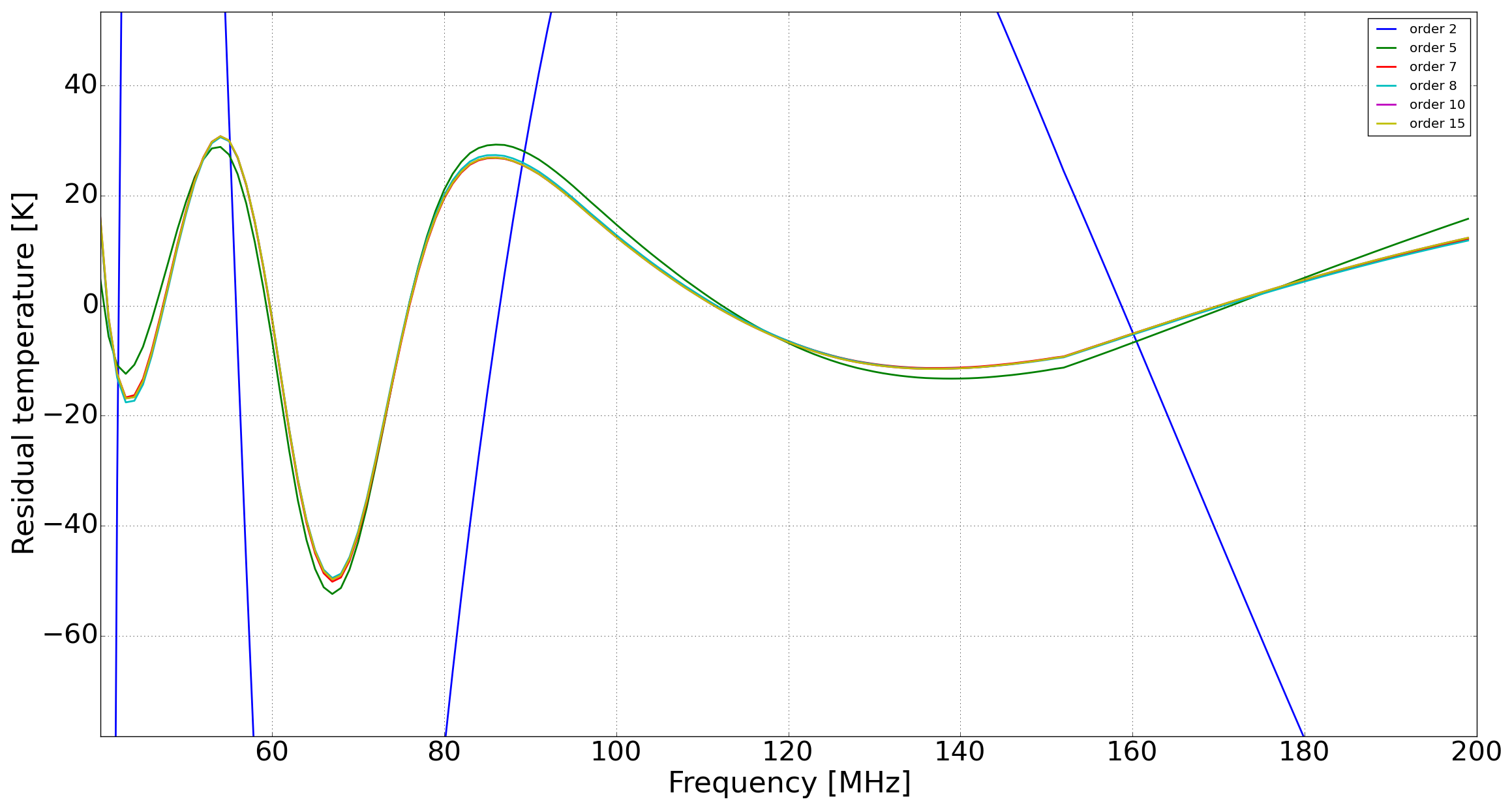} 
}
\end{minipage}
\caption[Mock spectrum generated with a frequency-independent cos$^2$ beam that observes the GMOSS sky, the CMB and the global EoR signal. Residuals obtained on fitting the spectrum with a Planck function plus MS functions of varying degrees]{(a) Mock spectrum generated on observing the GMOSS sky with a frequency-independent cos$^2$ beam. The spectrum also contains CMB with a distortion given by the vanilla model of the global EoR signal. (b) Residuals obtained on fitting the spectrum in panel (a) with a Planck function plus MS functions of degree 2, 5, 7, 8, 10, and 15. Once the MS function has fit out the bulk of the foreground and reduced its residual to well below the EoR signal,
the residual no longer changes on increasing the order of the MS function. This residual essentially retains a large part of the EoR signal, with all turning points preserved.}\label{fig:EoR_res_MS}
\end{figure}

\begin{figure}[t]
\centering
\begin{minipage}[b]{0.8\textwidth}
\subfloat[75\% confidence]{
\includegraphics[width=\columnwidth]{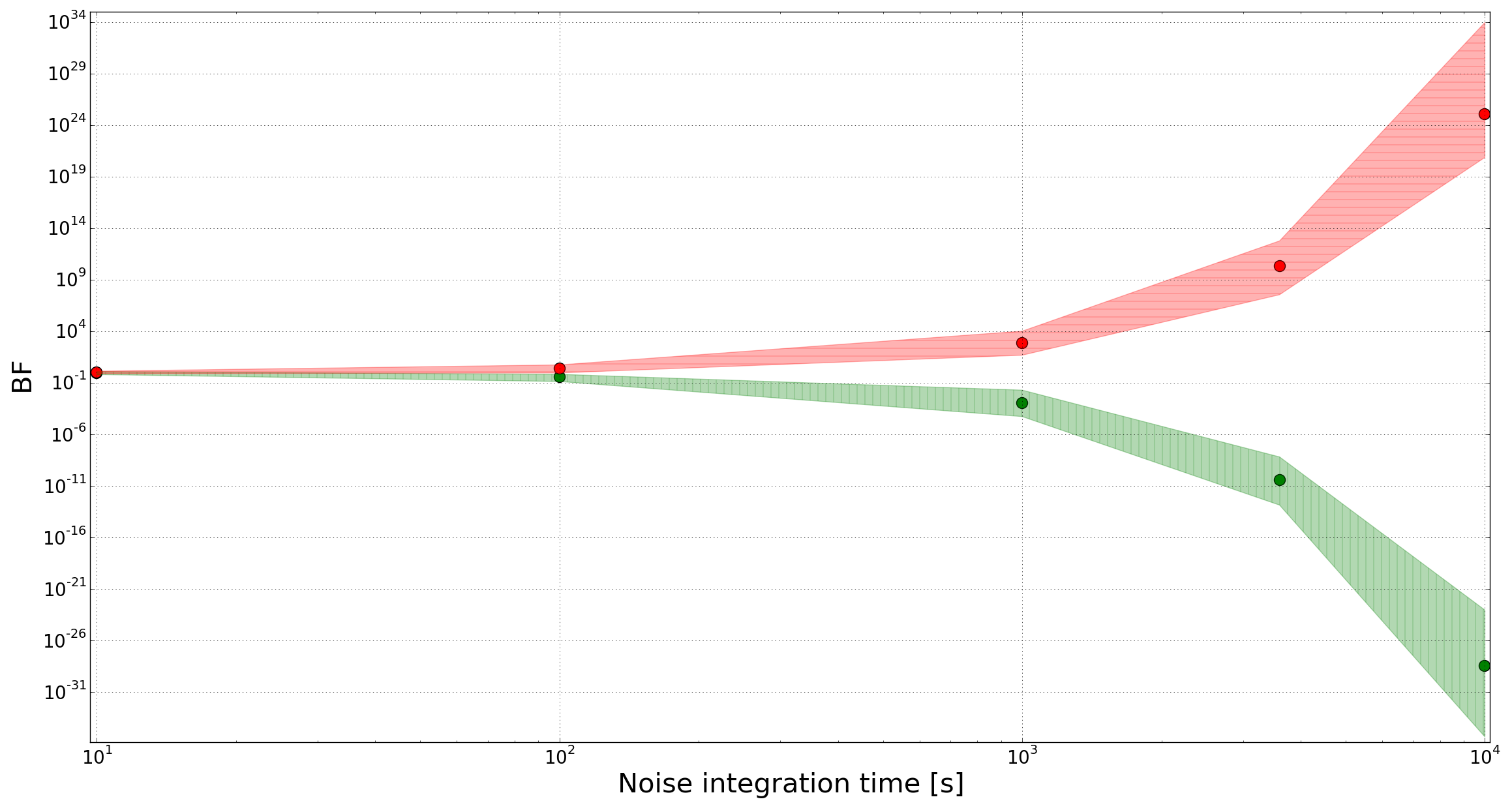} 
}
\end{minipage}
\begin{minipage}{0.8\textwidth}
\subfloat[95\% confidence]{
\includegraphics[width=\columnwidth]{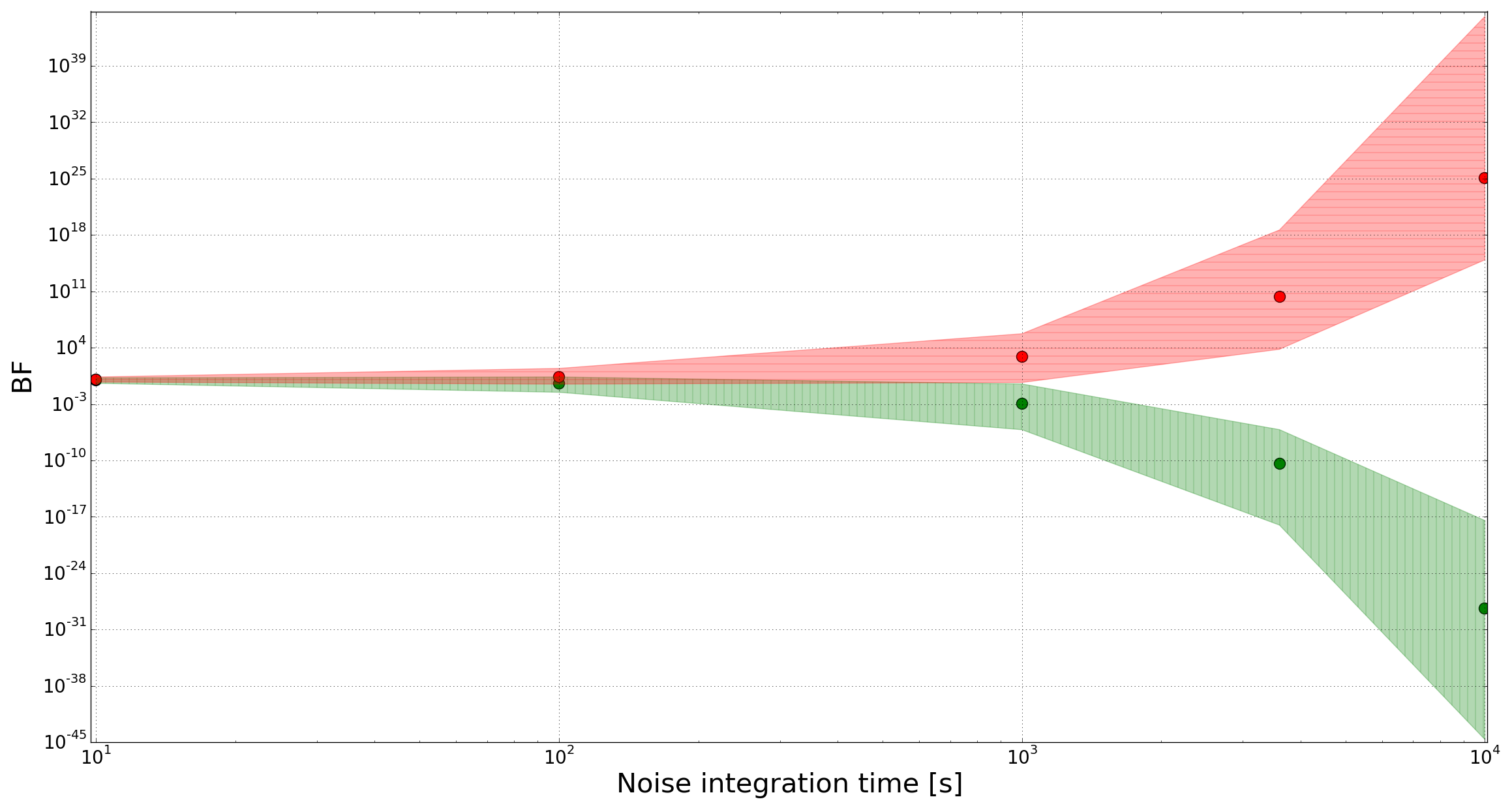} 
}
\end{minipage}
\caption[Distribution of Bayes factors versus integration time for spectra containing the EoR signal and spectra without the EoR signal, using MS functions to describe foregrounds]{Distribution of Bayes factors versus integration time for spectra containing the EoR signal (red shaded region) and spectra without the EoR signal (green shaded region).   The median Bayes factors are shown using filled circles.  The point at which the two shaded regions diverge represents the integration time required to distinguish between the presence and absence of the signal in a measurement with a certain confidence. The confidence is given by the width of the shaded regions about the median value. For detection with greater confidence, the regions diverge later, {\it i.e.}, at greater integration times.  It may be noted that the models used to describe the spectra for BF estimation employ MS functions of order eight to fit the foreground component.}
\label{fig:BF_MS}
\end{figure}

\begin{figure}[t]
\centering
\includegraphics[width=\columnwidth]{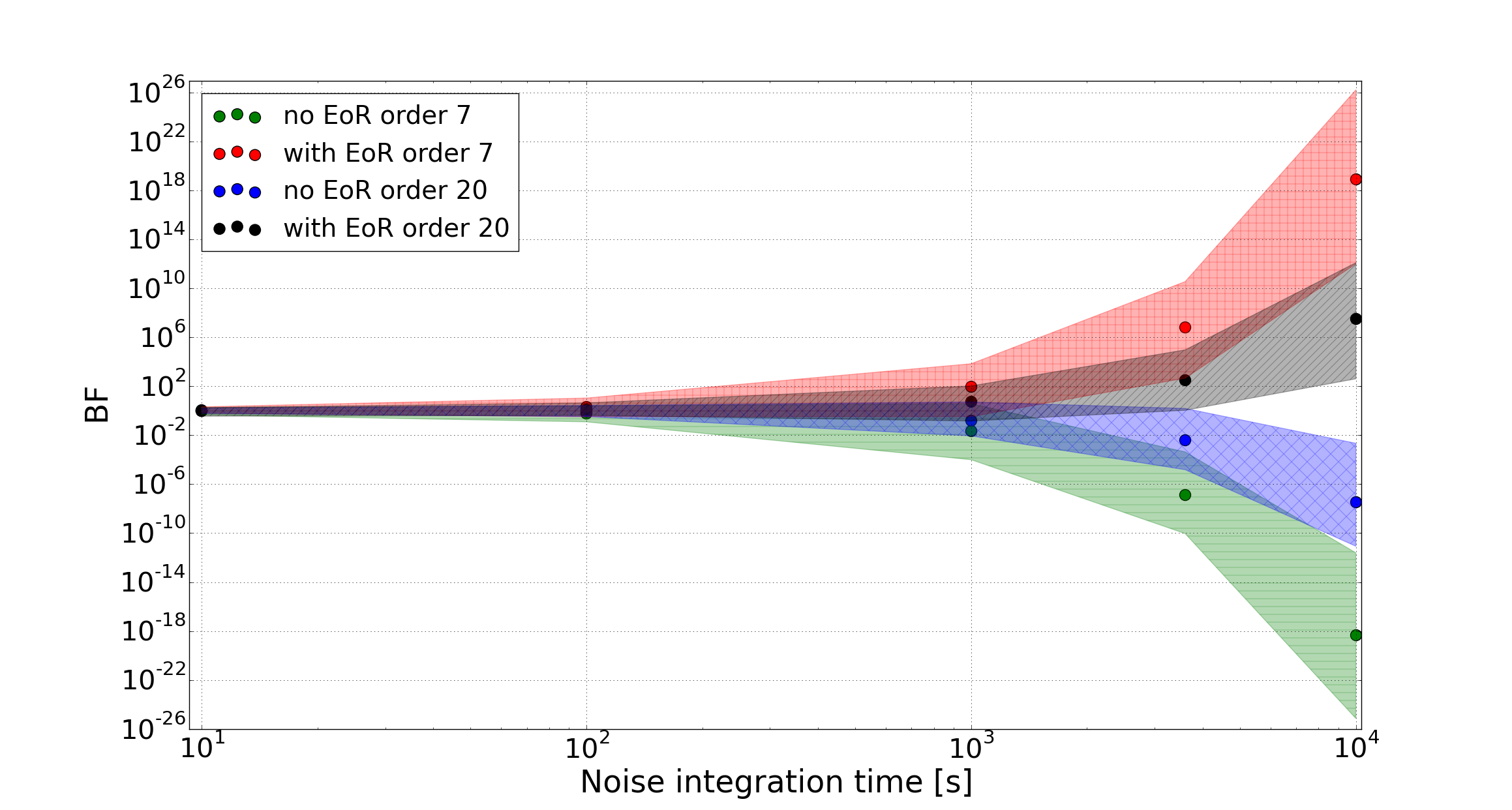} 
\caption[Distribution of Bayes factors versus integration time for spectra containing the EoR signal and spectra without the EoR signal, using polynomials to describe foregrounds]{Distribution of Bayes factors versus integration time for spectra containing the EoR signal and spectra without the EoR signal. The median Bayes factors are shown using filled circles.  The point at which the two shaded regions diverge represents the integration time required to distinguish between the presence and absence of the signal in a measurement with 95\% confidence. The red and green shaded regions represent Bayes Factor distributions on using a polynomial of order 7 to describe foregrounds. The black and blue regions denote Bayes Factors using a polynomial of order 20 to describe foregrounds. As expected, increasing the order of polynomial results in the two sets of Bayes Factors (with and without the EoR signal) diverging at larger integration times. This is because the signal-to-noise ratio degrades as more of the signal is erroneously subsumed in the foreground as the order of the fitting polynomial is increased. }
\label{fig:BF_poly7_20}
\end{figure}

A mock observation of the radio sky as observed by a cos$^2$ beam is shown in Fig.~(\ref{fig:EoR_res_MS}).   A GMOSS model for the sky has been adopted and CMB with a spectral distortion corresponding to generic global EoR has been added. We fit this mock observation of the sky spectrum with a function given by the sum of an MS function to model the foreground and a Planck spectrum to account for the undistorted CMB, as described by Equation~\ref{eq:fit_func}. We use the downhill simplex \citep{Nelder1965} optimization algorithm to iteratively fit this model to the synthetic spectrum, adopting a successive approximation strategy. The MS polynomial is expressed as a polynomial expansion about a frequency given by the parameter $p_{\rm 1}$.  The first iteration fits a function that has four parameters: the CMB temperature $p_{\rm 0}$, the pivot frequency $p_{\rm 1}$ and two coefficients that describe a single power law for the foreground. We successively include more terms using the optimized parameters from the previous iteration as initial guess and zero for the new coefficient introduced. The residuals on subtracting a model with a Planckian function and MS functions of degree 2, 5, 7, 8, 10, and 15 are also shown in Fig.~(\ref{fig:EoR_res_MS}). This may be compared to the residuals obtained above on fitting unconstrained polynomials to the same mock observation containing the EoR signal, which was shown in Fig.~(\ref{fig:EoR_res_poly}).  As noted earlier, for the case where the foregrounds were modeled as unconstrained polynomials, the residual amplitude gets progressively lower with increasing polynomial order and there are additional turning points introduced. On the other hand, while using MS functions, the residual remains unchanged once the MS function encounters a shape that cannot be described as smooth and thenceforth increasing the order of the MS function no longer changes the residual. This would continue to be the case on fitting the spectrum with MS functions of arbitrarily higher order. The residual recovered on fitting the mock spectrum with such components that may accurately model the undistorted CMB and foreground appears qualitatively similar to the adopted global EoR signal, which was included in the simulation. We now proceed to quantify the confidence in signal detection and the likelihood of false positives.
\\\\We adopt a Bayes Factor (BF) approach to quantify signal detection as described in \cite{apsera2015}. The BF is the Bayesian equivalent of the Maximum likelihood of detection in frequentist statistics. Given a dataset $D$, BF gives the more probable of two models $M_1$ and $M_2$ with parameters $\theta_1$ and $\theta_2$ respectively. BF is expressed as
\begin{equation}\label{eq:BF_def}
BF = \frac{P(D|M_1)}{P(D|M_2)} = \frac{\int P(\theta_1|M_1) P(D|\theta_1,M_1) d\theta_1}{\int P(\theta_2|M_2) P(D|\theta_2,M_2) d\theta_2}.
\end{equation}
For the case where both the models are equally likely, the ratio of priors is unity. It is assumed that it is as likely that an EoR model is present in an observation as not, and all plausible EoR models are {\it a priori} equally likely. 
If the outcome of the test is that model $M_1$ is more likely than $M_2$, then the BF is expected to be large (greater than unity) and if $M_2$ describes the data better than $M_1$, then BF is expected to be small (less than unity).\\\\
The likelihood for any model $M$ is given by
\begin{equation}\label{eq:likeli}
P(D|M) = \prod_{i=1}^{N}\frac{e^{-\frac{y_{\rm res}[i]^2}{2\sigma^2}}}{\sqrt{2\pi\sigma^2}},
\end{equation}
where $N$ is the number of independent points across the spectrum and $y_{\rm res}[i]$ is the residual spectrum following subtraction of the corresponding model from the data $D$.\\ 
The variance of the measurement noise, $\sigma^2$, is assumed to be half that estimated by differencing neighboring channels. The resulting expression is similar to the functional form used to describe the non-Gaussian phase noise in clocks, referred to as Allan Variance and is given by the relation
%the Allan Variance of the residual and is estimated from the data using the relation:
\begin{equation}\label{eq:av}
\sigma^2 = {1 \over 2(N-1)} \sum_{i=1}^{N-1}(y_{res}[i+1]-y_{res}[i])^2.
\end{equation}
Computing variance via Allen variance provides an estimate that is insensitive to residual signals that may be present following subtraction of the model.\\\\
In this work, for the purpose of demonstrating the value of MS function approach to modeling the foreground, we consider only the vanilla model for global EoR and use the BF test to compare between two models $M1$ and $M2$ that correspond, respectively, to a function describing foreground plus CMB plus EoR and a second function that models only the foreground and undistorted CMB:   
\begin{equation}\label{eq:M1}
M1:~~T(\nu) = \left( {\frac{h\nu}{k}} \right) / \left( {e^{\frac{h\nu}{kp_{0}}}-1} \right) + 10^{ \sum\limits_{i=0}^{N} \left[\log_{10}(\nu/p_1)\right]^i \,p_{i+2}} + y_{\rm t}(\nu)
\end{equation}
and
\begin{equation}\label{eq:M2}
M2:~~T(\nu) = \left( {\frac{h\nu}{k}} \right) / \left( {e^{\frac{h\nu}{kp_{0}}}-1} \right) + 10^{ \sum\limits_{i=0}^{N} \left[\log_{10}(\nu/p_1)\right]^i \,p_{i+2}}.
\end{equation}
Here, $y_{\rm t}(\nu)$ is the template of the EoR signal at frequency $\nu$ and the remaining terms are the same as in Equation~\ref{eq:fit_func}.\\\\
Consider two mock spectra that might separately represent the dataset $D$, one that contains the EoR signal and the other that does not.  For the two models $M1$ and $M2$ defined above, we would expect the BF for the spectrum that contains the EoR signal to be larger than unity and that for the spectrum that does not contain the signal to be less than unity. However, this may not necessarily be the case in the presence of noise. For large noise and poor signal to noise ratio, it is statistically possible that the particular mock spectrum that does not contain EoR erroneously yields a large value for BF.  This would lead to a mistaken conclusion that the EoR signal is indeed present in the data when there is none, thus giving a false positive detection.  On the other hand, it is also statistically possible for the spectrum containing the EoR to yield a small BF leading to the erroneous interpretation that the signal is not present in the measurement; a false negative. Both such incorrect inferences are best avoided by observing for sufficient time and attaining adequate sensitivity. An examination of the distribution of BFs for varying amounts of noise, or equivalently the integration time, can guide the interpretation of the BF by assigning appropriate confidence to outcomes of experiments.\\\\  
We generate two data sets `a' and `b' of 100 mock spectra each. While spectra in dataset `a' contain the vanilla model of the EoR signal added to them, spectra in dataset `b' do not. We also introduce Gaussian random noise in spectra of both datasets. The noise corresponds to that in a correlation spectrometer measuring a difference spectrum between the antenna temperature and reference load; we assume a scheme similar to that adopted in \cite{Patra2015}. The variance $\Delta T$ of the added Gaussian noise is given by 
\begin{equation}
\Delta T = \sqrt{ \left [ (5/4)(T_{\rm a}+T_{\rm ref})^{2} +(T_{a}+T_{ref}) (T_{\rm n1}+ T_{\rm n2}) + T_{\rm n1}T_{\rm n2} \right ] \over 2 \Delta \nu \Delta t}.
\end{equation}\label{eq:noise}
Here, $T_{\rm a}$ is the antenna temperature, $T_{\rm ref}$ is the temperature of a reference load that serves to provide a baseline against which the antenna temperature is measured,  $T_{\rm n1}$ and $ T_{\rm n2}$ give the effective amplifier noise temperatures in the two paths of the receiver that feed into the correlation spectrometer. We assume values $T_{\rm ref} = 300$~K, $T_{\rm n1} = T_{\rm n2} = 50$ K.  $T_{\rm a}$ in each channel of bandwidth $\Delta \nu = 1$ MHz is given by the noise-free antenna temperature of the mock spectrum in the corresponding channel. Spectra with varying amounts of noise are generated by varying the integration time $\Delta t$.\\\\ 
For different integration times, the simulation thus yields sets of 100 mock spectra each with and without EoR signal added, with different noise realizations.
For every spectrum from each of the two data sets we obtained two residuals by fitting models $M1$ and $M2$  given by Equations~\ref{eq:M1} and \ref{eq:M2} respectively. The corresponding likelihoods are computed using Equation~\ref{eq:likeli} and these are then used to compute the BF for the said spectrum using Equation~\ref{eq:BF_def}.  
We next determine the median and range in which the BFs for each integration time are expected to lie for a given confidence level.  For a given integration time, this range for the BFs is wider for a higher confidence level and narrower for a lower confidence level.\\ 
If the ranges of BFs for the data sets `a' and `b' for any integration time have significant overlap, then a result that is a BF from the overlap domain would be uncertain and results that are outside this overlap domain would be more conclusive.  If the detection strategy is a useful one, the BF ranges would be increasingly disjoint for decreasing noise and increasing integration time, thus giving a high confidence level of detection, high rejection of false positives, and low likelihood of false negatives.\\\\
The median and range in BFs for the pair of data sets were computed for three different integration times.  In Fig.~(\ref{fig:BF_MS}a) we show these ranges for detection with 75$\%$ confidence and delineate the divergence in the distributions with increasing integration time. The median values of the distributions are marked using filled circles about which the range is presented as a red shaded region for spectra from data set `a' and as green shaded region for spectra from data set "b." In order to represent regions with 75$\%$ confidence in this panel (a), the width is chosen to be the region in which 75$\%$ of the BF samples lie.  For greater confidence in any detection, we include a correspondingly greater fraction of samples in the delineated regions and the regions would diverge at correspondingly greater integration times.  For example, Fig.~(\ref{fig:BF_MS}b) is for detection with 95$\%$ confidence. Since increasing the MS function to arbitrarily high orders does not change the residuals, as demonstrated by Fig.~(\ref{fig:EoR_MS_poly}b), increasing the order of the MS function does not change the results presented in Fig.~(\ref{fig:BF_MS}).
The above approach that uses MS functions to model the foreground and BFs to arrive at an interpretation may be compared with the equivalent BF distribution plots for the case where the foreground component in the models M1 and M2 are described using unconstrained polynomials instead of MS functions. The order of the polynomials is fixed to be 7 as discussed in Section~\ref{sec:GMOSS}.   This is the necessary and sufficient order for polynomials to fit foregrounds that are generated using a GMOSS sky model. The resulting BF ranges for 95\% confidence when using polynomials of order 7 and 20 are shown in Fig.~(\ref{fig:BF_poly7_20}).\\\\ 
In both cases the ranges of BFs for data sets with and without the EoR signal diverge at integration times larger than those for the case when using MS functions. Firstly, fitting data with such polynomials degrades the SNR by fitting out a substantial part of the EoR signal itself along with the foreground. The functional form of the foreground spectrum is a priori unknown and using a polynomial of arbitrary order degrades the SNR. This downside is all the more severe when polynomials of high orders are adopted to describe the foreground. Clearly, the integration time required to distinguish between the presence or absence of the generic EoR signal using a BF test is larger when a polynomial of order 20 is used compared to the case when a polynomial of order 7 is used, which is in turn larger than the case when MS functions are used. Second, residuals of such fits with high-order polynomials may well mimic plausible EoR signal shapes, hence leading to degeneracy between the EoR signal and residuals produced on fitting spectra with polynomials. Thus the ability to distinguish between the presence or absence of the EoR signal is compromised by modeling the foregrounds as polynomials of high order as opposed to the more robust method of using MS functions to model foregrounds. 
The integration time beyond which the ranges for the BFs with and without EoR signal cease to overlap is the integration time required for detection of EoR with that confidence. If we assume that we use the BF method to detect the presence of the vanilla model of global EoR, and compare the two hypotheses corresponding to the presence of the vanilla model versus absence of any EoR signature, we may examine the confidence level at which the BF distributions diverge versus integration time. The resulting curve is shown as a solid line in Fig.~(\ref{fig:conf_int}).  We see that using a BF test with MS functions to model foregrounds, it is possible to detect the presence of the EoR signal with 95\% confidence in 10 minutes effective integration time using a correlation spectrometer with system parameters as described above. It may be noted here that this result may likely apply to other models that are similar in nature to the vanilla model of the EoR signal, which has multiple turning points in the frequency range of 40--200~MHz, and has a shape that is distinct and distinguishable from smooth foregrounds using MS functions.\\\\ 
The BF test proposed here considers a statistical detection of the signal over the entire bandwidth used, in this case 40-200 MHz. In other words, the BF test represents a cumulative likelihood for detection considering the entire observing band and all spectral channels in unison. This results in an observing time that is much smaller than estimated using the radiometer equation over the resolution bandwidth of the spectrometer used (1 MHz); the relevant bandwidth is the effective bandwidth of the signal and depends on the precise signal shape or distribution over the band.
\begin{figure}[ht]
\centering
\includegraphics[width=\columnwidth]{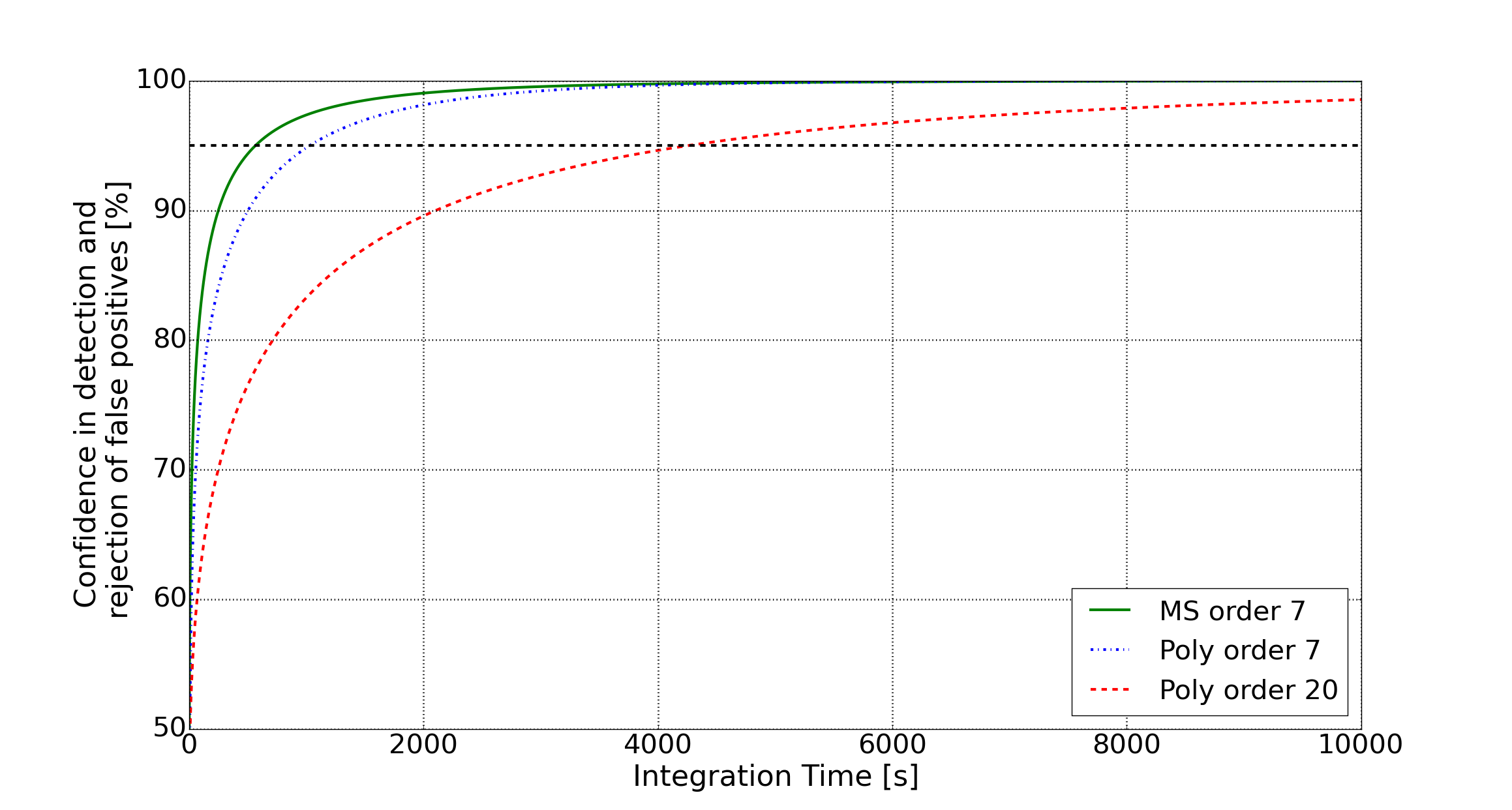} 
\caption[Confidence in detection of a signal using the BF test versus integration time]{The confidence in detection of a signal using the BF test versus integration time. The BF test that employs models using MS function to fit to the foreground component in spectra is shown as a solid green line. For noise as given by Equation~(\ref{eq:noise}), corresponding to a correlation spectrometer scheme described in \cite{Patra2015}, an effective integration time of 0.16 hours or 10 minutes results in a 95\% confidence detection of the vanilla model global EoR signal, along with the rejection of false positives with the same confidence. Also shown are confidence in detection using the BF test for models that use polynomials of order 7 (blue dashed-dotted line) and order 20 (dashed red line) to model the foreground component in spectra.  As seen from the figure, increasing the order of the polynomial used results in a larger integration time required to reach the same level of confidence; we interpret this as being due to the degradation in the signal-to-noise ratio since more of the global EoR signal is subsumed in the foreground model. For a polynomial of order 7, an integration time of 18 minutes results in a 95\% confidence detection and also confidence in rejection of false positives. For the case of an order 20 polynomial, this increases to 72 minutes. In both cases the integration time required is larger than when using MS functions to model the foreground.}
\label{fig:conf_int}
\end{figure}
For the corresponding case of modeling of foregrounds using polynomials of order 7 and 20, the confidence in detection of signals and in rejection of false positives are given as a dashed-dotted line and dashed line, respectively in Fig.~(\ref{fig:conf_int}). We see that if a BF test is done using a polynomial of order 7, an effective integration time of at least 18 minutes is required to detect the presence of the EoR signal with 95\% confidence, which is double the time required when using MS functions. This number increases to 72 minutes when using a polynomial of order 20. Thus, choosing a polynomial of different orders can result in different confidence in detection for the same integration time with the same instrument. The interpretation as signifying either the presence or absence of the EoR signal in a measurement set can critically depend on the order of the polynomials used to describe foregrounds or system parameters, unless supported by other methods to quantify signal detection. Additionally and more importantly, MS functions are robust in that even with increasing order they do not subsume EoR models with inflections; therefore, it is preferable to use MS functions to model foregrounds and distinguish the cosmological 21-cm signal.
\section{Conclusions and Summary}
Toward detection of the global redshifted 21-cm signal from reionization, we have done simulations of spectral radiometer observations in the 40--200~MHz band with wide-field antennas. The specific aim has been to study the spectral shapes expected for the foreground, whether these might confuse or limit detection of the expected wide-band cosmological signatures. This work is an improvement over previous studies in that we examined the efficacy of modeling the foreground as MS functions for the case of plausible spectral shapes as given by the physically motivated sky model GMOSS.  The new approach to global EoR detection has been compared with previous modeling of foregrounds that used polynomials in log($T$) versus log($\nu$) space.  Mock data corresponding to contrasting beam shapes, site latitudes, and the full range of observing sidereal times have been considered. In this work we do not consider complications of instrument systematics such as residual errors in bandpass calibration and mode coupling of sky structure to spectral confusion via a frequency dependent beam.\\\\ 
The work leads to the conclusion that wide-band wide-field sky spectra as measured by well-calibrated radiometers attached to frequency-independent antennas are expected to be MS to the precision necessary for detection of vanilla or generic predictions for the global EoR signal.  We have demonstrated that modeling foregrounds using Maximally Smooth functions is advantageous for the detection of global Epoch of Reionization signals that have multiple turning points in the observing frequency band.  The MS function can selectively fit to any smooth foreground and CMB with arbitrary precision by allowing for a sufficiently large order for the fitting function, despite the high order such a modeling would leave a residual that contains most of the global EoR signal. Most importantly, the residual following MS function fitting preserves turning points that define the critical epochs in the cosmological evolution of the baryons during First Light or the Cosmic Dawn and reionization.\\\\ 
The MS-function approach to modeling the foreground has been demonstrated to be amenable to a successive approximation type of solution for the model parameters.  The MS-function coefficients may be solved for iteratively by increasing the polynomial order successively while progressively approximating the foreground with greater accuracy.  There is no loss of signal in the residual if the order of the MS function were increased arbitrarily and hence the approximation may be continued to degrees well beyond what is needed and until the higher order fitting coefficients are found to be negligible.  The amplitude of the EoR features is thus largely recovered and, unlike in the case of the hitherto adopted polynomial fitting approach, the amplitude is undiminished with increasing order of the fitting function.\\\\ 
The MS-function modeling approach suggested herein may be viewed as a filter that progressively and in successive approximations removes or filters out the smooth component of the data while leaving behind the non-smooth component, for increasing orders of the MS function.  After having filtered out all the smooth components of the signal, further increasing the order of the MS function no longer affects the signal, resulting in a "saturated residual." The demonstrated success in reducing residual to below EoR signal detection levels by fitting foregrounds with MS functions, and the success in accurately modeling the foregrounds for mock observations corresponding to different sky directions, site latitudes, telescope beam types, and observing LSTs demonstrates the robustness of the MS-function approach as opposed to simple polynomial fits.\\\\ 
Our work that uses the physically motivated GMOSS sky model suggests that the foreground is indeed MS to the required precision for detection of global EoR.  
This also suggests a path forward for further study involving system design where bandpass distortions arising from internal reflections are constrained to be `smooth' and representable by MS functions. Nevertheless, we conclude by briefly discussing the consequence of non-smooth foreground components, which may manifest as spurious residuals following MS modeling of spectral data, particularly in observations made toward the Galactic Center and plane.  If the foreground spectrum is itself not smooth at a level above the EoR signal, fitting the sky spectrum with an MS function will fail to fit the non-smooth part of the foreground, and the EoR signal will appear confused in this foreground residual.   This consideration leads to the question that if an observation of the sky spectrum is fit with an MS function of a high order, and if the residual is not diminished any further by increasing the order of the MS function, then how do we distinguish between residuals that are  EoR signals from any non-smooth foreground?  A solution to this problem may be found in joint analysis of sky spectra observed separated in LST so that they contain different samplings of the sky, and hence different foreground spectra. Saturated residuals that differ between different regions of the sky is indicative of spectral structure in foreground spectra and a changing residual may be used to ``discover foreground components" and hence separate this non-smooth part of the foreground from global EoR \citep[]{Switzer2014}. However, if the saturated residuals toward different regions of the sky after fitting with MS functions of arbitrarily high order are the same in structure and amplitude, it would be highly suggestive that the residual is global in nature, making a case for EoR signal detection. Comparing the residual with smooth baseline subtracted templates, which may be obtained by subtracting MS functions from predictions for global EoR signals in various models, would provide the ability to distinguish between EoR scenarios.\\\\ 
We have presented a BF approach to estimate the integration time necessary to detect with varying levels of confidence the presence of global EoR signal in sky spectra, when MS functions are used to model the foreground.  We have also compared with the case where polynomials are used to model the foreground.  We have found that a fit to recorded sky spectra using an MS function provides the advantage of signal detection with improved signal-to-noise ratio as compared to the polynomial case.  
A receiver noise temperature of $\sim 50~$K, as has been assumed in our analysis above, is achievable. However, making a frequency-independent antenna over multi-octave bandwidths is non-trivial. Nevertheless, with such a well designed antenna and spectral radiometer, our analysis indicates that with about 10 minute effective integration the global EoR signal may be detected with $95\%$ confidence using MS functions. 
\small
\section{Acknowledgements}
J.C. is supported by the Royal Society as a Royal Society University Research Fellow at the University of Cambridge, U.K.
\bibliographystyle{apj}
\bibliography{MS_for_EoR}
\end{document}